\documentclass[manuscript]{aastex}

\pdfoutput=1

\usepackage{calc,longtable,pdflscape,multirow,graphicx,morefloats,url,apjfonts,color,amsmath,textcomp,caption,subcaption,alphalph,subscript,gensymb}
\def\littleprime{\ifmmode{\scriptscriptstyle \prime }
    \else{\hbox{$\scriptscriptstyle \prime$ }}\fi}
\def\littlecirc{\ifmmode{\scriptscriptstyle \circ }
    \else{\hbox{$\scriptscriptstyle \circ $ }}\fi}
\def\littless{\ifmmode{\scriptscriptstyle s }
    \else{\hbox{$\scriptscriptstyle s $ }}\fi}
\def\arcss{\raise .9ex \hbox{\littless}}
\def\arcsec{\raise .9ex \hbox{\littleprime\hskip-3pt\littleprime}}
\def\arcmin{\raise .9ex \hbox{\littleprime}}
\def\degre{\raise .9ex \hbox{\littlecirc}}
\def\arcsspoint{\hbox to 1pt{}\rlap{\arcss}.\hbox to 2pt{}}
\def\arcsecpoint{\hbox to 1pt{}\rlap{\arcsec}.\hbox to 2pt{}}
\def\arcminpoint{\hbox to 1pt{}\rlap{\arcmin}.\hbox to 2pt{}}
\def\degrepoint{\hbox to 1pt{}\rlap{\degre}.\hbox to 2pt{}}
\def\lsim{\setbox3=\hbox{$\sim$}\setbox4=\hbox{$<$}
           \mathrel{\hbox{\lower2pt\hbox{$\sim$}\kern-0.5\wd3
                           \kern-0.5\wd4\raise2pt\hbox{$<$}}}}
\def\gsim{\setbox3=\hbox{$\sim$}\setbox4=\hbox{$>$}
           \mathrel{\hbox{\lower2pt\hbox{$\sim$}\kern-0.5\wd3
                           \kern-0.5\wd4\raise2pt\hbox{$>$}}}}

\slugcomment{Accepted in Astrophysical Journal}

\shorttitle{A brown dwarf census from the SIMP survey}
\shortauthors{Robert et al.}

\begin{document}


\title{A BROWN DWARF CENSUS FROM THE SIMP SURVEY}


\author{Jasmin Robert\altaffilmark{1,7}, Jonathan Gagn\'e\altaffilmark{2,3}, \'Etienne
  Artigau\altaffilmark{4}, David Lafreni\`ere\altaffilmark{4,7}, Daniel Nadeau\altaffilmark{1},  Ren\'e Doyon\altaffilmark{4}, Lison Malo\altaffilmark{5},
Lo\"{i}c Albert\altaffilmark{4}, Corinne Simard\altaffilmark{1}, Daniella C. Bardalez Gagliuffi\altaffilmark{6}, Adam J. Burgasser\altaffilmark{6}}
\affil{D\'epartement de physique and Observatoire du Mont-M\'egantic, Universit\'e de Montr\'eal, Montr\'eal, QC H3C 3J7, Canada}




\altaffiltext{1}{D\'epartement de Physique, Universit\'e de Montr\'eal,
  C.P. 6128 Succ. Centre-ville, Montr\'eal, Qc, H3C 3J7, Canada; jasmin@astro.umontreal.ca}
\altaffiltext{2}{Carnegie Institution of Washington. 5241 Broad Branch
  Road, Washington, DC 20015, USA }
\altaffiltext{3}{Sagan Fellow}
\altaffiltext{4}{Institut de Recherche sur les Exoplan\`etes (iREx), Universit\'e de Montr\'eal, D\'epartement de Physique,
C.P. 6128 Succ. Centre-ville, Montr\'eal, QC H3C 3J7, Canada}
\altaffiltext{5}{Canada-France-Hawaii Telescope, 65-1238 Mamalahoa Hwy,
  Kamuela, HI 96743, USA }
\altaffiltext{6}{Center for Astrophysics and Space Sciences, University
  of California San Diego, 9500 Gilman Dr., Mail Code 0424, La Jolla,
  CA 92093, USA}
\altaffiltext{7}{Visiting Astronomer at the Infrared Telescope Facility,
  which is operated by the University of Hawaii under contract
  NNH14CK55B with the National Aeronautics and Space Administration.}


\begin{abstract}

We have conducted a near-infrared (NIR) proper motion survey,
the Sondage Infrarouge de Mouvement Propre, in order to
discover field ultracool dwarfs (UCD) in the solar
neighborhood. The survey was conducted by imaging
$\sim28\%$ of the sky with the Cam\'era PAnoramique Proche-InfraRouge
both in the southern hemisphere at the Cerro Tololo
Inter-American Observatory 1.5 m telescope, and in the northern
hemisphere at the Observatoire du Mont-M\'egantic 1.6 m
telescope and comparing the source positions from these
observations with the Two Micron All-Sky Survey Point Source Catalog
(2MASS PSC). Additional color criteria were used to further discriminate
unwanted astrophysical sources. We present the results of an NIR
spectroscopic follow-up of 169 M, L, and T dwarfs. 
Among the sources discovered are 2 young field
brown dwarfs, 6 unusually red M and L dwarfs, 25 unusually
blue M and L dwarfs, 2 candidate unresolved L+T binaries, and 24 peculiar
UCDs. Additionally, we add 9 L/T transition dwarfs (L6--T4.5) to
the already known objects.

\end{abstract}


\section{INTRODUCTION}

Ever since the discovery of the high proper
motion of Barnard's star (\citealt{bar16}), 
proper motion has proved to be an effective method to identify faint
stars in the solar neighborhood. Even thin disk stars, which are
comoving with the Sun, have residual velocities of the order of tens of
km\,s$^{-1}$. This arises mainly from the fact that, as stars age, they
randomly interact with interstellar cloud complexes and
spiral arms, which progressively increases their velocity
dispersion. This leads to proper motions of about 100\,mas\,yr$^{-1}$
for stars up to $\sim$50\,pc.

With the advent of sensitive near-infrared (NIR) hybrid mosaic detectors, the
search for low-mass stars and substellar objects has become
possible. Hundreds of ultracool dwarfs (UCDs; spectral type  M7 or
later; \citealt{kir95}) have been discovered through large-scale
red-optical (0.6--1.0$\,\mu$m) surveys, such as the Sloan Digital Sky
Survey (SDSS; \citealt{yor00}), the Canada--France Brown Dwarf Survey
(\citealt{del08}), the Pan-STARRS1 3$\pi$ Survey (PS1;
\citealt{kai10}), along with large-scale and all-sky NIR
(1.0--2.5$\,\mu$m) surveys, e.g.,~the DEep Near Infrared Survey (DENIS;
\citealt{epc97}), the Two Micron All Sky Survey (2MASS;
\citealt{skr06}), the UKIRT Infrared Deep Sky Survey (UKIDSS;
\citealt{law07}), the VISTA Hemisphere Survey (VHS; \citealt{mcm12}) and
the all-sky mid-infrared (3.4--22$\,\mu$m) {\it Wide-Field Infrared Survey
Explorer} ({\it WISE}; \citealt{wri10}) survey.

Most of these surveys used color-based criteria to identify UCDs
(\citealt{del97}; \citealt{kir99}; \citealt{bur02}; \citealt{haw02};
\citealt{kir11}). While this is an effective way of finding UCDs, it may
  overlook objects with peculiar colors, such as very low metallicity
  halo objects (subdwarfs; \citealt{bur03}; \citealt{siv09}).

Furthermore, an increase in opacity of the CH$_4$, H$_2$O, and H$_2$ molecules
in the atmosphere of L/T transition dwarfs (L6--T4.5) leads to NIR colors
similar to those of M and early-L dwarfs (\citealt{leg00}; \citealt{chi06}).
The ability to select transition dwarf candidates for
spectroscopic follow-up is important to uncover a large sample of these
objects and better understand the processes causing condensate cloud disruption
(\citealt{ack01}; \citealt{bur02b}; \citealt{kna04}), which is
known to happen in their photosphere as they evolve and cool
down. This effect has been demonstrated in several
variability studies (\citealt{cro14} and references therein).
While recent studies were able to discover some
L/T transition dwarfs (\citealt{bes13}; \citealt{sch14}), it is
likely that several more are still hiding in surveys such as 2MASS and
{\it WISE}.

This could also be said of UCDs very close to the Sun.
Previous studies (\citealt{met08}; \citealt{burn10}; \citealt{rey10};
 \citealt{kir12}) have suggested that the space density 
of brown dwarfs is below that of stars, with possibly a minimum at
the L/T transition (\citealt{day13} and references therein). However,
the recent finding of the brown dwarf binary WISE~J104915.57$\relbar$531906.1
(\citealt{luh13}), and of the coldest brown
dwarf known to date, WISE~J085510.83$\relbar$071442.5 (\citealt{luh14}), outclassing
Wolf 359 and Lalande 21185 as the third and fourth closest systems to
the Sun, provides evidence that the search for UCDs in the solar
neighborhood is far from over. The proper motion method is the most efficient way of
uncovering nearby ($\lesssim$ 50\,pc) UCDs in a color-unbiased search,
just as it is for warmer stars.

While some surveys include multi-epoch
observations that can lead to a detection of 
proper motion (\citealt{loo08}; \citealt{kir10}), the small timespan
(approximately one to two years) between individual observations is
better suited for the detection of very large proper motions. This is
the case for AllWISE (\citealt{kir14}), released to track
down high-proper motion objects in the {\it WISE} catalog. However, as noted
by \citet{kir14}, AllWISE measures only the apparent motion on the sky,
because most objects were observed
only at two epochs separated by six months. This interval maximizes the
effect of the parallax on the motion of the sources,
hindering the calculation of their true proper motion.

Some studies have combined the different epochs of two or more
of those surveys to search for high proper motion objects: DENIS and 2MASS
(\citealt{art10}), 2MASS and SDSS (\citealt{met08}; \citealt{she09};
\citealt{gei11}), 2MASS and UKIDSS (\citealt{dea09}), and more recently
2MASS and {\it WISE} (\citealt{bih13}; \citealt{per14}) or 2MASS and AllWISE
(\citealt{gag15}) and Pan-Starrs and AllWISE (\citealt{bes13}). Others
used several of those surveys to identify new candidates
(\citealt{abe11}; \citealt{sch12}) or added a second epoch of
$K$-band astrometry, such as the UKIDSS Galactic Plane Survey (GPS;
\citealt{luc08}), in which a new T5 dwarf and several other L and T
dwarf candidates were recently discovered in the galactic plane
(\citealt{smi14}).

It is in this context that the large area NIR proper motion
survey Sondage Infrarouge de Mouvement Propre (SIMP) was initiated using
the 2MASS Point Source Catalog (PSC\footnote{see
  \protect\url{http://irsa.ipac.caltech.edu/}}) as a first epoch. This
brings the possibility of discovering UCDs in the solar neighborhood
that were missed in other searches, such as L/T transition dwarfs or
UCDs with peculiar photometric colors. This survey is also
suited for the discovery of wide binaries and members of young
moving groups. Discovering more of these objects will help constrain the
substellar mass function and the brown dwarf space density in the solar
neighborhood. Furthermore, cataloging a large number of high-proper motion brown 
dwarfs will be useful for predicting microlensing events, enabling an
independent and direct measurement of their masses (\citealt{eva14}).

Since 2MASS and SIMP use similar filters and pixel scales and reach photometric
comparable depths (see Section~\ref{SIMPsurvey}), cross-matching sources
is straightforward. This survey has already enabled the discovery
of several UCDs: a bright, nearby T2.5 dwarf (SIMP~J013656.5+093347;
\citealt{art06}), five very low-mass binaries (SIMP~J1619275+031350AB $\&$
SIMP~J1501530$\relbar$013506AB; \citealt{art11}, 2MASS~J10432513$\relbar$1706065,
2MASS~J11150150+1607026 $\&$ 2MASS~J12594167+1001380; \citealt{bar15}), and a
low-gravity L5~$\beta/\gamma$ candidate member of the Argus association
(SIMP~J21543454$\relbar$1055308; \citealt{gag14, gag15b}). This paper
presents the remaining UCDs discovered so far through the SIMP
survey.\footnote{A list of all known ultracool dwarfs is
  maintained at \protect\url{http://www.astro.umontreal.ca/~gagne/listLTYs.php}.}

The SIMP survey is described in detail in Section~\ref{SIMPsurvey}, and the
spectroscopic and photometric follow-ups are presented in
Section~\ref{spectrocand} and Section~\ref{PhotoObs}, respectively. We
describe the method used to classify the candidates in
Section~\ref{spectral_class} and we present the results in
Section~\ref{Results}. We summarize and conclude in
Section~\ref{Summary}.

\section{THE SIMP SURVEY}\label{SIMPsurvey}

The SIMP survey was initiated in early 2005, when a newly developed
wide-field NIR camera, the Cam\'era PAnoramique Proche
Infra-Rouge (CPAPIR; \citealt{art04}), was installed at the Cerro Tololo
Inter-American Observatory (CTIO) 1.5 m telescope, which is operated by
the SMARTS consortium. CPAPIR uses a HAWAII-II 2048$\times$2048 HgCdTe
detector and is designed to image a large portion of the sky at once,
making it one of the most efficient NIR survey instruments for
small-class (up to 2 m) telescopes. At the CTIO, it has a pixel scale of
1\arcsecpoint03, yielding a field of view of
34\arcminpoint8$\times$34\arcminpoint8. Two filter wheels contain 10
broad- and narrowband filters, within its 0.8--2.5$\,\mu$m
bandpass. CPAPIR was used at CTIO from 2005 February to 2007 March.

From 2007 August to 2009 November, the SIMP survey was conducted with
CPAPIR at the Observatoire du Mont-M\'egantic (OMM) 1.6 m telescope
(\citealt{rac78}). The OMM and CTIO survey procedures were
similar (see below), except that the OMM field of view decreased to 
$30'{\times}30'$, with a pixel scale of 0\arcsecpoint89.

The SIMP observations were obtained in the MKO $J$ band ($\lambda_{c}=1.25\,
\mu$m and $\Delta\lambda=0.16\,\mu$m) for two reasons: (1) to
minimize the sky background level, which is the main source of noise for
faint objects, and (2) to identify the T dwarfs, which have blue
$J-H$ colors. At each sky position, three to five individual
8\,s exposures were obtained, with a 7$\arcsec$dither pattern
along the declination axis. The survey was divided into stripes
extending over 20$\degree$ in declination, to minimize
telescope dithering time, and it was highly automated with computer
scripts. The number of exposures obtained at each sky position was
adjusted in function of the observing conditions to reach a
$5\sigma$ depth of $J=17.2$.

\begin{figure}
  \centering
  \begin{subfigure}{\textwidth}
    \includegraphics[width=0.99\textwidth]{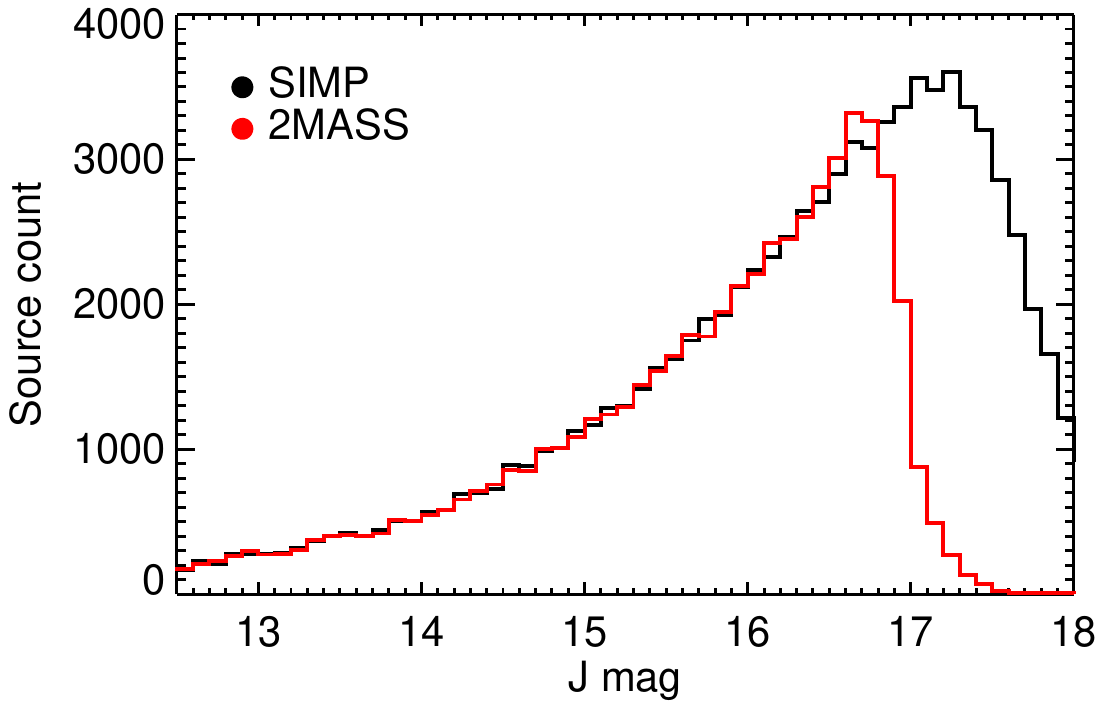}
  \end{subfigure}
  \caption{Distribution of $J$ magnitude for the sources from the
    2MASS PSC (red) and the SIMP images (black) for a 6.7 square
  degree area of the SIMP survey.}\label{f1}
\end{figure}

   Figure~\ref{f1} shows the source count as a function of $J$
  magnitude for both the 2MASS and SIMP photometry of a 6.7 square
  degree area of the sky corresponding to one night of SIMP data. The
  SIMP and 2MASS source counts agree precisely up to $J=16.4$. For
  $16.4<J<16.8$, the number of sources appears slightly but
  systematically larger in 2MASS than in SIMP. In the $16.6<J<16.8$
  range, the SIMP source count is $94.0\%$ of the 2MASS source count,
  but a SIMP counterpart is found for $97.2\%$ of the 2MASS sources in
  that magnitude range, within 2$\arcsec$ (or 1 SIMP 
  FWHM). The source count discrepancy is probably due to
  uncertainties on the magnitudes, and the plot is consistent
  with the SIMP source count being complete to $J=17.2$. The SIMP survey
  is thus limited by the depth of the 2MASS Survey.

The SIMP survey has covered $\sim28\%$ of the sky; the region surveyed
was restricted to absolute galactic latitudes larger than 20\degre
(except for a few specifically targeted nearby stars) to avoid source
confusion when cross-matching with 2MASS.

The final SIMP survey contains more than 115,000 raw frames; a custom
IDL pipeline was developed to reduce these data. Each individual frame
is divided by a flat-field image obtained from on and off dome-flats
taken either at the beginning or at the end of each night. A bad
pixel mask is applied and a running median of 11 exposures is used
to substract sky emission before individual images are median combined.

\begin{figure}
  \centering
  \begin{subfigure}{\textwidth}
    \includegraphics[width=0.99\textwidth]{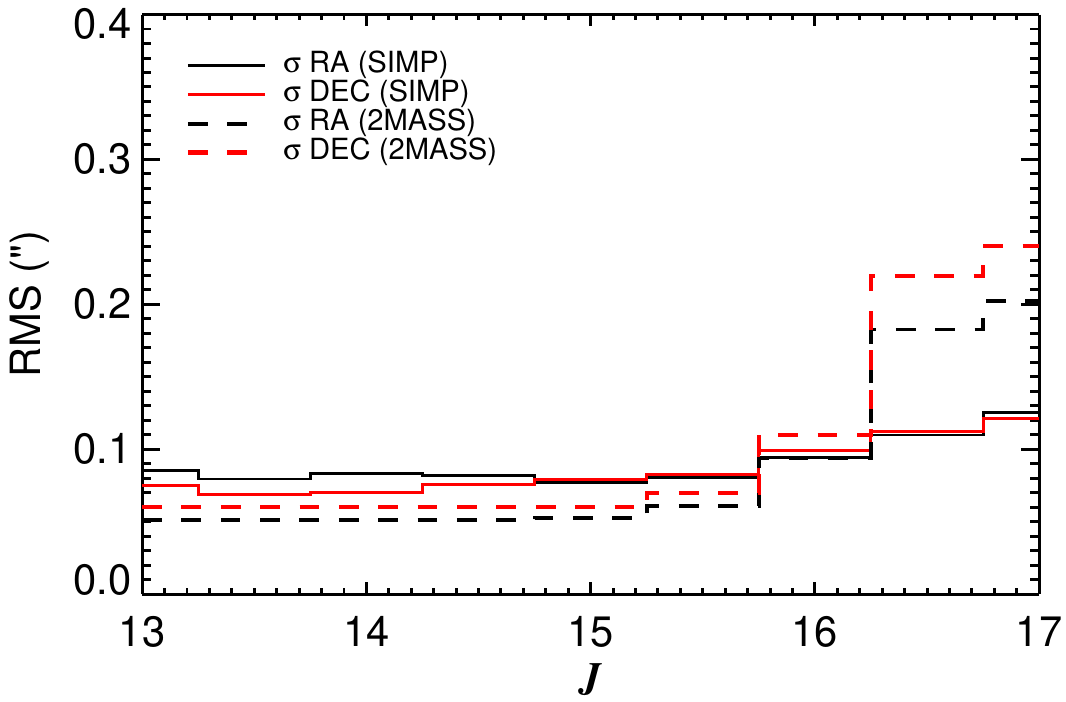}
  \end{subfigure}
  \caption{Mean R.A. (black) and decl. (red) uncertainties for SIMP
    (solid line) and 2MASS (dashed line) versus $J$ magnitude for the
  same data subset as in Figure~\ref{f1}.}\label{f2}
\end{figure}

 An astrometric solution is obtained for each SIMP image based on the
2MASS PSC, and the SIMP images are corrected for spatial distortion
  using a second-order polynomial in both directions, in order to
  minimize the residual errors between the PSC and SIMP
  positions. Figure~\ref{f2} presents the positional uncertainty in
  R.A. and decl. as a function of $J$ magnitude, for the
  same data subset as in Figure~\ref{f1}. The 2MASS uncertainties
  are plotted as given in the PSC. They amount to $\sim$0\arcsecpoint06
  in each direction for $J<15.7$, increasing to $\sim$0\arcsecpoint2 in
  each direction at $J=16.7$. The SIMP curves show the $1\sigma$
  positional uncertainty in R.A. and decl. obtained from the difference in
  position measured for those sources that were observed twice due to
  frame overlap. The SIMP uncertainties are $\sim$0\arcsecpoint08 in each
  direction for $J<15.7$, increasing to $\sim$0\arcsecpoint11 at
  $J=16.7$.

\subsection{Candidate Selection} \label{candselec}

 The SIMP survey being roughly half a magnitude more sensitive
than the typical $J\sim16.7$ limit of 2MASS, cross-matching between SIMP
and 2MASS is done by searching in the SIMP images for a counterpart to
the PSC sources that meet the following criteria: they must have a
  $J$-band photometric quality (ph$\textunderscore$qual) of A, B, or C, 
  an $H$-band photometric quality of A, B, C, or D, they must not be
  flagged as a minor planet (mp$\textunderscore$flg = 0), and there must
  be no confusion (cc$\textunderscore$flg = 0) or galaxy
  contamination (gal$\textunderscore$contam = 0). A match is made with
  the nearest source in the SIMP image that is within $\Delta J=0.6$
  of the 2MASS source, is not already matched to another source, has a
  profile sharpness consistent with a gaussian, has a long to short axis
  ratio less than three, and is within a matching radius of 27$\arcsec$.

The proper motion detection limit of the survey was derived from the
astrometric uncertainties in both 2MASS and SIMP. Since 2MASS started in
1997 June and ended in 2001 February, the difference in epochs between
SIMP and 2MASS varies between 4 and 12 years. For an average timespan of
eight years, proper motions down to $\mu=88\,\hbox{mas yr}^{-1}$ can be
measured to $5\sigma$ for $J<15.7$, increasing to $\mu=200\,\hbox{mas
  yr}^{-1}$ at $J=16.7$.  For the 27$\arcsec$ matching radius, the 
  typical upper bound limit on proper motion is $\mu\simeq3400\,\hbox{mas
    yr}^{-1}$.

The use of a photometric color criterion makes it possible to select UCD
candidates with proper motions down to $\mu=100\,\hbox{mas yr}^{-1}$
with minimal contamination by other sources. For this purpose, we used
$I$-band photometry from the SuperCosmos Sky Survey (SSS;
\citealt{ham01}) or, if available, $ugriz$ photometry from SDSS. 
If neither of these surveys were available, follow-up $i$ and $z$-band
observations were obtained with MegaCam (\citealt{bou03}) at the
Canada-France-Hawaii Telescope (CFHT) by taking single 200\,s
exposures. The MegaCam photometry is in the AB system, like SDSS, and
 no color-dependent corrections were applied (see
  \citet{del08} for photometric corrections from SDSS to
  MegaCam). The resulting $i$ magnitude for those candidates, which were
confirmed spectroscopically, is presented in Table~\ref{t5} of
Section~\ref{Results}.

A target is selected as a candidate for further observation if any
one of these criteria is met:

1. $\mu>100\,\hbox{mas yr}^{-1}$ and detected in $I$ with $I-J > 3.5$,

2. $\mu>100\,\hbox{mas yr}^{-1}$, with a $3\sigma$ upper
limit on $I$ ($I_{3\sigma}$) such that $I_{3\sigma}-J > 3.0$, or

3. $\mu>200\,\hbox{mas yr}^{-1}$ and undetected in $I$.

For a typical velocity dispersion of $\sim$25\,km\,s$^{-1}$, stars with
$\mu>100\,\hbox{mas yr}^{-1}$ will be within 50\,pc of the Sun. Dwarfs
of type M8\,V and earlier have $M_{J} < 11.5$ and $ I-J < 3.4$ (\citealt{pec13}).
If such a star were closer than 50\,pc to the Sun, it would be detected
in the $I$ band and rejected due to its blue $I-J$ color. Nevertheless,
some M dwarfs were still selected because they were found to meet the third
criterion (see Section~\ref{Results} for more detail).

\begin{figure}
  \centering
  \begin{subfigure}{\textwidth}
    \includegraphics[width=0.99\textwidth]{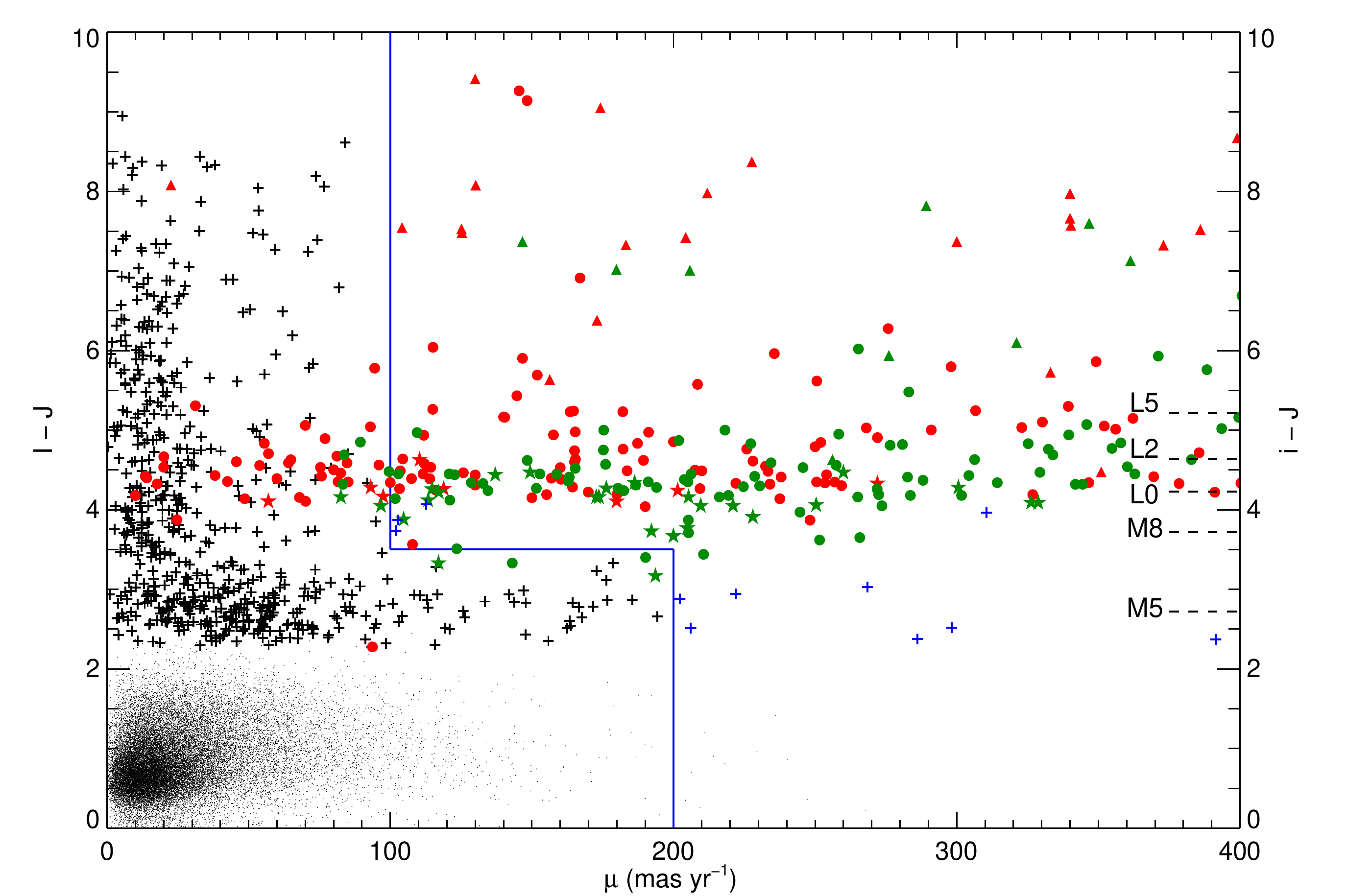}
  \end{subfigure}
  \caption{UCD candidate selection criteria represented as a blue
    line on a plot of $I-J$ color versus proper motion for the SIMP survey
    sources from the same data subset as in
    Figure~\ref{f1}. These are shown as small black dots if they have a
    measured $I$ magnitude. If undetected in $I$, they are shown as
    black crosses, or blue for the 11 sources meeting the selection
    criteria, and are located on the plot by assigning the value
    $I=19.5$ as a lower limit. Shown in green are the UCDs discovered in
    this work and in red other UCDs with
    measured $i-J$ color and proper motion (from
    \protect\url{http://www.astro.umontreal.ca/~gagne/listLTYs.php}),
    M dwarfs as stars, L dwarfs as dots, and T dwarfs as triangles. On
    the right-hand scale is indicated the $i-J$ color of spectral types
    from M5 to L5 (\citealt{haw02}).}\label{f3}
\end{figure}

  In Figure~\ref{f3}, the selection criteria are represented as
  a line on a plot of the $I-J$ color versus proper
  motion, for UCDs discovered in this work, other known UCDs with
  measured $i$ magnitude and proper motion, and the SIMP survey sources
  from the same data subset as in Figure~\ref{f1}. For legibility,
  UCDs with $I-J>10$ or a $\mu>400\,\hbox{mas yr}^{-1}$ are omitted from
  the graph. For the data subset, the $I$-band magnitude was obtained
  from SSS, and if the source is undetected in the $I$ band, an $I-J$
  color is calculated using the SSS survey limit, $I=19.5$. For the
  UCDs discovered in this work and other known UCDs, the $i$-band
  magnitude was obtained from SDSS or Megacam when available. As shown
  in Figure~\ref{f3}, for a data subset typical of the whole survey, UCD
  selection criteria eliminate all but 11 candidates among the 27644
  SIMP survey sources. However, all except one of the UCDs identified by
  other surveys, with measured $i-J$ color and proper motion, satisfy
  the color criteria. As expected, the SIMP survey will miss the low
  proper motion UCDs. A few SIMP UCDs are found outside the selection
  criteria as their proper motion or color, given in Table~\ref{t5}, have
  been updated with data posterior to the SIMP survey.

\section{SPECTROSCOPIC OBSERVATIONS} \label{spectrocand}

Once a considerable list of candidates had been extracted from the
SIMP survey, a spectroscopic follow-up to confirm their UCD nature and
determine their spectral type, was conducted at four different telescopes,
with as many NIR instruments, as described in the following subsections.

Limits on telescope time imposed a selection of the candidates to be
observed spectroscopically, based on the magnitude of their proper motion,
their optical-NIR colors, and airmass considerations.
Table~\ref{t1} presents a log of all spectroscopic observations, in
order of increasing R.A. of the targets. Their 2MASS
designation is given in Column~1; hereafter, the coordinate will be
abbreviated to the first four digits of the R.A. and decl.
(i.e. SIMP J23185497$\relbar$1301106 becomes J2318$\relbar$1301).

Columns~2-9 list the instruments used, the UT date of observation, the
integration time in minutes, the mode used for the instrument (when
applicable), the standard star observed and its spectral type, 
along with the weather, the mean airmass or any difficulty encountered
during the observation.

In total, 197 spectra were obtained for 169 targets. In the next
subsections, we detail the observing setups and data reduction processes
for every instrument used. All spectra were corrected for telluric
absorption by observing an A-type star immediately before or after the
science target.

\subsection{SIMON at OMM} \label{SIMON}

The spectra of 50 candidates were obtained at the OMM
1.6 m telescope during 81 nights, from 2006 April to 2007 November,
with the Spectrom\`etre Infrarouge de Montr\'eal (SIMON; \citealt{alb06}),
featuring a HAWAII-I 1024$\times$1024 HgCdTe detector with a
0\arcsecpoint46 pixel$^{-1}$ scale.
All observations were obtained with the Amici prism and
a 1\arcsecpoint1 wide slit (2.4 pixels) with a north--south orientation,
resulting in a spectral resolution of $R\sim40$ over the wavelength range
from 0.9 to 2.4$\,\mu$m. The spectra were obtained in series of
nine exposures using a $10''$ spacing five-position nod pattern along the
slit.

An IDL pipeline was used first to divide each
image by a flat-field and correct for bad pixels. Sky emission was
removed by taking the difference of images obtained at adjacent nod
positions. The spectra were extracted from each difference image,
weighted according to their signal-to-noise ratio (S/N), and
combined. White light recorded through the Paschen~$\beta$ or
Brackett~$\gamma$ filters was used for wavelength calibration, in
combination with the previously calibrated dispersion of the Amici
prism. Some spectra with low-S/Ns were smoothed using a second order
Savitzky--Golay polynomial filter (\citealt{pre92}), with a width of
five points, to help with the visual comparison of the spectra.

\subsection{Gemini Near-Infrared Spectrograph (GNIRS) at Gemini South} \label{GNIRS}

Spectra for 56 of the fainter SIMP candidates were obtained
at the Gemini South telescope from 2006 August to 2007 April
with GNIRS (\citealt{eli06}). The cross-dispersed mode (XD), covering
the wavelength range from 0.85 to 2.5$\,\mu$m across five
orders, was used with the 32 l mm$^{-1}$ grating and a
0\arcsecpoint3-wide slit, yielding a spectral resolution of $R\sim1700$. A
total of 8 to 33 individual exposures of 60--120\,s were acquired for
each target, using a $5''$ ABBA nod pattern along the slit.

A custom IDL pipeline, detailed in \citet{alb11}, was used to reduce all
GNIRS data. An image acquired with a pinhole mask is used to correct for the
distortion and to straighten up the cross-dispersed spectrum. Pairs of
adjacent exposures from the ABBA pattern
are subtracted from each other and the spectra extracted from
each difference image are median combined.
The resulting spectrum is wavelength calibrated using an Ar lamp,
and the calibration is fine-tuned with OH lines from sky emission
(\citealt{rou00}).

\subsection{NIRI at Gemini North} \label{NIRI}

Observations of 20 of the fainter SIMP candidates were obtained
at the Gemini North telescope from 2008 February to 2008 November with
the Near InfraRed Imager and Spectrometer (NIRI; \citealt{hod03}). The
$f/6$ camera and the 6-pixel ($\sim$0\arcsecpoint7) centered slit were used
with the KRS5 $J$, $H$, and $K$ grisms, to obtain a spectral resolution
of $R\sim500$, sufficient to resolve the 1.25$\,\mu$m K~I doublet. The $J$,
$H$, and $K$ bands of a target spectrum  were obtained at different
times, and sometimes on different nights, leading to large uncertainties
in the relative calibration of the band fluxes. Using the 2MASS
photometry to normalize the fluxes in each band was also not reliable
since the uncertainties on the 2MASS photometry for most targets were
higher than 0.1 mag. This led us to normalize each of the $J$, $H$ and
$K$ bands independently to a standard spectrum (at 1.00--1.35$\,\mu$m,
1.40--1.80$\,\mu$m, and 1.95--2.40$\,\mu$m, respectively) in order to
facilitate the visual classification. A total of 4 to 30 exposures of
60\,s were obtained in each band with a $5''$ nod along the slit. The
reduction pipeline is described in \citet{del08}.

\subsection{SpeX at InfraRed Telescope Facility (IRTF)} \label{SPEX}

Spectroscopic observations of 71 targets were obtained
at the NASA IRTF for a total of nine nights grouped in three observing
runs, in 2008 February--March, 2008 September, and 2009 April.
The medium-resolution infrared spectrograph SpeX 
(\citealt{ray03}) was used either in
the cross-dispersed ShortXD mode (SXD, 0.8--2.46$\,\mu$m coverage)
with a 0\arcsecpoint8$\times$15${}''$ slit, 
yielding a spectral resolution of $R\sim750$, 
sufficient to resolve the K~I doublet at 1.25$\,\mu$m,
or in the low-resolution prism mode
(PRISM, 0.8--2.5$\,\mu$m coverage) with $15''$-long slits
of width 0\arcsecpoint8 ($R\sim75$),
0\arcsecpoint5 ($R\sim120$),
or 0\arcsecpoint3 ($R\sim200$).
The choice of resolution was based on target
brightness and observing conditions.
The orientation of the slit was maintained near the parallactic angle.
For each candidate, a total of 6 to 22 exposures of 60--240\,s were
acquired using a $7\arcsecpoint5$ ABBA nod pattern along the slit.

The SpeX data were reduced with version 3.4 of the SpeXtool
package\footnote{Available at \protect\url{http://irtfweb.ifa.hawaii.edu/~spex/}.}
  (\citealt{cus04}). The adjacent image subtraction ${\rm A}-{\rm B}$
  reduction mode was used together with optimal spectrum extraction
(\citealt{hor86}). The resulting spectra were wavelength calibrated
using Ar lamp calibration exposures. Telluric correction and flux
calibration were achieved with the \emph{xtellcor} procedure that is
included in the SpeXtool package (\citealt{vac03}).

\section{PHOTOMETRIC OBSERVATIONS}\label{PhotoObs}

A photometric follow-up in the $J$, $H$ and $K$ bands was conducted in
2007 and 2008 for 28 targets with low precision 2MASS photometry. These
observations were obtained at the OMM during the spectroscopic follow-up
and SIMP survey runs, using SIMON and CPAPIR, respectively, to optimize
telescope time when it would allow short photometric measurements, but
was insufficient for the longer spectroscopic or survey observations. 
The photometry also provided supplementary astrometric epochs to
refine the proper motion measurements of our targets.

The raw SIMON and CPAPIR images were processed using a custom IDL
pipeline similar to the SIMP survey pipeline described in Section~\ref{SIMPsurvey}. 
Aperture photometry was performed, calibrated on 2MASS stars in the
field of view, while masking the potentially saturated bright stars.
Since the NIR colors of UCDs can vary by up to $\sim$0.4 mag depending on
the photometric system used (\citealt{ste04}) and the filters in SIMON
and CPAPIR conform with the Mauna Kea (MKO) system (\citealt{sim02};
\citealt{tok02}), the 2MASS magnitudes were converted to the MKO system
using polynomial relations from \citet{leg06}. Table~\ref{t2}
presents a log of the photometric observations and the results are
presented in Table~\ref{t5}.

\section{SPECTRAL CLASSIFICATION} \label{spectral_class}

All of the observed spectra were classified by the use of spectral
indices as a first step, followed by visual comparison with NIR 
standard spectra, as described in the following subsections.\footnote{All
  spectra are publicly available at
  \protect\url{http://dx.doi.org/10.5281/zenodo.58501}.}

Visual comparison is particularly useful for
the low-resolution SIMON spectra, for some low S/N ($\sim$5) spectra,
and for the identification of peculiar sources such as young 
or low-metallicity objects, and unresolved binaries.
The spectral indices are most useful when a visual classification is
ambiguous.

\subsection{Spectral Indices} \label{indices}

Over the years, multiple sets of NIR spectral indices
have been defined to classify UCDs. 
Most of them measure the flux ratio in narrow bands subject to
molecular absorption (H${}_2$O, CH${}_4$, etc.) within the $J$,
$H$ or $K$ bands. As any molecular absorption feature varies 
over a limited range of atmospheric conditions, most indices are
restricted to a range of spectral types and become inaccurate
outside this range. The broad spectral range of our objects, 
and the variety of instruments used for this campaign,
led to the use of two sets of NIR spectral indices.

The first set is made of four indices determined by \citet{all13} to be
spectral type sensitive for a broad range of M and L dwarfs while being
insensitive to surface gravity. These indices and their respective valid
spectral type range are H${}_2$O (M5--L4), H${}_2$OD (L0--L8),
H${}_2$O-1 (M4--L5), and H${}_2$O-2 (M4--L2).

The second set of indices is from \citet[with spectral type
relations from \citealt{bur07}]{bur06b} and is useful for the L and T
dwarfs. These indices and their respective valid
spectral type range are H${}_2$O-$J$ (L0--T8), H${}_2$O-$H$ (L0--T8),
CH${{}_4}$-$K$ (L0--T7), CH${{}_4}$-$J$ (T0--T8), and
CH${{}_4}$-$H$ (T1--T8).

Table~\ref{t3} lists a compilation of the indices for our
targets, along with the calculated spectral type if within the index
useful range. A median type is also given along with its uncertainty,
both rounded to the nearest half-integer. The median type is calculated
by iterative rejection of the types (put in brackets) that fall too
far from the median, until it converges or the rounded uncertainty
equals 0.5. This resulted in the most likely spectral type 
from all the useful indices. Nevertheless, for some candidates,
the indices give a broad range of spectral types or a spectral type that
is far from the visual type, due to a lower quality spectrum, or to a
peculiar object. These cases are discussed in Section~\ref{Results}.

\subsection{Comparison with standard spectra} \label{Standardscomp}

The most straightforward and accurate method of classifying stellar
spectra is direct comparison with standard spectra (MK process;
\citealt{mor43}; \citealt{mor73}). We therefore
compare our $J$, $H$ and $K$ band spectra with the NIR spectral standard
sequences listed in Table~2 of \citet{bur06b}, for the T dwarfs, and in
Table~4 of \citet{kir10}, for the M and L dwarfs.
The NIR standard spectra extend from 0.7 to 2.5$\,\mu$m and
they are all available in the SpeX Prism Spectral
Libraries\footnote{\protect\url{http://www.browndwarfs.org/spexprism}}
(\citealt{bur14}), except for M6 and earlier types. For those types,
instead of using the NIR standards for the comparison, we used the NIR
spectra of the equivalent optical standards, which are Gl 213 (M4), Gl
51 (M5), and Gl 406 (M6). Since only a handful of mid-M dwarfs were
found, the use of optical standards should not increase the
uncertainties. We have also used the M6--L5 spectral templates
  defined by \citet{gag15b} to identify low-gravity UCDs, since their
  set of templates includes both field ($\alpha$, implied) and
  low-gravity ($\beta$ and $\gamma$) sequences, extending in the NIR the
  optical sequence of \citet{cru09}, which uses the nomenclature scheme
  introduced by \citet{kir05} and \citet{kir06}.

The method of choice to assign a spectral type is to visually compare
each candidate spectrum with five neighboring spectra of the
standard sequence. The parts of the candidate spectrum that exhibit a
very low S/N are masked to avoid interfering with the comparison. The
regions of strong telluric absorption (1.15--1.18$\,\mu$m, 1.35--1.45$\,\mu$m,
and 1.8--1.95$\,\mu$m) are also masked.
If the candidate and comparison spectra differ in spectral resolution,
the higher resolution data are smoothed to the lower resolution. The
candidate and standard spectra are normalized to the same average value
over the waveband from 1.235 to 1.305$\,\mu$m (or over the
complete range of each individual band in the case of NIRI spectra).
A half-subtype is assigned to a target spectrum when it falls in between
two adjacent standard spectra.

Following \citet{kir10}, the M and L spectra are classified by
matching only the $J$ band, while the $H$ and $K$ bands are used to
ascertain whether the candidate has a blue or red NIR spectrum or one
that shows other peculiarities. For the L dwarfs, the spectral types
were also determined by a method developed by K. Cruz et al. (2016, in
preparation) that uses, for each
optically anchored spectral type, median-combined spectra reflecting the
diversity of spectral morphologies present in each type. This method
visually classifies the spectra using all three NIR bands, normalized
separately to minimize the difference in NIR colors that a given type
can display.  The spectral types obtained with both methods agreed
within their uncertainties and only the classification with the
Kirkpatrick method is presented in Figure~\ref{f4}.

The visual comparison method yields a typical
uncertainty of one subtype, though in some cases
where the S/N is low or the spectrum displays
peculiar features, the uncertainty increases to
two subtypes. On the other hand, if the spectral index method
agrees with the visual comparison within a 0.5 subtype,
and the candidate spectrum matches the standard very closely,
the uncertainty of the adopted spectral type is estimated to be a 0.5
subtype.

\section{RESULTS} \label{Results}

The spectral type of each target, resulting from the classification
scheme described in Section~\ref{spectral_class} is presented in
Table~\ref{t4}. The object name is given in column~1, and a code
to the instrument used for the observations is given in
column~2. Results from the observations of an object with different
instruments are presented as separate entries. Column~3 gives the
spectral type obtained with the spectral indices, column~4 the type
obtained from visual comparison, and column~5 the adopted spectral type
based on both methods, and weighted according to their respective
accuracy. A ``:'' is used for an adopted spectral type with an
uncertainty of $\pm$1 subtype, while ``::'' indicates a larger
uncertainty. Spectral types based on previous observations in the
optical or the NIR are also presented in columns~6 and~7,
respectively, along with the reference in column~8. The spectral type
obtained from optical data may differ from the infrared spectral type
because the two wavelength ranges probe different layers of the
atmosphere and/or because the source is an unresolved binary.

The properties of all targets are presented in
Table~\ref{t5}, which is subdivided into eight parts (normal UCDs, young
UCDs, Hyades Candidates, unusally red UCDs, subdwarf, unusally blue
UCDs, binary UCDs and peculiar UCDs).
In each part, the targets are ordered according to their spectral type
(column~2; from column~5 of Table~\ref{t4}). The adopted proper motion and
its uncertainty are presented in columns 3 and 4. 
In order to get the most accurate proper motion possible, it was recalculated after
the SIMP survey by including the object's position in as many different
surveys as available (column~5). The different epochs used are from
2MASS-PSC, SIMP, SDSS-DR9, CFHT Follow-up, OMM Follow-up, and AllWISE. To
ensure that every position corresponds to the same moving
object at different epochs, a visual inspection was performed in each
survey. The adopted proper motion is the median of the values obtained in a
$10^5$-step Monte Carlo simulation by linear regression over source
positions randomly chosen around the astrometric measurements, according
to a Gaussian distribution of width equal to each measurement
uncertainty. The standard deviation of the simulated distribution
determines the uncertainty on the adopted proper motion. 

 A photometric distance ($d$\textsubscript{phot}), calculated
  with the {\it WISE} $W2$ band (or $H$ if $W2$ was unavailable) and
  polynomial fits for $M_{W2}$ (or $M_{H}$) from \citet{dup12}, is
  listed in column~7. A tangential velocity ($V$\textsubscript{tan}),
 derived from the photometric distance and the proper motion, is listed in
  column~6. Source photometry from each available survey is
presented in columns 8 to 19. This includes $i$ and $z$ from SDSS
or CFHT; $J$, $H$, and $K_s$ from 2MASS; MKO $J$, $H$, and $K$ from
OMM/SIMON and OMM/CPAPIR (this paper); and finally W1 and W2 from
AllWISE. The different colors obtained from these measurements are
plotted as a function of spectral type in Figure~\ref{f5}.

The 169 targets that were observed spectroscopically range from earlier
than M0 to T6.5, including 16 normal M dwarfs, 62 normal L dwarfs, and 17
normal T dwarfs. The remainder of this section presents a detailed
discussion of several unusual objects: 2 young M dwarfs, 3 Hyades
candidates, 6 unusually red M and L dwarfs, 1 subdwarf, 30 unusually
blue M and L dwarfs, 4 unresolved binaries, and 28 peculiar objects,
which are discussed in subsections~\ref{youngdwarf}--\ref{peculiar}
respectively. 

One object, J0136$+$0933 was discovered previously as
part of the SIMP survey by \citet{art06}. This paper presents
an unpublished GNIRS spectrum. The adopted
classification remains unchanged (T2.5$\pm$0.5).

\citet{kir10} suggest to classify as peculiar (blue or red) any object whose
1.5--2.5$\,\mu$m continuum is significantly bluer/redder than expected and
the cause is not obviously low metallicity for blue objects or
low gravity for red objects, whereas \citet{fah09}, who did not obtain
spectra for the objects studied, used a photometric color criterion only:
the 2MASS color $J\relbar K_s$ (or MKO $J\relbar K$ converted to the
2MASS system) must be off by more than $2\sigma$ from
the mean color of known field objects with the same spectral type.
 
We use both spectral and color information. For an object to be defined
as a ``pec red'' or ``pec blue,'' in addition to the 1.5--2.5$\,\mu$m
continuum criterion, its 2MASS $J\relbar K_s$ color (or the MKO
photometric observations if observed with better precision) must be off
by more than $1\sigma$ from the mean color of field objects of the same
type, as can be seen in Figure~\ref{f5}. Objects with a
peculiar spectrum but an uncertain color, or objects with a definite
abnormal color but an otherwise normal spectrum, are labelled as pec$^1$
(possibly red) or as pec$^2$ (possibly blue) and are discussed in
subsection~\ref{peculiar}.

\subsection{Young Candidates} \label{youngdwarf}

Young late-M and L dwarfs are still contracting; they are larger and
less massive than older dwarfs of the same spectral type and their
spectra display peculiar features, such as weaker alkali and molecular
absorption lines, a flatter 2.18--2.28$\,\mu$m continuum and
a triangular-shaped continuum in the $H$ band due to a decrease of
collision-induced absorption of molecular hydrogen, as a
consequence of their low gravity. The properties of the young candidates
found in the SIMP survey are presented in Table~\ref{t5}b. 
As displayed in Figure~\ref{f6}, their spectra were compared with
NIR spectral templates of the $\beta$ (intermediate-gravity) and
$\gamma$ (very low-gravity) classes for M and L dwarfs, constructed by
K. Cruz et al. (2016, in preparation) and \citet{gag15b}. This visual
  comparison alone was used to identify an object as a
  young candidate. We also verify that the adopted spectral type is
consistent with the four \citet{all13} indices that are
gravity-insensitive, and we calculate their gravity class and score with
the indices that are sensitive to surface gravity.

\textit{J0120$+$1518}: this object, with a red $i\relbar J$ color (SDSS:
1.84$\pm$0.06; MegaCam: 1.64$\pm$0.09), and a red $J\relbar K_s$ color
(2MASS: 1.82$\pm$0.18), displays features of a young brown dwarf that 
match quite well (within one subtype) the M9$\,\gamma$ standard;
J0120$+$1518 displays a triangular $H$ band, as well as a flatter $K$
band and weaker alkali lines than the field standard. While the
M9:$\,\gamma$ spectral type agrees with the type obtained from the
gravity-insensitive indices of \citet[M9.5]{all13}, the gravity score
yields an FLD-G (field) classification. This object was previously
classified by optical spectroscopy (\citealt{zha09}) as an M9.

\textit{J0357$\relbar$1937}: this young candidate has a spectrum that fits
in-between the M7$\,\gamma$ and M8$\,\gamma$ types, leading us to adopt a
classification of M7.5$\,\gamma\pm$1. The INT-G gravity
score suggests a $\beta$ class, the lower S/N possibly explaining the
difference with the visual classification.

\subsubsection{Hyades Member Candidates} \label{hyades}

The SIMP survey recovered three objects that were previously
identified as candidate members of the Hyades cluster, which has an
estimated age of 625$\pm$50\,Myr (\citealt{per98}). While this is
  young compared to the Sun and field stars in general, it makes the
  Hyades older than about 85\% of the open clusters in the solar
  neighborhood (\citealt{cox00}). The age of the Hyades is much higher
  than the age estimate for the $\,\gamma$ gravity-class (log(age) (years)
  $\approx$ 7), and even at the higher limit of the age estimate of the
  $\,\beta$ class (log(age) (years) $\approx$ 8)(\citealt{all13}).

  The space motion of the Hyades cluster members is consistent with
  parallel motion with an internal velocity dispersion of
  $0.3$\,km\,s${}^{-1}$ (\citealt{per98}), and an important test of
  cluster membership is provided by the comparison of a candidate SIMP
  proper motion with the cluster convergence point (CP) and velocity.

  Figure~\ref{f7} shows the proper motion trajectories of stars
  selected as members of the Hyades by \citet{per98}, as calculated from
  their Hipparcos proper motion (\citealt{van07}), giving a good measure of
  the location and extent of the area of convergence. Superimposed on this
  plot are the trajectories of the three SIMP Hyades cluster candidates,
  discussed below.
  
  The cluster velocity divided by the magnitude of a candidate proper
  motion gives its kinematic distance ($d$\textsubscript{kin};
  \citealt{lep09}). The Hyades cluster has a tidal radius of 10\,pc and
  a central distance of 46.34$\pm$0.27\,pc (\citealt{per98}) and a
  better test of membership may result from a comparison of the
  kinematic distance with the photometric distance (column~7 of
  Table~\ref{t5}) than with the cluster central distance.

\textit{J0358$+$1039}: this Hyades candidate L dwarf was discovered by
\citet{hog08} and observations with the 10 m Keck II telescope revealed
it to be a tight binary (\citealt{duc13}). \citet{lod14} obtained an
optical spectrum, classified it as an M5 (known as candidate Hya05),
and rejected it as a Hyades member. We obtained the first NIR spectrum
of the unresolved system, which leads to a spectral type of L2.5$\pm$1,
in sharp contrast to the optical classification. The SIMP proper
  motion path is in very good agreement with the Hyades CP, as shown
  in Figure~\ref{f7}, and the kinematic distance obtained from this
  proper motion (31.8$\pm$0.6\,pc) is consistent with the photometric
  distance of the candidate (36.2$\pm$5.9\,pc). Based on this evidence,
  J0358$+$1039 may well be a member of the Hyades.

\textit{J0410$+$1459}: the SpeX spectrum of this object yields a
classification of L1 with no sign of youth (and a gravity score of
FLD-G), but it is redder than the standard, which matches the 2MASS
$J\relbar K_s$ color ($1.580\pm0.093$). This object, discovered by
\citet{hog08}, is located $\sim$1\arcsecpoint9 from another source whose low proper
motion suggests that it is a background object (\citealt{duc13}).
\citet{lod14} classified J0410$+$1459 as an L0.5 (known as candidate Hya03) using
optical spectroscopy, and confirmed it as a member of Hyades. The
SIMP proper motion measurement ($\mu_{\rm
  \alpha}\cos\delta=100.1\pm9.9$\,mas\,yr${}^{-1}$, $\mu_{\rm
  \delta}=-18.3\pm10.1$\,mas\,yr${}^{-1}$) gives a trajectory in
excellent agreement with the Hyades CP, as shown in
Figure~\ref{f7}, and the kinematic distance (46.0$\pm$1.0\,pc)
agrees with the cluster distance and is within $1\sigma$ of the
photometric distance (53.6$\pm$8.7 pc). This evidence supports the
Hyades membership of J0410$+$1459.

\textit{J0417$+$1634}: our proper motion measurement ($\mu_{\rm
\alpha}\cos\delta=83.5\pm7.4$\,mas\,yr${}^{-1}$ $\&$ $\mu_{\rm
\delta}=-60.1\pm7.3$\,mas\,yr${}^{-1}$) gives a value near the center of
the range of proper motions for members of the 
Hyades (see Fig.~4 of \citealt{bou08}), and outside the range of
proper motions for members of the Taurus--Auriga association
(\citealt{fri97}). Accordingly, the kinematic distance, assuming
  J0417$+$1634 shares the Hyades velocity, is 45.5$\pm$0.7\,pc, in
  agreement with the distance to the center of the Hyades
  cluster; if the Taurus velocity is assumed, the kinematic distance
  comes out as 34.2$\pm$0.6\,pc, far from the 137$\pm$20\,pc distance to
  Taurus (\citealt{tor07}). On the other hand, the proper motion path
  falls in between the two cluster CPs, $2.8\sigma$ from the Hyades CP
  (Figure~\ref{f7}) and $2.6\sigma$ from the Taurus CP
  (Figure~\ref{f8}). As shown on Figure~\ref{f8}, some Taurus members
  have proper motion paths similar to that of J0417$+$1634. As shown in
Figure~\ref{f6c}, this object has a spectrum similar to the template for
the L0$\,\gamma$ type. The \citet{all13} gravity score, calculated from the SpeX
spectrum, yields INT-G, which agrees better with the intermediate age
$\beta$ class than the younger $\gamma$ class. An L0:$\,\gamma$
classification is consistent with the previous identification of this
object by \citet{giz99} as a probable young M8 UCD, and with
its classification by \citet{luh06} as a likely intermediate age UCD.
If confirmed, a young age would tend to exclude this object from
membership in the 625\,Myr old Hyades. If J0417$+$1634 were a field
  object, its photometric distance would be 30$\pm$3\,pc on the basis of
  the L0 absolute magnitudes determined by \citet{dup12}. A difference
  of 3.1 magnitudes is found, however, between the absolute magnitudes
  for a field M9 object (\citealt{dup12}) and those of the pre-main
  sequence M9 star LH 0429$+$17 (\citealt{rei99}) for which a parallax
  was obtained by \citet{har99}. This magnitude difference increases the
  photometric distance to 125$\pm$12\,pc, in agreement with the distance
  to Taurus. With the data presently available, it is not possible to
  determine if J0417$+$1634 is a member of either cluster.

\subsection{Unusually Red Objects} \label{reddwarf}

UCDs that show enhanced flux in the $H$ and $K$ bands as compared with
the NIR standard, when their spectrum is normalized in the $J$ band, and
have unusually red 2MASS $J\relbar K_s$ or MKO $J\relbar K$ colors (see
Figure~\ref{f5}), but lack the characteristics of low gravity
UCDs, are considered to be peculiar (red) and their properties are
presented in Table~\ref{t5}d. Objects that satisfy one but not both
of the spectral and photometric criteria are classified as peculiar and
possibly red, and they are discussed in Section~\ref{peculiar}.

Table~\ref{t5}d presents 6 pec (red) objects that were recovered in
the SIMP survey. We list here the spectral type, 2MASS $J-K_s$ color,
and MKO $J-K$ color when MKO photometry was obtained, for the following
objects: \textit{J0026$\relbar$0936} (M9, $1.35\pm0.14$),
\textit{J0432$\relbar$0639} (L2,$1.65\pm0.07$),
\textit{J1307$\relbar$0558} (L3::, $2.10\pm0.21$, $1.82\pm0.10$), and
\textit{J2245$+$1722} (L0.5:, $1.72\pm0.12$). The remaining two objects
are presented below.

\textit{J0307$+$0852}: this source is classified as an 
L1.5: pec (red) UCD. The MKO and 2MASS colors are red for this spectral
type ($J\relbar K=1.72\pm0.04$, $J\relbar K_s=1.73\pm0.14$). Its spectrum
features a small bump on the blue side of the $H$ band but otherwise
fits in between  the L1 and L2 standards. Although such an $H$-band
feature could suggest the presence of an unresolved T-type companion,
the combination of two standard UCD spectra, as described in
Section~\ref{binarydwarf}, did not improve the fit to its spectrum
peculiarities. It was also discovered in a proper motion survey by
\citet{schn16} to be a candidate UCD but no spectrum was obtained.

\textit{J0421$\relbar$1133}: this source is classified as an L3: pec
(red). It displays extreme flux enhancements in the $H$ and $K$ bands
when compared with early L standards, as well as a red 2MASS $J-K_s$
color of $1.78\pm0.14$. The \citet{all13} gravity score is INT-G,
but we do not classify it as a $\beta$ dwarf because its spectrum
matches the field standard better than an intermediate age spectrum.

\subsection{Subdwarfs} \label{subdwarf}

Subdwarfs are UCDs with a low metallicity and a higher surface gravity
and atmospheric pressure. This causes an increase of the
collision-induced absorption by H$_2$ in the $H$ and $K$ bands,
hence an unusually blue $J-K_s$ color. 
They also have stronger hydride bands (CaH, CrH, FeH), since the
metal--metal molecules (TiO, VO) are less abundant, leaving the
absorption by hydride molecules more prominent. Metal (\ion{Ti}{1} and \ion{Ca}{1})
and alkali (\ion{Na}{1}, \ion{K}{1}, \ion{Cs}{1}, and \ion{Rb}{1}) \ion{Na}{1} lines also remain important
at lower temperatures since condensate formation is delayed due to the
less frequent interactions of the rarefied atoms. Finally, as part of
the galactic halo, they move with high transverse velocities
($\sim200$\,km\,s${}^{-1}$) and are easily detectable in proper motion
surveys.

One such object, J1158$+$0435, was recovered in the SIMP
survey. This blue ($J\relbar K_s=1.17\pm0.08$), very high
proper motion object has an adopted spectral type of sdL7$\pm$2, its
spectrum displaying strongly suppressed $H$ and $K$ bands and a strong
FeH band around 1.24$\,\mu$m. This subdwarf was first identified by
\citet{zha09} as an  L3 candidate through SDSS colors. Our result agrees
with the sdL7 classification by \citet{kir10}, based on both
optical and NIR spectra.

\subsection{Unusually Blue Objects} \label{bluedwarf}

Objects with a 2MASS $J\relbar K_s$ color
significantly bluer than the color of field UCDs (see
Figure~\ref{f5}), and a 1.5--2.5$\,\mu$m continuum significantly
bluer than the best-matching field standard, while not showing obvious
signs of low metallicity such as hydride bands, are classified as
peculiar (blue) and are presented in Table~\ref{t5}f. Objects that
have either a blue color or a blue spectrum, but not both, are flagged
as peculiar and possibly blue and are presented in
Section~\ref{peculiar}.

Table~\ref{t5}f presents 30 pec (blue) objects that were recovered in the
SIMP survey.  We list here the spectral type, 2MASS $J-K_s$ color, and
MKO $J-K$ color when MKO photometry was obtained, for the following
objects: \textit{J0357$+$1529} (L1::, $1.11\pm0.08$),
\textit{J0421$\relbar$1950} (M9:, $1.00\pm0.13$), \textit{J0429$+$0607}
(L0:, $0.89\pm0.11$, $0.99\pm0.03$), \textit{J0430$+$1035} (M9::,
$1.28\pm0.23$, $0.94\pm0.11$), \textit{J0946$+$0922} (M9.5::, $<1.5$,
$0.89\pm0.11$), \textit{J0956$\relbar$1910} (L6.5:, $1.36\pm0.07$),
\textit{J1014$+$1900} (L0.5::, $0.92\pm0.12$),
\textit{1041$\relbar$0429} (L2.5:, $1.34\pm0.18$, $1.33\pm0.14$),
\textit{J1052$+$1722} (L2:, $1.18\pm0.20$), \textit{J1058$+$1339} (L1.5:,
$0.55\pm0.26$, $1.09\pm0.06$), \textit{J1130$+$2341} (L2.5:,
$0.90\pm0.22$), \textit{J1218$\relbar$1332} (L4::, $1.08\pm0.19$),
\textit{J1308$+$0432} (M9.5::, $0.49\pm0.26$),
\textit{J1510$\relbar$1147} (L4:, $1.40\pm0.10$), \textit{J1708$+$2606}
(L4:, $1.05\pm0.19$), \textit{J1811$+$2728} (L3:, $1.31\pm0.14$),
\textit{J2203$\relbar$0301} (M9:, $0.96\pm0.17$),
\textit{J2248$\relbar$0126} (L1.5:, $0.87\pm0.27$), and
\textit{J2257$\relbar$0140} (L1::, $0.99\pm0.20$). The remaining 11
objects are presented individually below.

\textit{J0148$+$3712}: we adopt an L1: pec (blue) spectral type for
this source. Its spectrum displays a small bump on the blue side of the $H$
band and overall suppressed flux in both the $H$ and $K$
bands. This led us to investigate a possible binary nature of the
source, but no pair of UCD spectra could reproduce its spectrum
satisfactorily. 
With a blue 2MASS $J\relbar K_s$ color of $1.03\pm0.13$, it is
similar to 2MASS J1440$\relbar$1303, which is classified as an L1 pec
(sl. blue) by \citet{kir10}.

\textit{J0921$\relbar$1534}: we adopt a spectral type of L5:: pec
(blue) for this source; its MKO $J\relbar K=1.37\pm0.07$ is bluer than a
standard L5. Its spectrum has features in its $H$ and $K$ bands that
could be due to an unresolved binary, but it could not be fit by a
combination of two UCD spectra.

\textit{J1004$\relbar$1318}: we classify this object as an L3: pec
(blue) UCD, since its spectrum displays peculiarities in the $H$ and $K$
bands as well as a slightly blue continuum, consistent with its 2MASS
$J\relbar K_s$ color of $1.33\pm0.06$. It was first classified by
\citet{mart10} as an L0 dwarf based on optical spectroscopy at the
Nordic Optical Telescope. \citet{mar13} classify it as an L1 dwarf from
a fit, over the 1.2--2.3$\,\mu$m range, to a spectrum obtained with the
Ohio State Infrared Imager/Spectrometer at the 4.1 m Southern
Astrophysical Research telescope.

\textit{J1036$+$0306}: the adopted classification for this object is
L3.5: pec (blue). Our MKO photometry ($J\relbar K=1.40\pm0.07$) and the
2MASS $J\relbar K_s$ color ($1.15\pm0.16$) are bluer than field
L3-L4 dwarfs. There is an unresolved discrepancy between our SIMON
spectrum, which has all the characteristics of an L3 standard, and our GNIRS
spectrum, which is similar to the L3 standard but for a suppressed flux
in the $H$ and $K$ bands. This object was discovered by
\citet{luh14b} through its high proper motion in AllWISE data, and they classified it
as a blue L4 dwarf based on a SpeX spectrum.

\textit{J1150$+$0520}: we have obtained NIR spectra from both 
NIRI and SpeX, and classify this object as an L5.5: pec
(blue), as its spectral continuum is blue, and so is its 2MASS
$J\relbar K_s=1.24\pm0.23$. It was discovered by \citet{zha09}
and classified with optical SDSS spectroscopy as an L5.5 dwarf, while
\citet{sch10} determined a spectral type of L6, also using SDSS
spectroscopy.

\textit{J1343$\relbar$1216}: our classification, based on three distinct
spectra and a blue 2MASS color $J\relbar K_s=1.05\pm0.20$, yields a
spectral type of L5.5: pec (blue). \citet{luh14b} classified this
object as an L6.5$\pm$2 pec (blue) and found it to be a good match to
2MASS J11263991$\relbar$5003550 (L6.5 pec; \citealt{bur08b}).

\textit{J1411$+$2948}: our adopted classification for this object is
L6:: pec (blue). The peculiar spectrum that we obtained
corresponds to a visual type of L5$\pm$2, but an index-based type of
L8.0$\pm$2. Its blue continuum is consistent with its 2MASS
color $J\relbar K_s=1.10\pm0.14$. It was previously classified as an
L3.5 by \citet{zha09} from SDSS colors. Some features in the spectrum
could be due to an unresolved binary, and an L3+T5 pec binary solution
yields a match comparable to a single L5.

\textit{J1452$+$1114}: we infer a spectral type of L2.5 pec
(blue), from the visual match of the $J$ band of its SpeX spectrum to
the L2 ($\pm1$) standard, its index-based spectral type of L2.5$\pm$0.5,
and a 2MASS color $J-K_s=1.18\pm0.12$ that is blue for this spectral
type. It was found by \citet{dea09} to be a UKIDSS candidate and classified
optically by \citet{zha10} as an L2 dwarf. \citet{bar14} assigned a
spectral type of L3 to this object from a SpeX spectrum.

\textit{J1756$+$2815}: the adopted spectral type for this object is
L1.5: pec (blue) since its 2MASS color ($J\relbar K_s=0.90\pm0.05$)
is bluer than that of L1.5 field dwarfs. \citet{kir10} discovered this object
through its large proper motion and obtained optical (sdL1) and NIR spectroscopy
(L1 pec (blue)), not classifying it as a subdwarf in the NIR because the
FeH bands have a strength similar to that of the L1 standard.

\textit{J2250$+$0808}:  we classify this UCD as an L3.5::
pec (blue). The SpeX spectrum, of low S/N, is visually
most similar to the L3 standard and the calculated indices correspond to
a spectral type of L4.0$\pm$1.5. Its blue continuum is
consistent with its 2MASS color $J\relbar K_s=1.03\pm0.10$, as well as
the UKIDSS color $J\relbar K=0.99$ presented by \citet{day13}. These
authors visually classified this source (identified as BRLT317) as an
L1, from a NIR spectrum obtained with the X-shooter (\citealt{ver11}) at the Very Large
Telescope, though the \citet{bur06b} spectral indices yielded spectral
types between L4.5 and L6.

\textit{J2322$\relbar$1407}: our adopted classification of L2: pec
(blue) for this UCD results from an index-based spectral type of
L4.0$\pm$1 and L2$\pm$2.0 for the NIRI and SpeX spectra, respectively,
and a visual classification of L1.5$\pm$2 for both spectra. This object
has a blue 2MASS color $J\relbar K_s=0.89\pm0.15$. It is classified as an
L1 pec (blue) dwarf by \citet{luh14b}.

\subsection{Unresolved Binary Candidates} \label{binarydwarf}

The follow-up of our candidates revealed several spectra displaying
peculiarities that could be attributed to the combined emission from two
unresolved UCDs. In particular, the combined spectrum of a T dwarf and an
earlier-type UCD features bumps where the flux of the T dwarf emerges
between the strong H$_{2}$O and CH$_{4}$ absorption bands, the most
prominent being around $1.58\,\mu$m. We searched for such
binaries in our sample using the visual inspection method
described in \citet{bar14}, which combines spectra from the SpeX library
and performs a chi-square per degree of freedom
 fit to see if the combination of
two spectral types results in a better match than any single object
spectrum. This section presents the cases where a two-object
combination yields a better match to the observed spectrum, but we
emphasize that these sources remain binary candidates until this
hypothesis is confirmed either by high-resolution imaging or a radial
velocity follow-up. This is especially the case for
  J0006$\relbar$1319 and J0006$\relbar$0852, where the T dwarf spectrum
  is at the noise level, leading to a small ($<$1) chi-square.

The adopted spectral types listed in Table~\ref{t4} take into
account the comparison with a two-object combination, as seen in
Figure~\ref{f9}, while the visual classification listed in the same
table is based only on the comparison with single standards.

\textit{J0006$\relbar$1319}: the two spectra that we obtained yield a
single-standard visual classification of L6$\pm$2 pec
with some features hinting at an unresolved binary. We analyzed the
GNIRS spectrum with the method described above and found that an
L6: and T7: combination ($\chi^2=0.16$) is a slightly better match to our
spectrum than the best single object fit (L8,
$\chi^2=0.22$). \citet{luh14b} previously classified this object as an
L5 pec and mentioned that its peculiar spectrum could be due to an
unresolved binary, comparing it to 2MASS J17114573$+$2232044, a strong
L5.5+T5.5 binary candidate (\citealt{bard15} and references
therein).

\textit{J0006$\relbar$0852}: this binary candidate has a spectrum well
matched by a combination of an M9: 
and a T6: pec ($\chi^2=0.29$), while the match to a single object
gives an M9 classification but with a $\chi^2=0.51$. It was
previously confirmed as an M8.5+T5 binary by radial velocity variability
(named SDSS J0006$\relbar$0852~AB; \citealt{bur12}). It may 
form a triple system with the M7 dwarf LP 704-48, on the basis of a
common proper motion (\citealt{bur12}).

\textit{J0013$\relbar$1816}: the strange shape
of this object SpeX spectrum, and the fact that the single-object
index-based classification (L3.5$\pm$1.5) differs significantly from the
visual classification (L6.5 pec) led us to investigate the possible
binary nature of this source. Indeed, a better match is obtained with
two objects of spectral types L3: and T4: ($\chi^2=1.04$). \citet{bar15}
discovered this object by its common proper motion with NLTT 687
(M3$\pm$0.5), obtained a GMOS spectrum, and classified
J0013$\relbar$1816 as an L1$\pm$0.5 in the optical. It was also
classified in the NIR as an L5 pec with SpeX spectroscopy by
\citet[named PSO J003.4$\relbar$18]{bes15}, and identified it as a
possible early L + early T binary.

\textit{J2249$\relbar$1628}: this source
spectrum displays peculiarities that make it a strong binary
candidate. The combined spectra of an L4 and a T1 give the best match to
the observed spectrum ($\chi^2=1.07$) and this match is much better than
any fit to a single spectrum of the SpeX libraries (L4:, $\chi^2=4.31$)
or the standard sequence (L5.5$\pm$1). \citet{bes15} previously
classified this object (named as PSO J342.3$\relbar$16) as an L5: and
mentioned that the morphology of its spectrum could be due to an
unresolved L+T binary.

\subsection{Other Peculiar Objects} \label{peculiar}

Objects that do not fit in any of the previous categories while still
presenting anomalies are listed in Table~\ref{t5}h. This includes
objects classified as pec$^1$ (possibly red) and pec$^2$ (possibly
blue), that did not meet all the criteria required to be listed in
subsection~\ref{reddwarf} or \ref{bluedwarf}.

\textit{J0006$\relbar$2158}: this source is classified as an M9:: pec
because the $H$ and $K$ band NIRI spectra differ from the M8-9 standards
(bump on the blue side of the $H$ band, suppressed flux in the center of
the $K$ band), while its 2MASS color $J\relbar K_s=1.39\pm0.25$ is
red for an M9.

\textit{J0532$\relbar$3253}: the adopted classification for this source
is M9.5: pec$^2$ due to a blue 2MASS color $J\relbar K_s=0.89\pm0.26$
which is unconfirmed by its GNIRS spectrum. It features a blue slant in
the $H$ band, but a binary analysis could not find a better match to the
spectrum.

\textit{J0552$+$0210}: this peculiar object is probably a reddened K dwarf,
given its strange spectrum, rather low proper motion, and red NIR colors
(2MASS: $J\relbar K_s=1.12\pm0.29$; MKO: $J\relbar K=1.50\pm0.10$).

\textit{J0858$+$2710}: this object has a classification of L1.5: pec
because of the suppressed blue part and enhanced red part of the $J$
band, the weaker CO drop at $\sim2.3\,\mu$m and the blue slant of the $H$
band, but a normal 2MASS color $J\relbar K_s=1.38\pm0.07$. A
binary analysis was performed on this object but yielded no good
match. \citet{wes08} used an optical SDSS spectrum automatically
processed with the HAMMER stellar spectral-typing facility
(\citealt{cov07}), which is limited to the M0--L0 spectral types, to
classify this object as an L0. It was also classified as an L0 with SDSS
spectroscopy by \citet{sch10}, and \citet{zha10} found it to be a
candidate wide binary with LP 312-49 (M4). \citet{bar15} obtained a GMOS
spectrum and classified this object as an L0$\pm$1 in the optical, while
confirming that it is comoving with LP 312-49.

\textit{J1039$+$2440}: the spectrum of this source is peculiar and
somewhat similar to the L5$\,\gamma$ template; however, its $H$ band is not as
triangular as the template and, while its \ion{Na}{1} and \ion{K}{1}
lines are weaker than usual, its FeH lines are normal. Since this
candidate cannot be confidently confirmed as young, we adopt a spectral
type of L5:: pec.

\textit{J1118$\relbar$0856}: our spectra from three different
facilities are all consistent with an L6.5
classification. The SpeX spectrum is blue, while the SIMON
spectrum shows a normal continuum. Because of this ambiguity
and the fact that the 2MASS $J\relbar K_s=1.56\pm0.12$ is within the
1$\sigma$ blue limit of the mean ($J\relbar K_s=1.73\pm0.22$), this
object is classified as pec$^2$. It was previously discovered
by \citet{kir10} and was classified as an L6 pec (blue) with a SpeX
spectrum.

\textit{J1132$\relbar$3809}: we adopt a spectral type of L8: pec
for this source. The blue side of its $H$ band displays enhanced flux
compared to the standard L8--L9. An analysis for binarity
could not identify a good match to this spectrum.

\textit{J1143$+$1905}: the peculiar spectrum obtained here,
especially in the $J$ band, is classified visually as an L8$\pm$2. We
mainly used the $H$ and $K$ bands for this classification, but the spectral
indices yield an earlier type of L5.5$\pm$1.0, resulting in an adopted NIR
classification of L7:: pec. This object was classified by \citet{zha09}
as an L9 on the basis of its SDSS colors.

\textit{J1256$\relbar$1002}: this object's peculiar spectrum displays a
$K$ band flux level higher than the standards, and a very peculiar $J$
band, leading to a classification of M8:: pec. It also has a very
red 2MASS color $J\relbar K_s=1.81\pm0.22$. The \citet{all13} gravity
score of VL-G ($\gamma$ type) indicates that it could be young, though
the spectrum does not match any of the young M templates of
\citet{gag15b}.

\textit{J1455$+$2619}: we classify this object as an L2: pec.
With its 2MASS color $J\relbar K_s=1.27\pm0.08$ and its suppressed flux
in the $H$ and $K$ bands (which match the L1 standard almost perfectly),
it could be a blue L dwarf; however, the blue part of its $J$ band is suppressed,
not enhanced, compared to the L2 standard. \citet{sch10} previously
classified it as an L1 UCD using SDSS spectroscopy.

\textit{J1631$\relbar$1922} this peculiar object, with red NIR colors (2MASS
$J\relbar K_s=1.53\pm0.08$) is probably a reddened M dwarf.

\textit{J2351$+$3010}: the adopted classification for this source is
L5.5:. Our SpeX spectrum matches almost perfectly the L6 standard,
except for a slightly enhanced $K$ band flux, while the spectral
indices yield a type of L5$\pm$1.5. \citet{schn14} classified this
object as an L6.5 dwarf. However, \citet{kir10} and \citet{gag15b}
classified it as an L5 pec (red) and L5 pec respectively.

\section{SUMMARY} \label{Summary}

A sample of 169 spectra have been obtained
from different observatories for UCD candidates found through proper
motion detection by comparing source position from the SIMP survey to
2MASS. From this sample, 164 UCDs, with spectral types from M8 to T6.5,
were found, including 2 young field brown dwarfs, 6 unusually red M and
L dwarfs, 25 unusually blue M and L dwarfs, 2 candidate unresolved L+T binaries
and 24 other peculiar UCDs. We also discovered 9 L/T transition dwarfs
(L6-T4.5), which helps better populate this interesting range of
spectral type, often the target of variability studies.

This work demonstrates that many undiscovered UCDs remain to be found in
preexisting surveys, including several peculiar objects. Proper motion
has proved to be an efficient method to uncover those missed
objects. Furthermore, most objects found are relatively bright ($J <
16$), enabling follow-up studies (parallax, variability, binarity, etc.).

\acknowledgments

We thank the referee for suggestions that led to significant improvement
of the manuscript. This work was supported in part through grants from the Fonds
de Recherche Qu\'eb\'ecois-Nature et Technologie and the
Natural Science and Engineering Research Council of Canada.
This research has benefited from the SpeX Prism Spectral
Libraries, maintained by Adam Burgasser at \protect\url{http://pono.ucsd.
edu/~adam/browndwarfs/spexprism}, as well as the M, L, T,
and Y dwarf compendium housed at \protect\url{http://DwarfArchives.org}
and maintained by Chris Gelino, Davy Kirkpatrick, and Adam
Burgasser, whose server was funded by a NASA Small
Research Grant, administered by the American Astronomical
Society. This research made use of the SIMBAD database and
VizieR catalog access tool, operated at the Centre de Donn\'ees
astronomiques de Strasbourg, France (\citealt{wen00});
data products from the Two Micron All Sky Survey (2MASS;
\citealt{skr06}), which is a joint project of the University of
Massachusetts and the Infrared Processing and Analysis Center
(IPAC)/California Institute of Technology (Caltech), funded by the
National Aeronautics and Space Administration (NASA) and the National
Science Foundation; data products from the {\it Wide-field Infrared Survey
Explorer} ({\it WISE}; \citealt{wri10}), which is a joint project of the
University of California, Los Angeles, and the Jet Propulsion Laboratory
(JPL)/Caltech, funded by NASA; the NASA/IPAC Infrared
Science Archive (IRSA), which is operated by JPL, Caltech,
under contract with NASA; the Infrared Telescope Facility
(IRTF), which is operated by the University of Hawaii under
Cooperative Agreement NNX-08AE38A with NASA, Science
Mission Directorate, Planetary Astronomy Program; the
Database of Ultracool Parallaxes maintained by Trent Dupuy
(\citealt{dup12}). Based on observations
obtained at the Gemini Observatory through program numbers
GS-2006B-Q19, GS-2007A-Q-28, GN-2008A-Q-51, and GN-2008B-Q-80. The
Gemini Observatory is operated by the Association of
Universities for Research in Astronomy, Inc., under a
cooperative agreement with the National Science Foundation
(NSF) on behalf of the Gemini partnership: the NSF (United
States), the National Research Council (Canada), CONICYT (Chile), the
Australian Research Council (Australia), Minist\'erio 
da Ci\^encia, Tecnologia e Inova\c{c}\~{a}o (Brazil), and Ministerio de
Ciencia, Tecnolog{\'{\i}}a e Innovaci\'on Productiva (Argentina). Some
data were acquired through the Canadian Astronomy Data
Center. This publication uses observations obtained at IRTF through
program numbers 2008A085, 2008B054, and 2009A065. The authors recognize and
acknowledge the very significant cultural role and reverence that the
summit of Mauna Kea has always had within the indigenous Hawaiian
community. We are most fortunate to have the opportunity to conduct
observations from this mountain.

{\it Facilities:} \facility{IRTF (SpeX)}, \facility{Gemini:South
  (GNIRS)}, \facility{Gemini:Gillett (NIRI)}, \facility{OMM (SIMON)},
  \facility{OMM (CPAPIR)}, \facility{CTIO:1.5 m (CPAPIR)}, \facility{CFHT (MegaCam)}.

\clearpage

\clearpage

\clearpage

\clearpage

\begin{figure}
\centering
     \begin{subfigure}{\textwidth}
     \includegraphics[trim = 15mm 28mm 40mm 16mm, clip, width=0.33\textwidth]{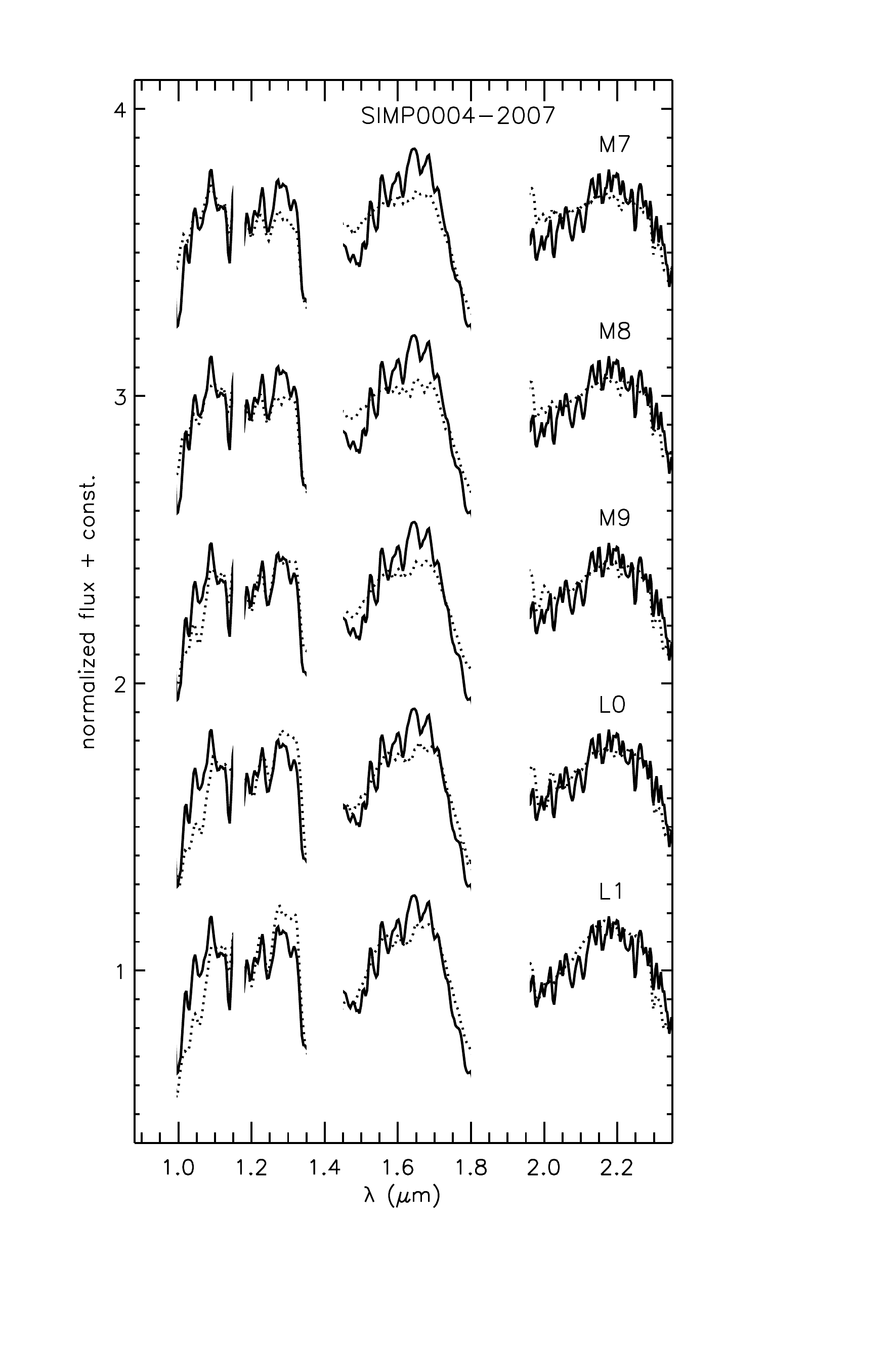}
     \includegraphics[trim = 15mm 28mm 40mm 16mm, clip, width=0.33\textwidth]{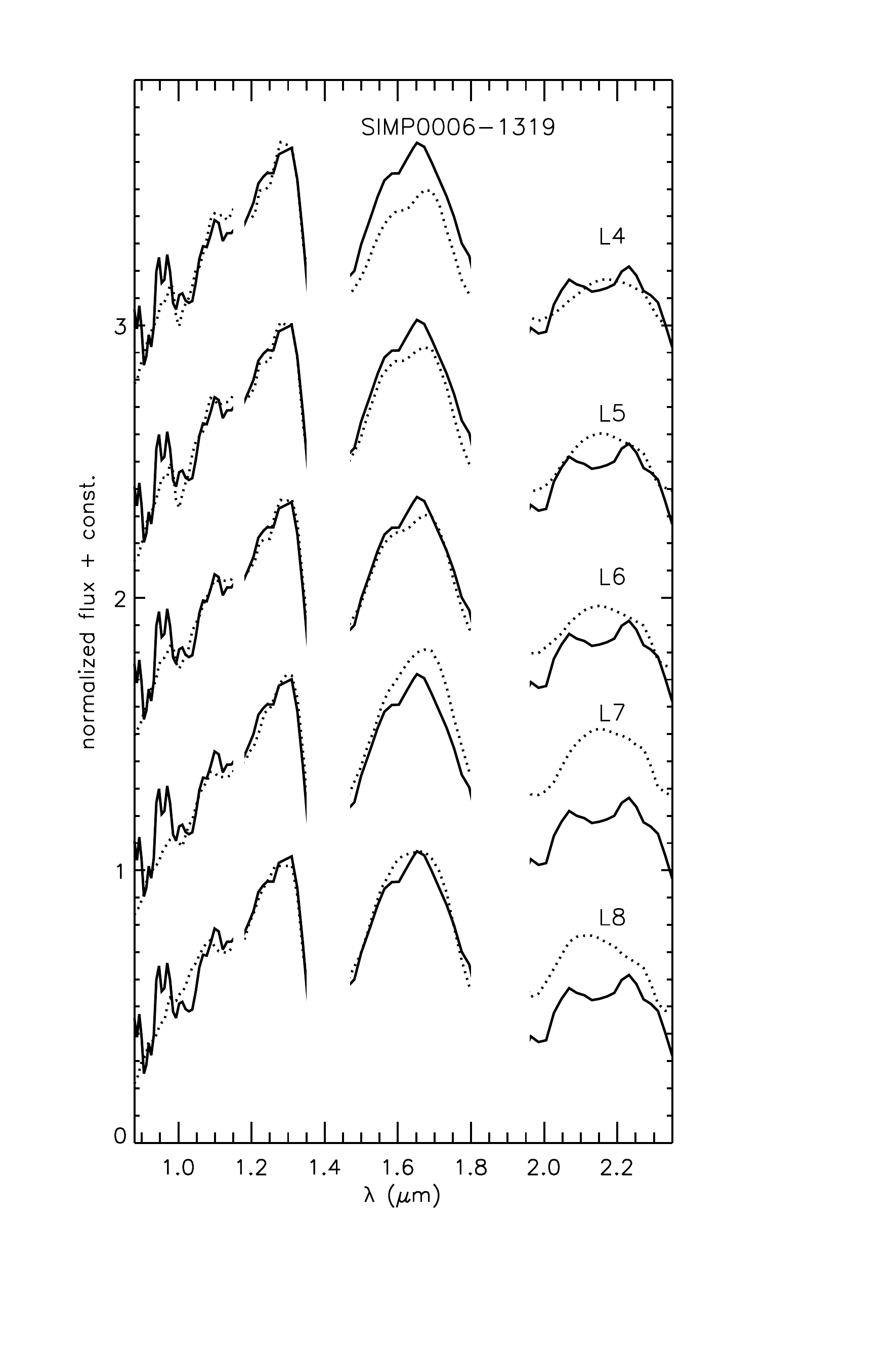}
     \includegraphics[trim = 15mm 28mm 40mm 16mm, clip, width=0.33\textwidth]{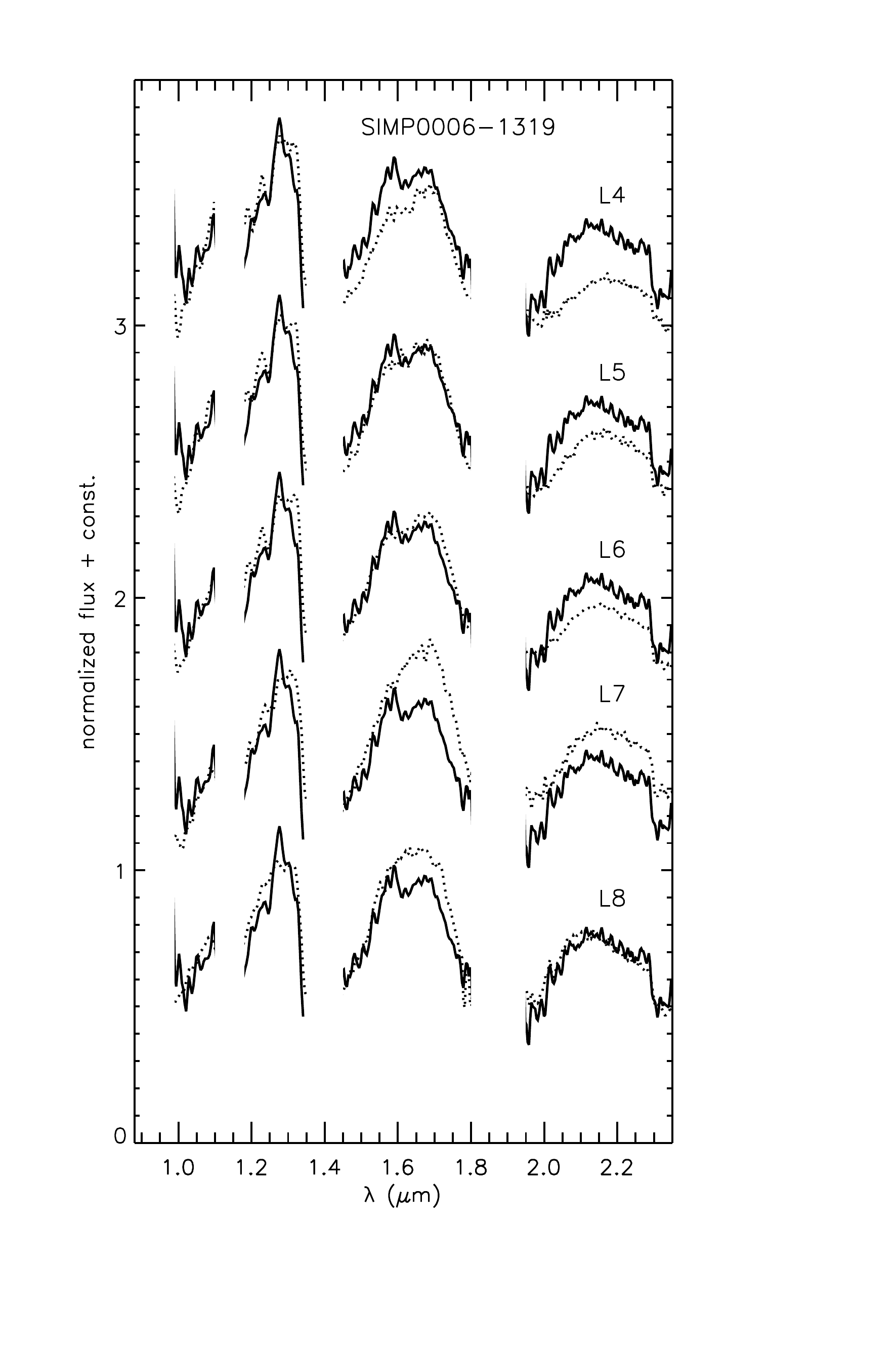}
     \caption{0004-2007 (NIRI: M9$\pm$2 pec) \& 0006-1319 (SIMON: L6$\pm$2 pec; GNIRS: L6$\pm$2 pec)}\label{f4a}
    \end{subfigure}
     \begin{subfigure}{\textwidth}
     \includegraphics[trim = 15mm 28mm 40mm 16mm, clip, width=0.33\textwidth]{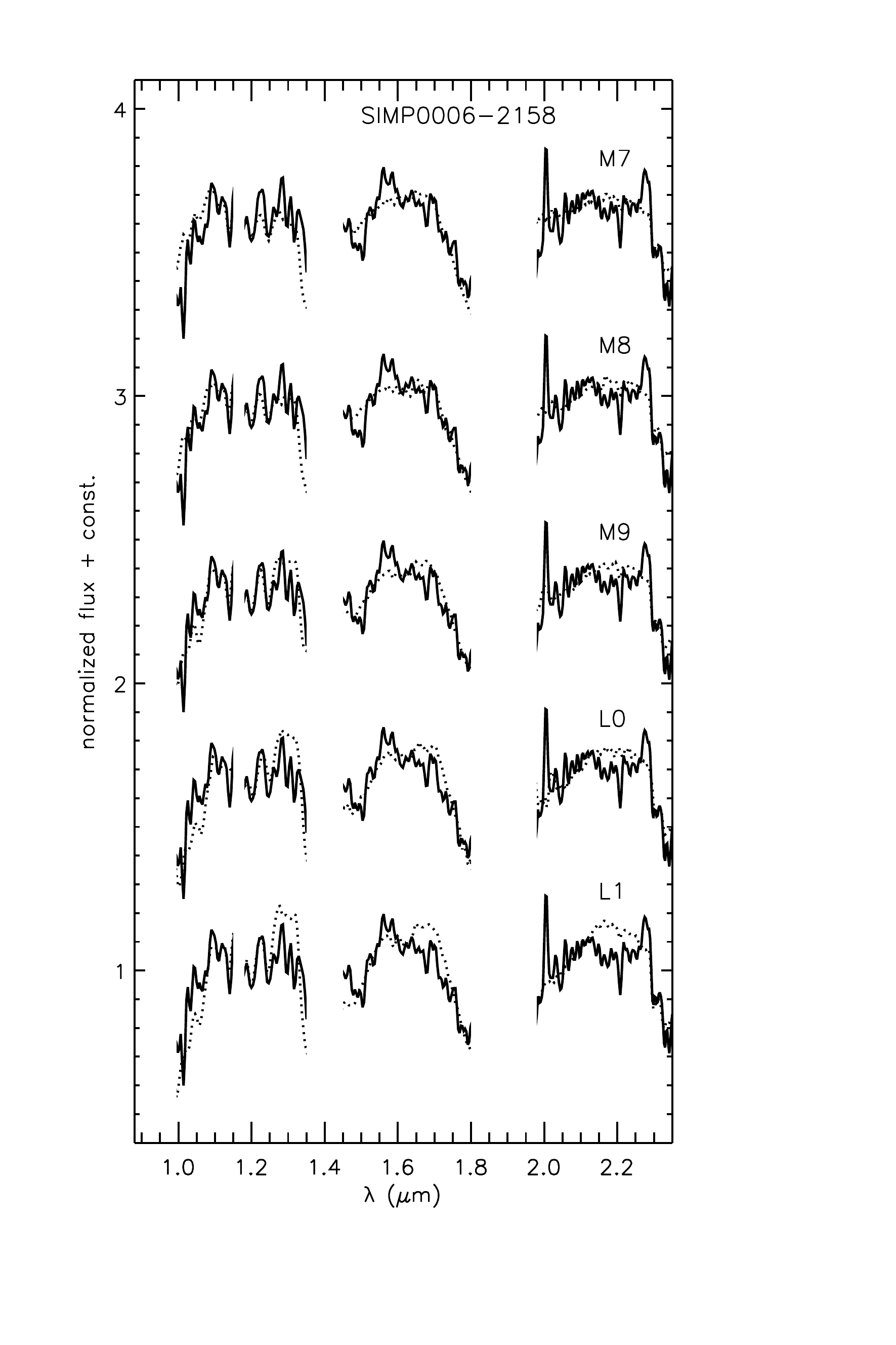}
     \includegraphics[trim = 15mm 28mm 40mm 16mm, clip, width=0.33\textwidth]{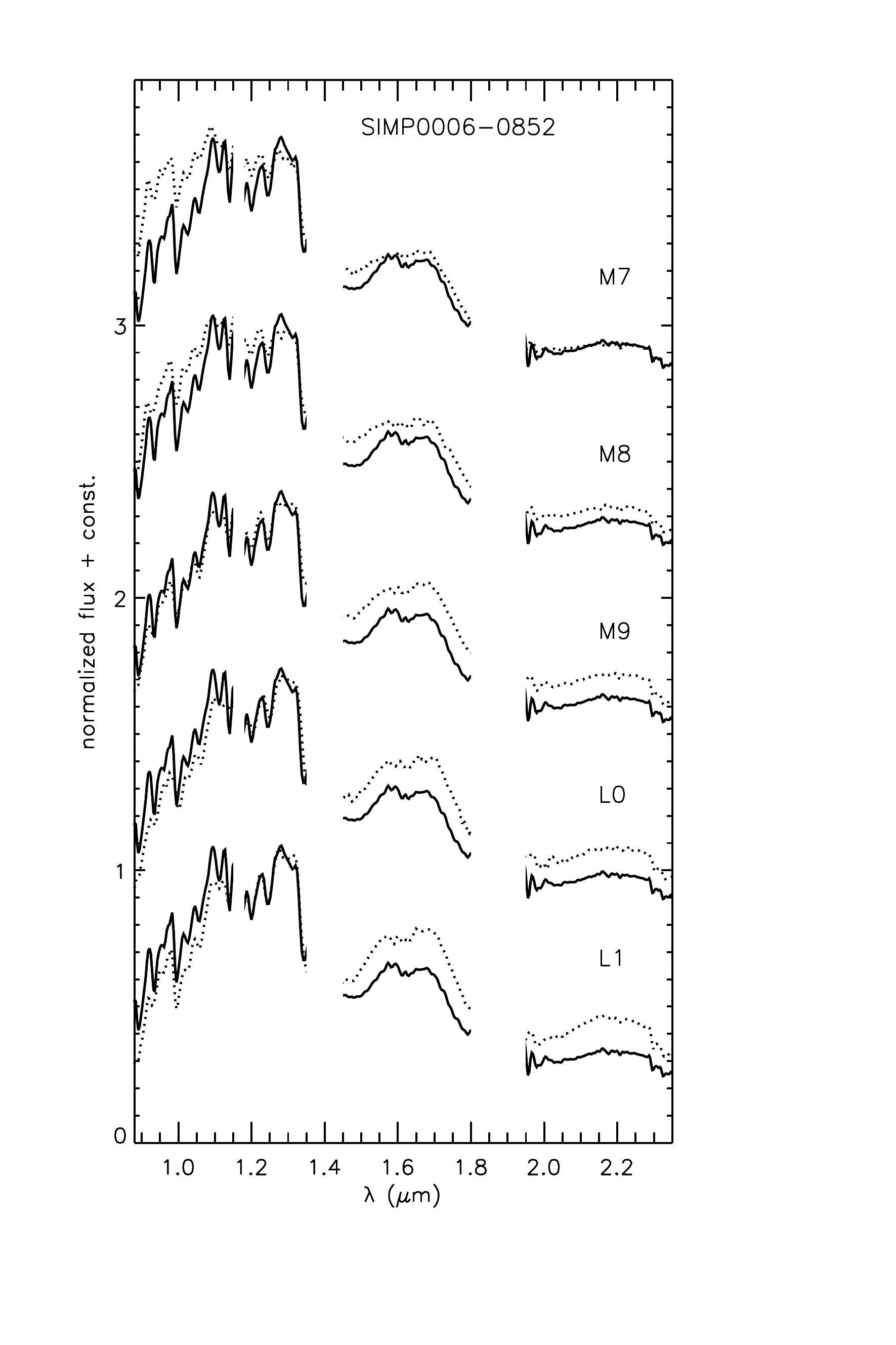}
     \includegraphics[trim = 15mm 28mm 40mm 16mm, clip, width=0.33\textwidth]{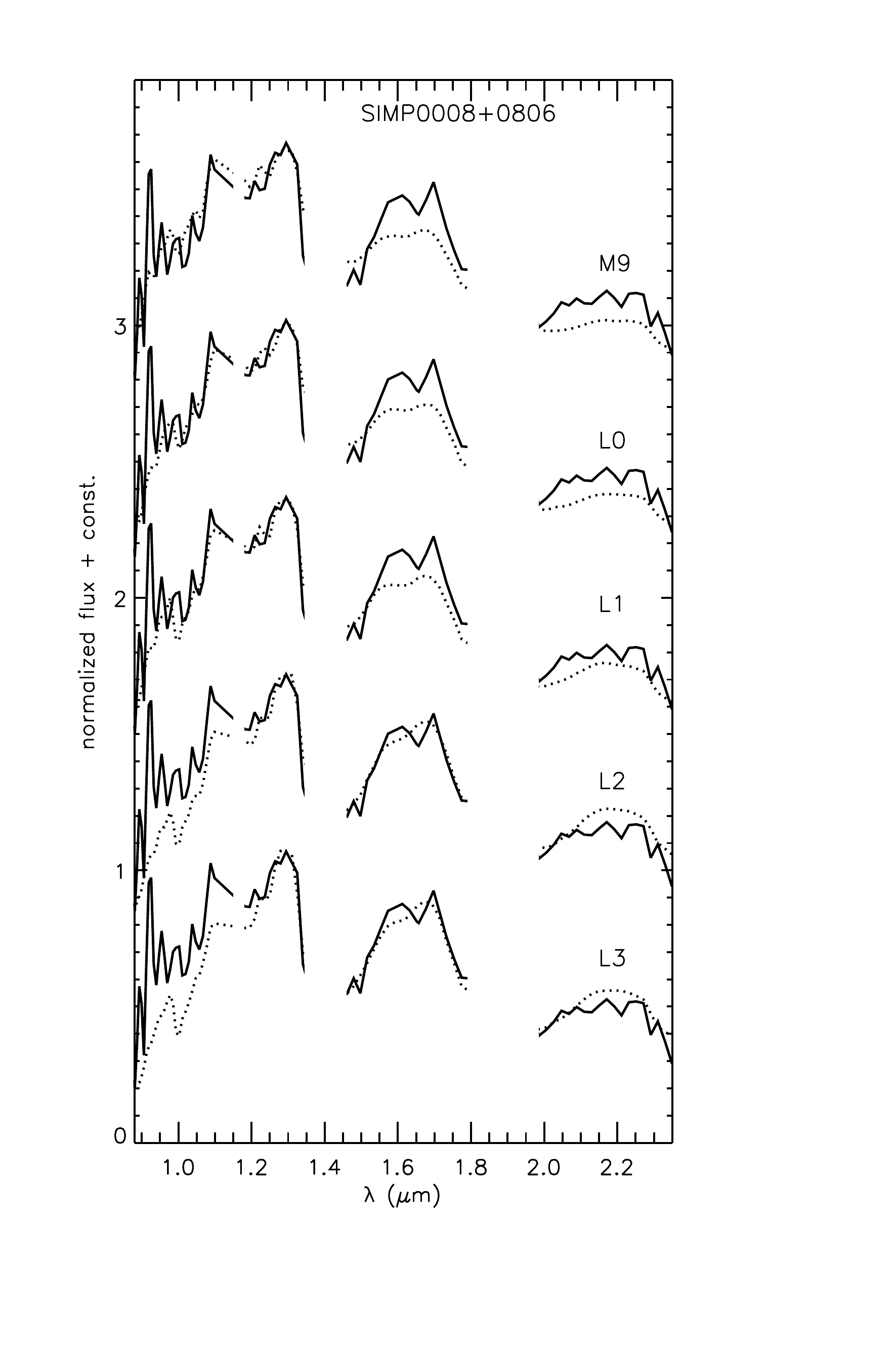}
     \caption{0006-2158 (NIRI: M8.5$\pm$2), 0006-0852 (GNIRS: M9$\pm$1 pec) \& 0008+0806 (SIMON: L0.5$\pm$2)}\label{f4b}
    \end{subfigure}
\caption{Comparison of the spectra of the targets (solid
 lines) to the M, L, or T dwarf standards (dotted lines). All
 spectra (publicly available at
 }\label{f4}
\end{figure}

\clearpage

\begin{figure}
\ContinuedFloat
\centering
\caption{\protect\url{http://dx.doi.org/10.5281/zenodo.58501}) are normalized to
 the average flux in the 1.235--1.305$\,\mu$m region (or over the
 complete range of each individual band in the case of NIRI
 spectra) and the target spectrum is offset by aconstant in each panel. The spectral standards are VB~8 (M7;
  \citealt{bur08}), VB~10 (M8; \citealt{bur04}), LHS~2924 (M9;
  \citealt{burM06}), 2MASSP J0345432+254023 (L0; \citealt{burM06}),
  2MASSW J2130446$\relbar$084520 (L1; \citealt{kir10}), Kelu--1 (L1;
  \citealt{bur07b}), 2MASSW J1506544+132106 (L3;
  \citealt{bur07}), 2MASS J21580457$\relbar$1550098 (L4;
  \citealt{kir10}), SDSS J083506.16+195304.4
  (L5; \citealt{chi06}), 2MASSI J1010148$\relbar$040649   (L6;
  \citealt{rei06}), 2MASSI J0103320+193536 (L7; \citealt{cru04}),2MASSW
  J1632291+190441 (L8; \citealt{bur07}), DENIS-P J0255$\relbar$4700 (L9;
  \citealt{bur06b}), SDSS J120747.17+024424.8 (T0; \citealt{loo07}), SDSS J015141.69+124429.6 (T1; \citealt{bur04}),
  SDSSp J125453.90-012247.4 (T2; \citealt{bur04}),
  2MASS J12095613-1004008 (T3; \citealt{bur04}), 2MASSI J2254188+312349
  (T4; \citealt{bur04}), 2MASS J15031961+2525196 (T5; \citealt{bur04}),
  SDSSp J162414.37+002915.6 (T6; \citealt{bur06}), 2MASSI
  J0727182+171001 (T7; \citealt{bur06}), and 2MASSI J0415195-093506 (T8;
  \citealt{bur04}).\\
  (The complete figure set (197 images) is available in the online journal.)}
\end{figure}

\clearpage

\begin{figure}
\centering
         \begin{subfigure}{\textwidth}
          \includegraphics[width=0.495\textwidth]{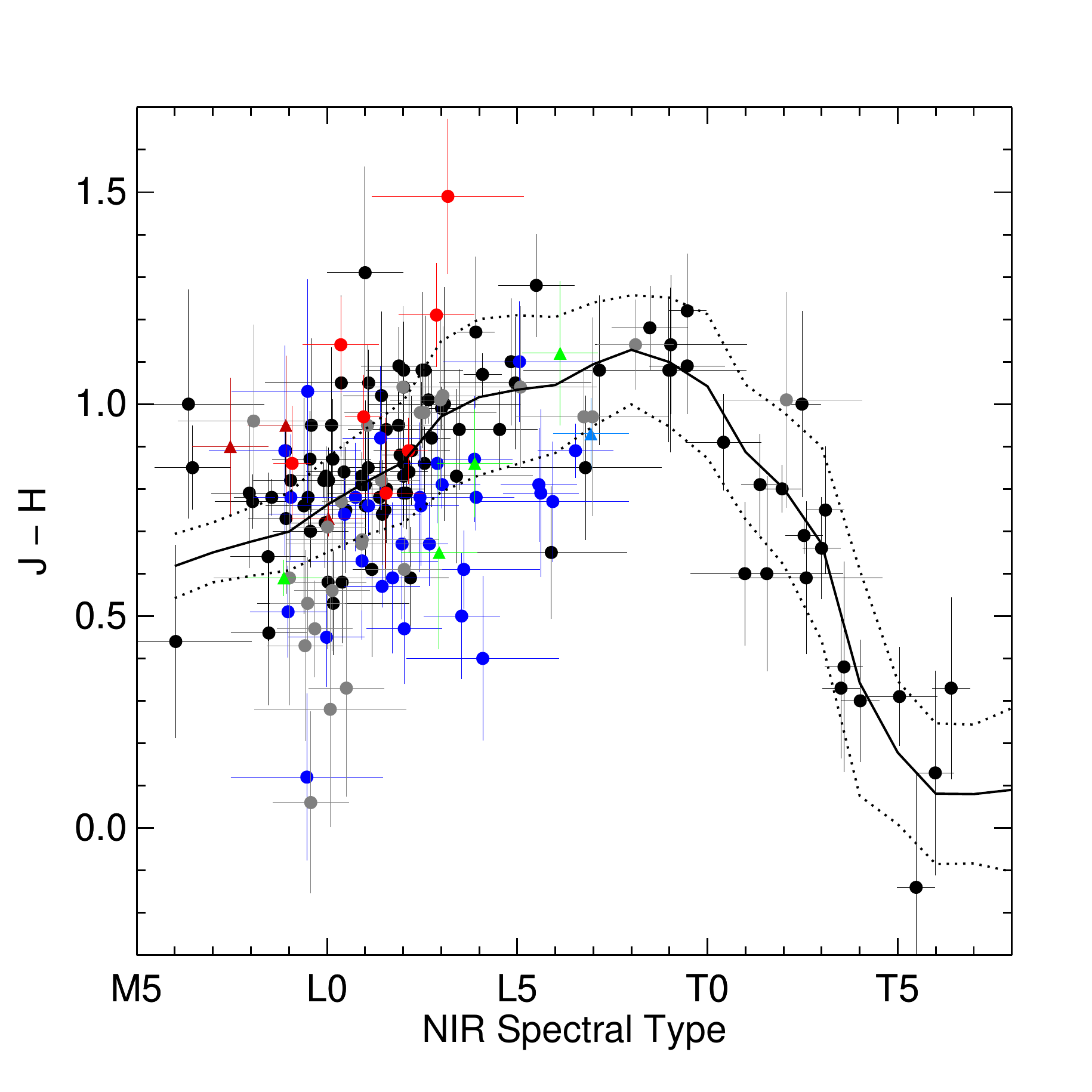}
          \includegraphics[width=0.495\textwidth]{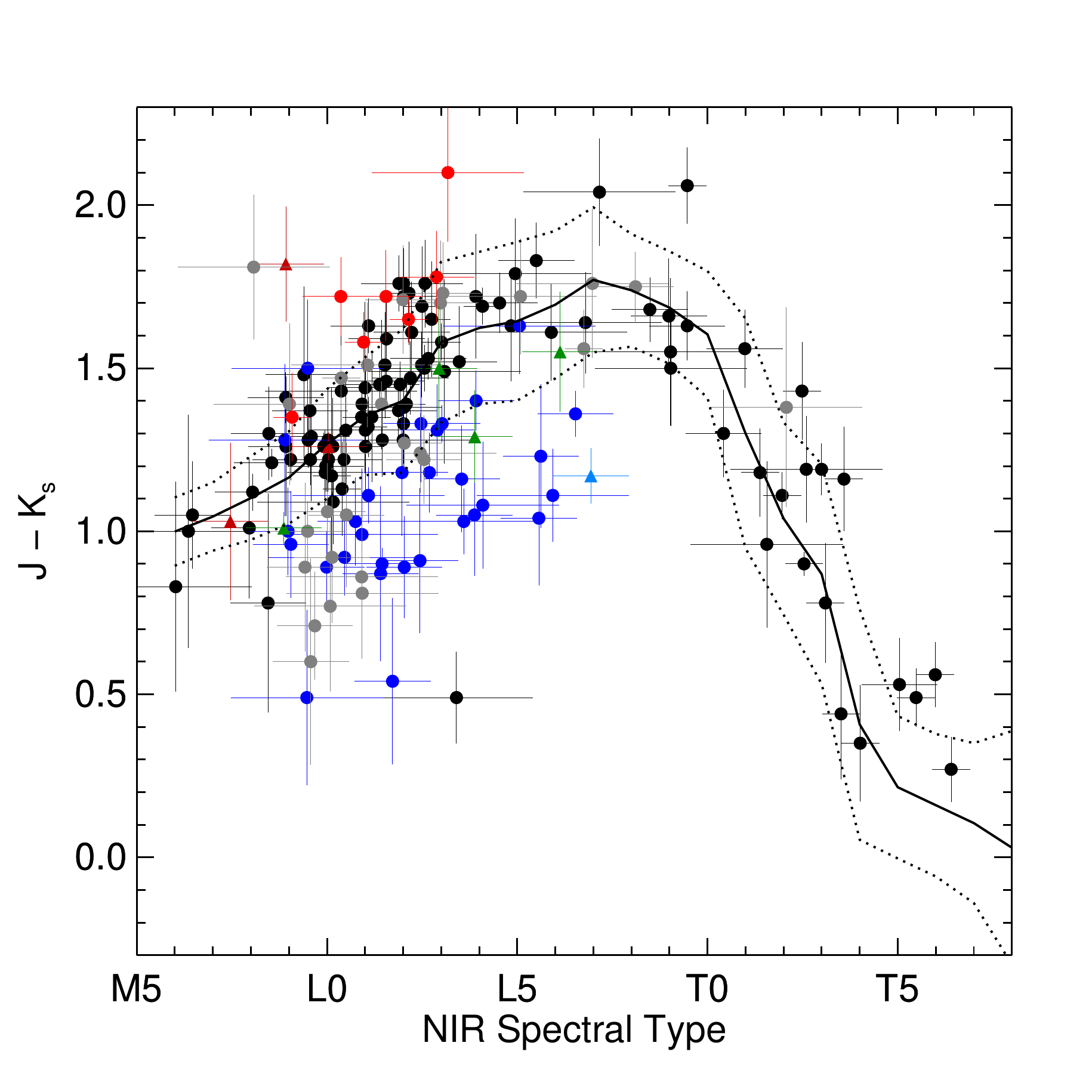}
          \caption{$J-H$ and $J\relbar K_s$ vs spectral type}\label{f5a}
        \end{subfigure}
         \begin{subfigure}{\textwidth}
          \includegraphics[width=0.495\textwidth]{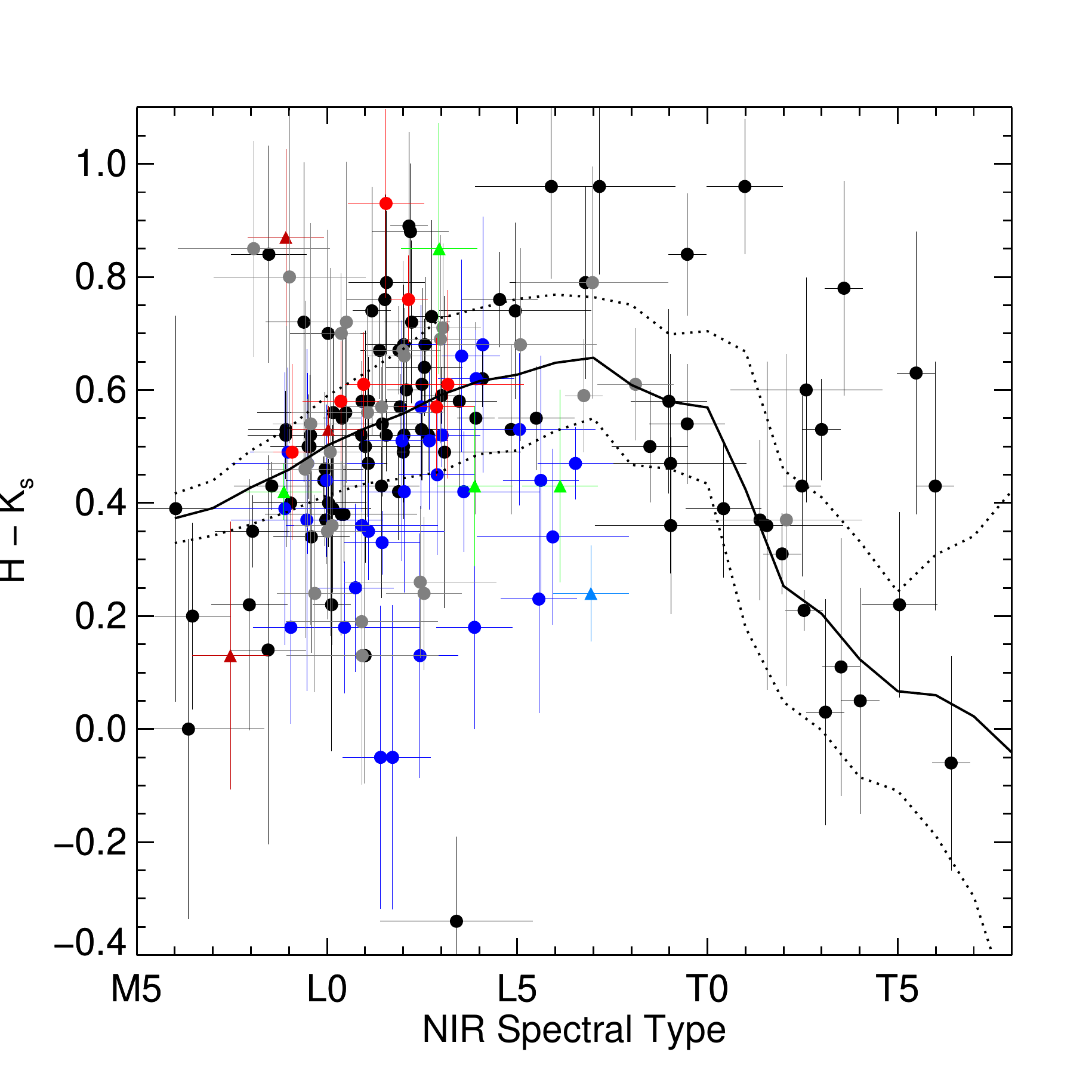}
          \includegraphics[width=0.495\textwidth]{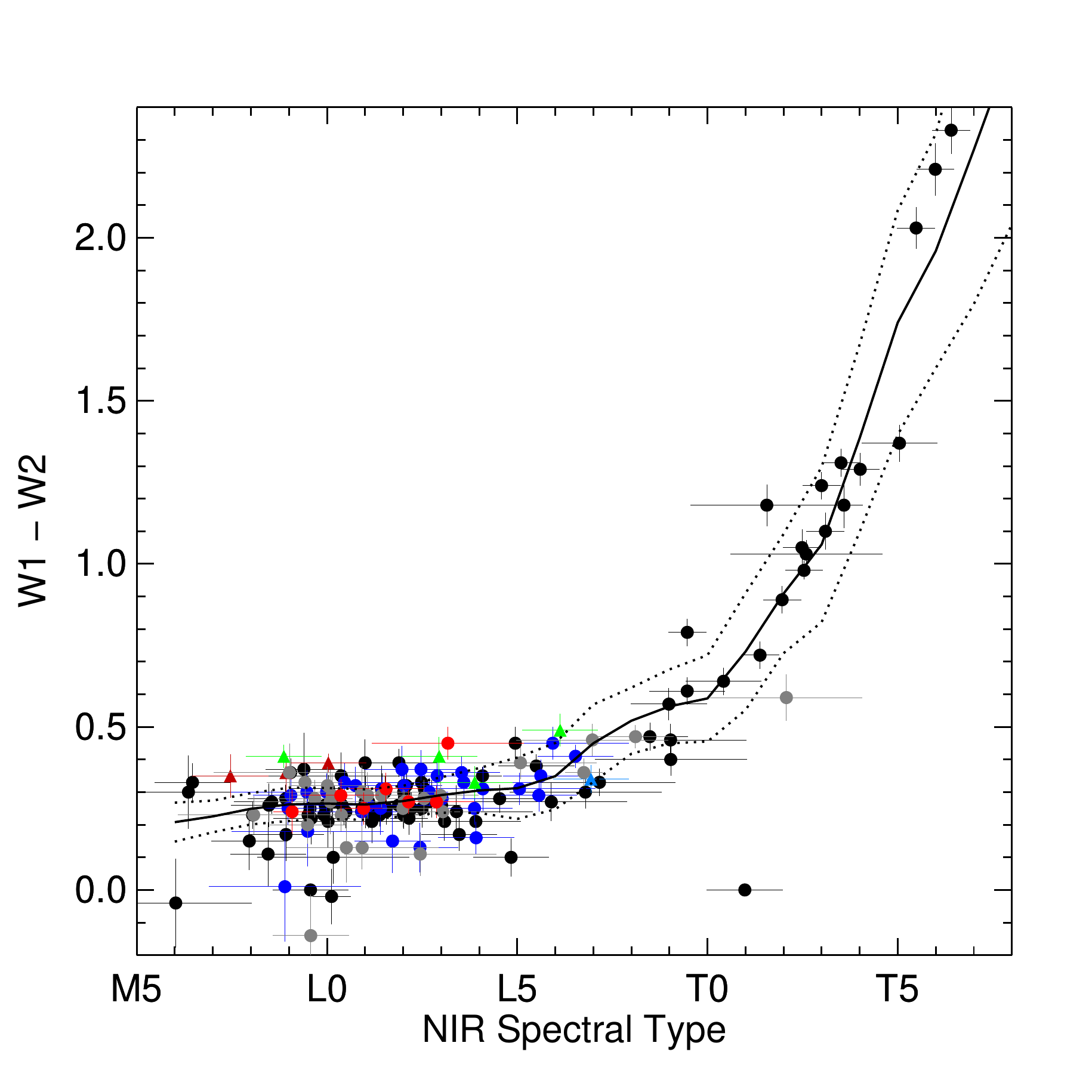}
          \caption{$H-K$ and $W1-W2$ vs spectral type}\label{f5b}
        \end{subfigure}
\caption{Color-NIR spectral type sequences for UCDs found in this work
  (black circles for non-peculiar; blue circles for blue dwarfs; light blue
  triangle for the subdwarf; red circles for red dwarfs; dark red triangles for
  young dwarfs; dark green triangles for binary dwarfs and gray circles for
  other peculiar dwarfs). Random noise smaller than 0.25 subtype was
  added to the spectral types for visibility. The uncertainties on
  spectral type were taken as $\pm$1 for ``:'' and $\pm$2 for ``::.'' The
  color distribution (black line) and $1\sigma$ dispersion (dotted line)
  were calculated from the median color}\label{f5} 
\end{figure}
         
\begin{figure}
\ContinuedFloat
\centering
\begin{subfigure}{\textwidth}
          \includegraphics[width=0.495\textwidth]{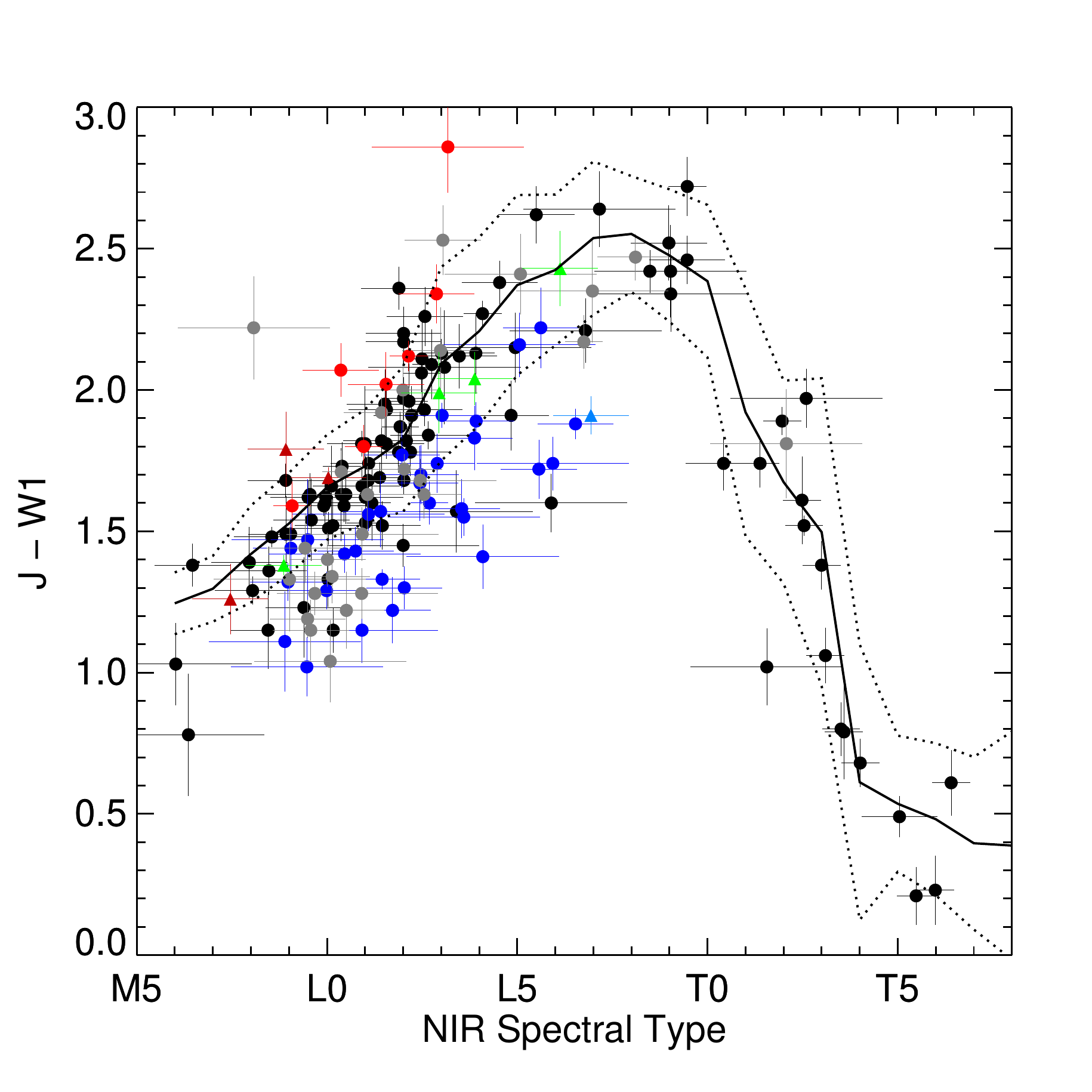}
          \includegraphics[width=0.495\textwidth]{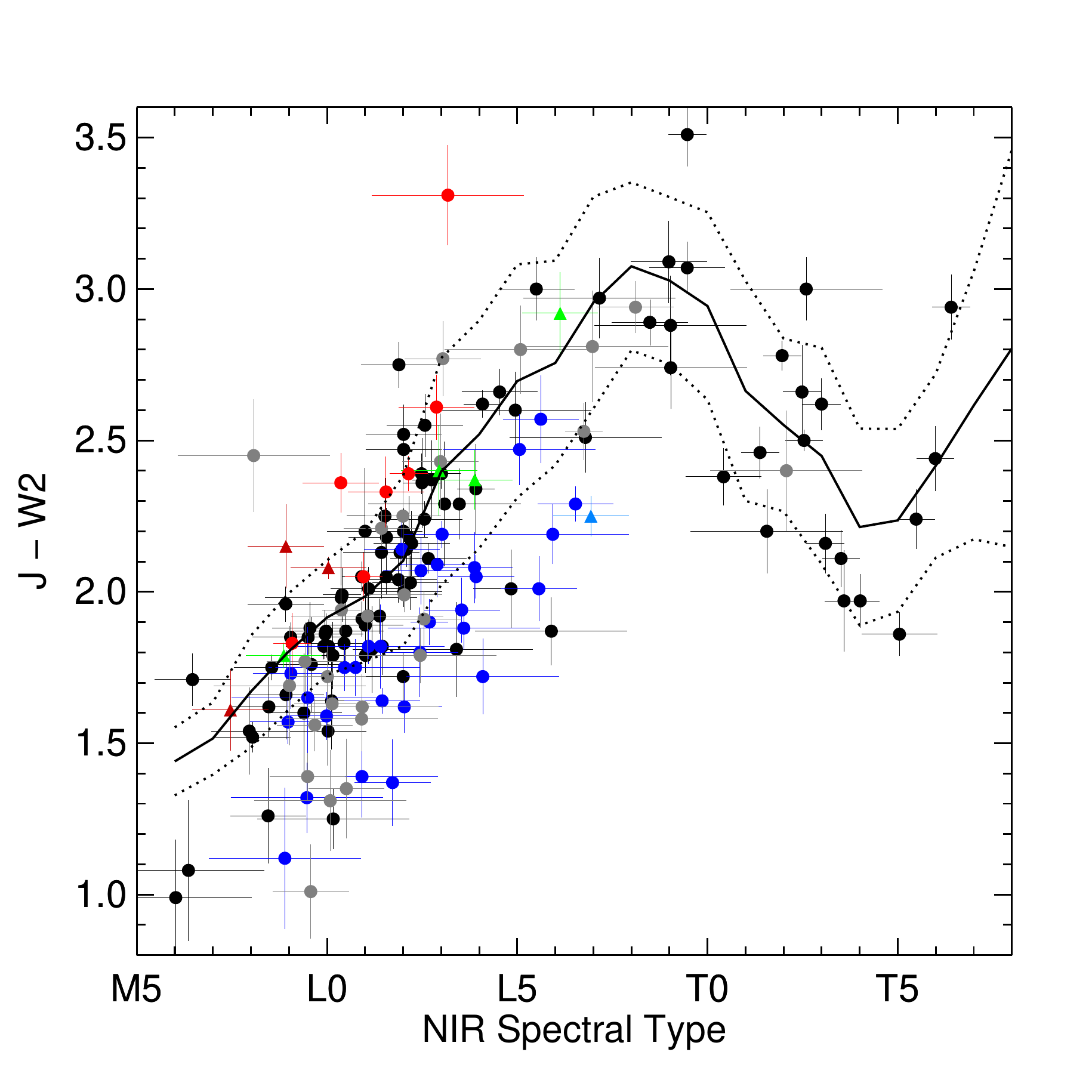}
          \caption{$J-W1$ and $J-W2$ vs spectral type}\label{f5c}
        \end{subfigure}
         \begin{subfigure}{\textwidth}
          \includegraphics[width=0.495\textwidth]{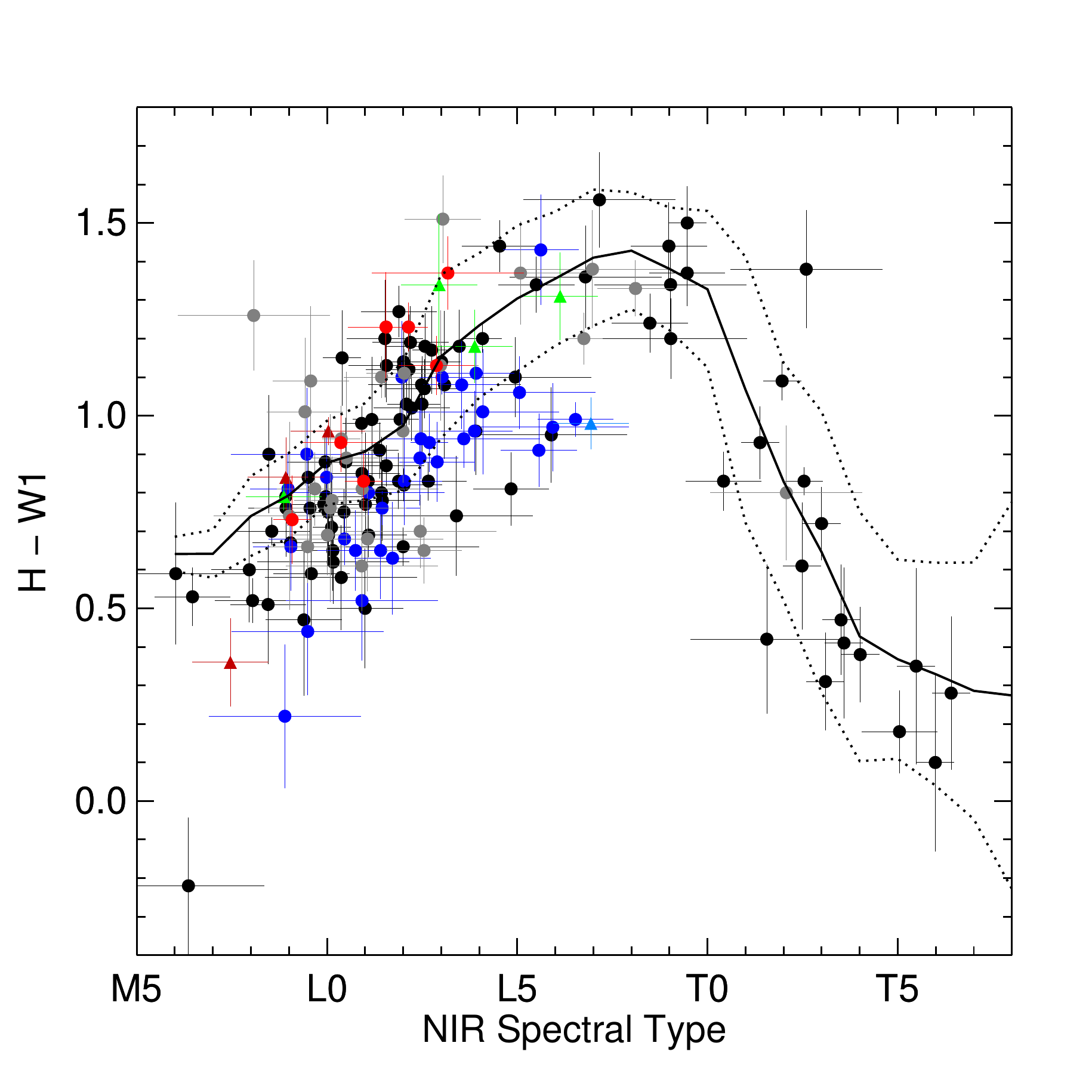}
          \includegraphics[width=0.495\textwidth]{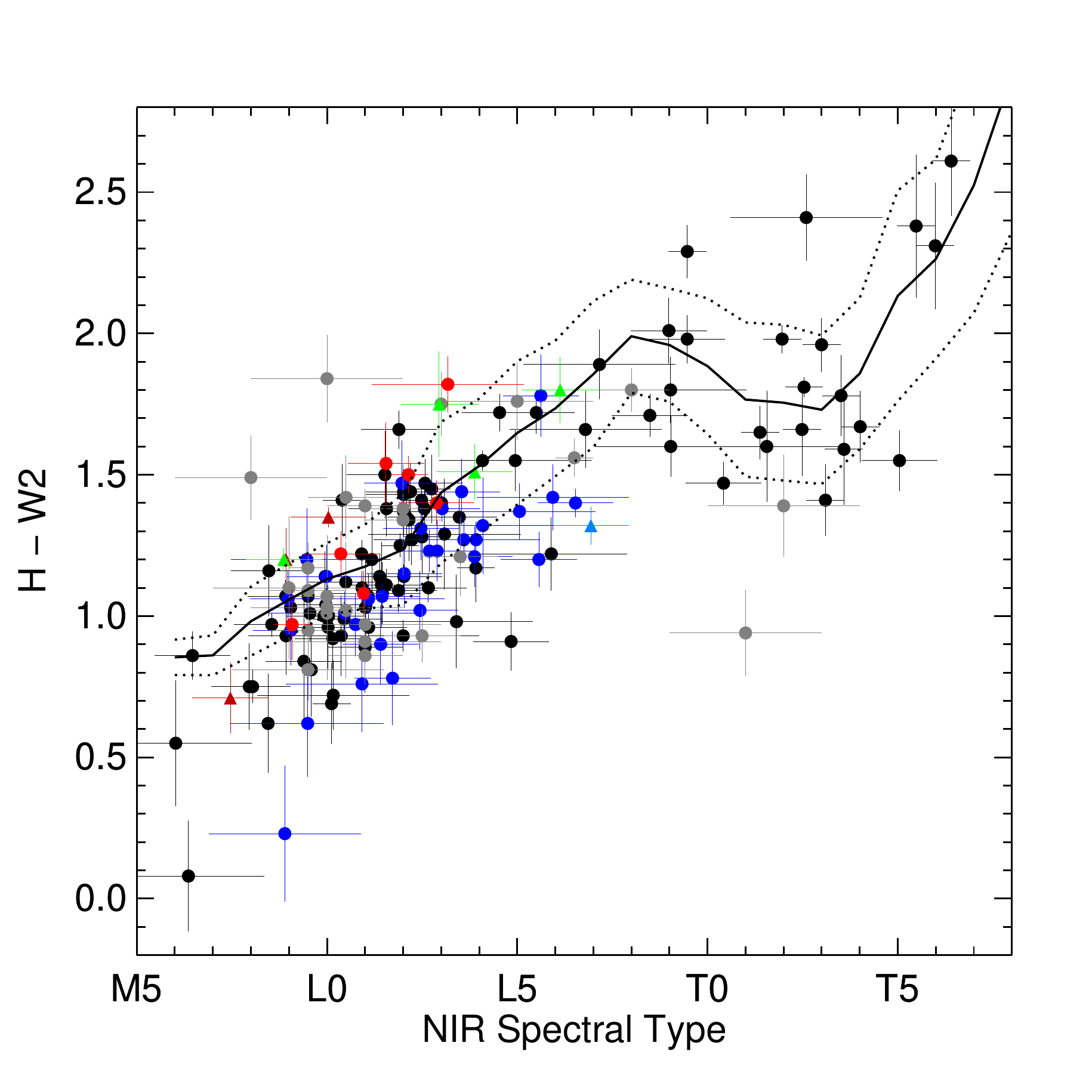}
          \caption{$H-W1$ and $H-W2$ vs spectral type}\label{f5d}
        \end{subfigure}
\caption{(Continued) and dispersion within 1 spectral
  type of non-peculiar known UCDs using 2MASS data taken from
  \protect\url{http://www.astro.umontreal.ca/~gagne/listLTYs.php}.}
\end{figure}

\begin{figure}
\ContinuedFloat
\centering
\begin{subfigure}{\textwidth}
          \includegraphics[width=0.495\textwidth]{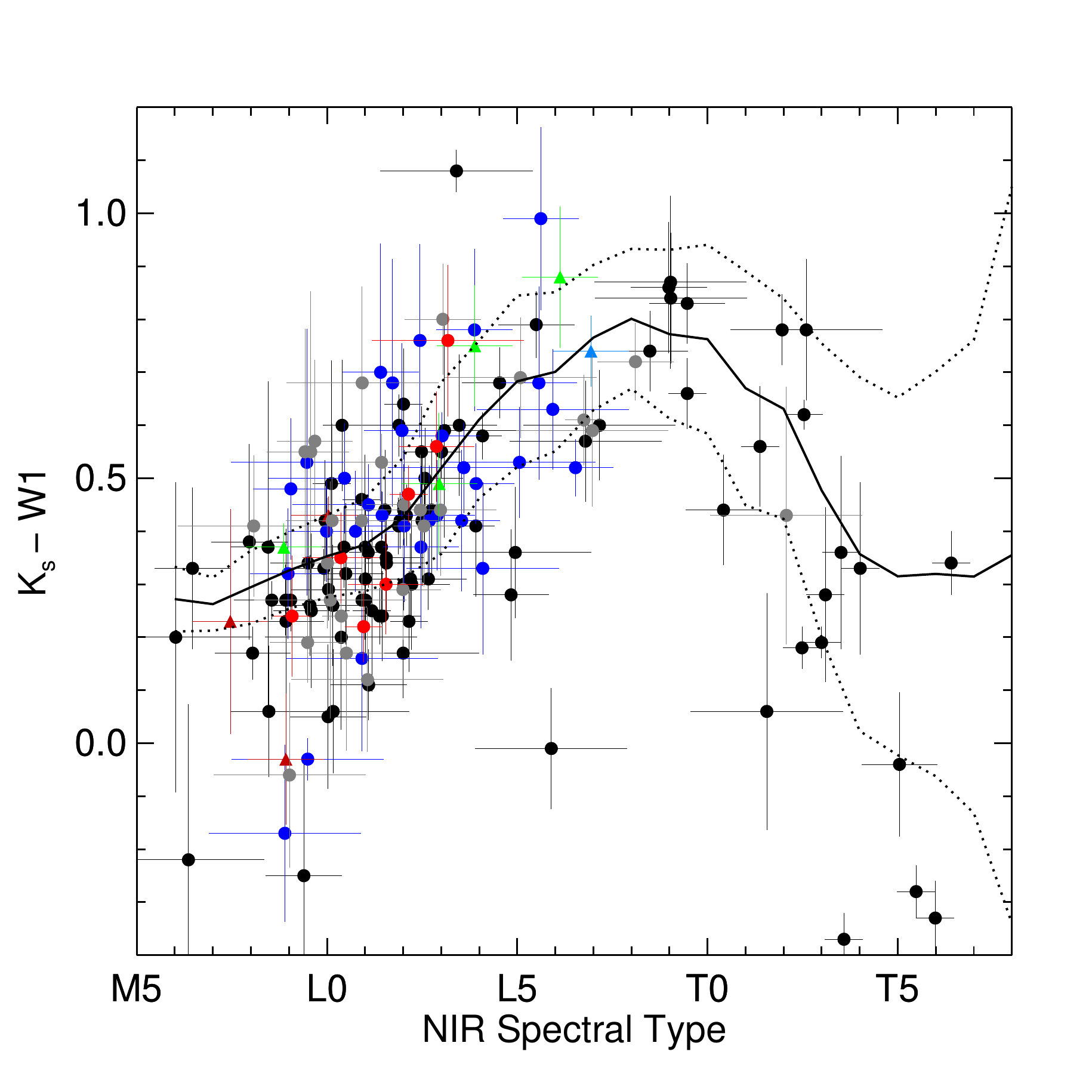}
          \includegraphics[width=0.495\textwidth]{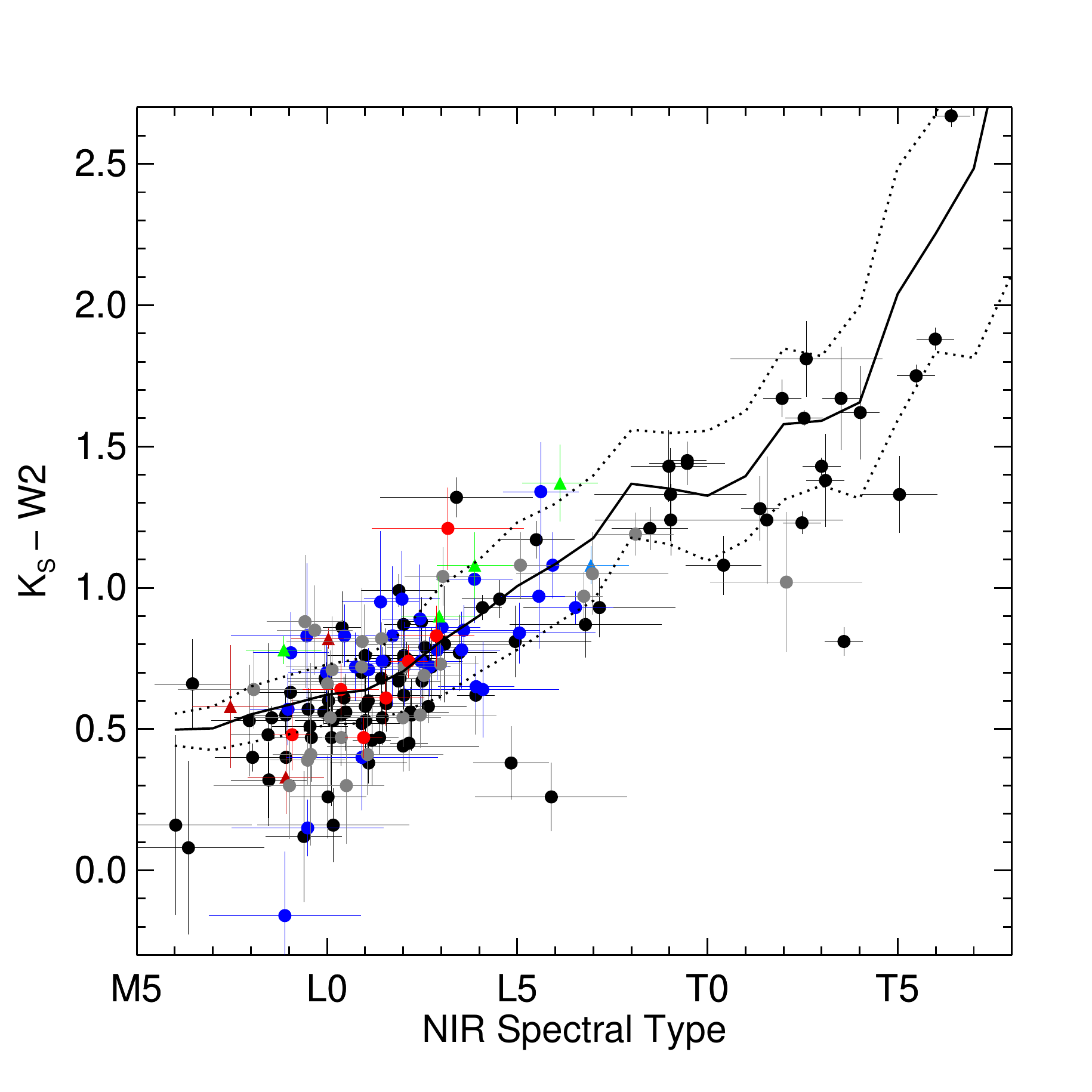}
          \caption{$K_s-W1$ and $K_s-W2$ vs spectral type}\label{f5e}
        \end{subfigure}
\caption{(Continued)}
\end{figure}

\clearpage

\begin{figure}
\centering
     \begin{subfigure}{\textwidth}
     \includegraphics[width=0.995\textwidth]{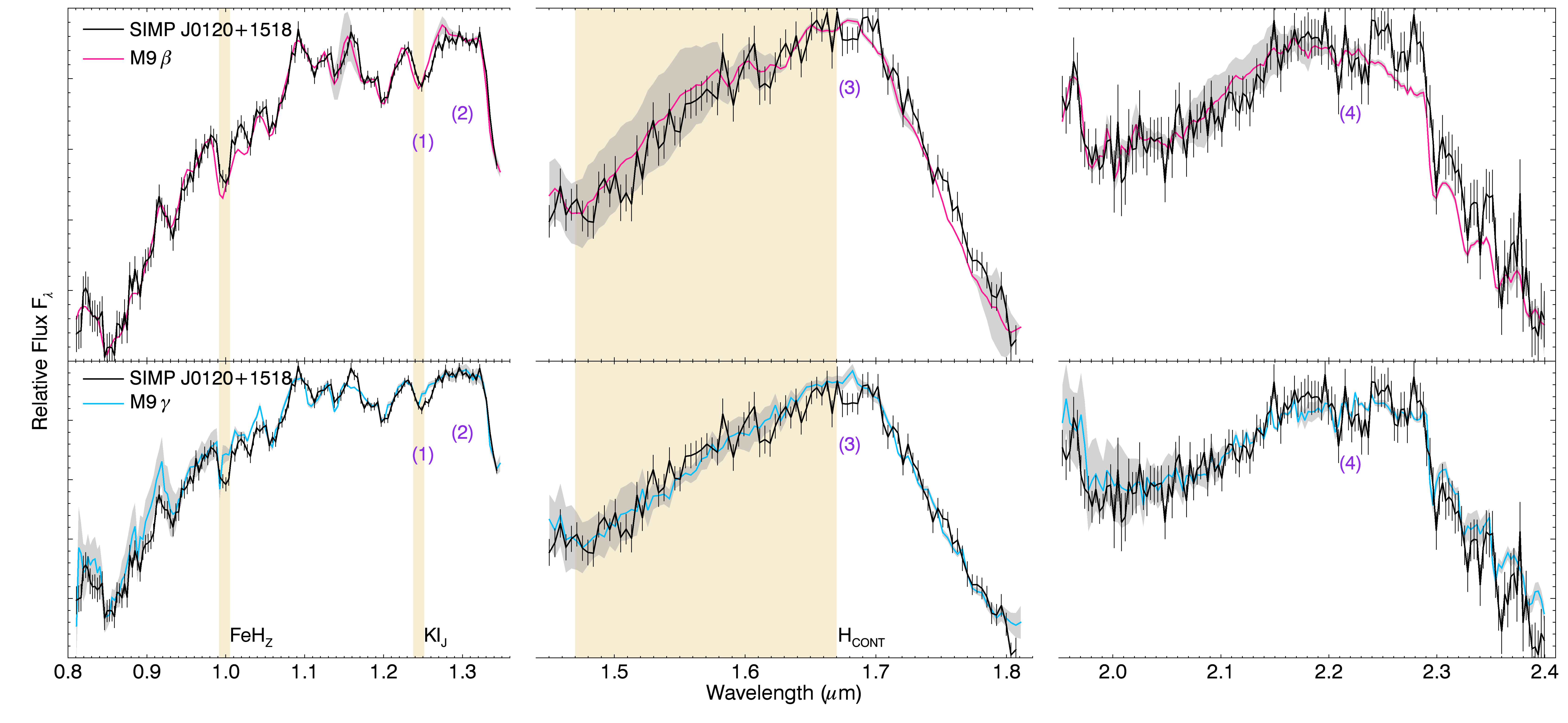}
     \caption{J0120$+$1518 (M9:$\,\gamma$)}\label{f6a}
    \end{subfigure}
 \begin{subfigure}{\textwidth}
     \includegraphics[width=0.995\textwidth]{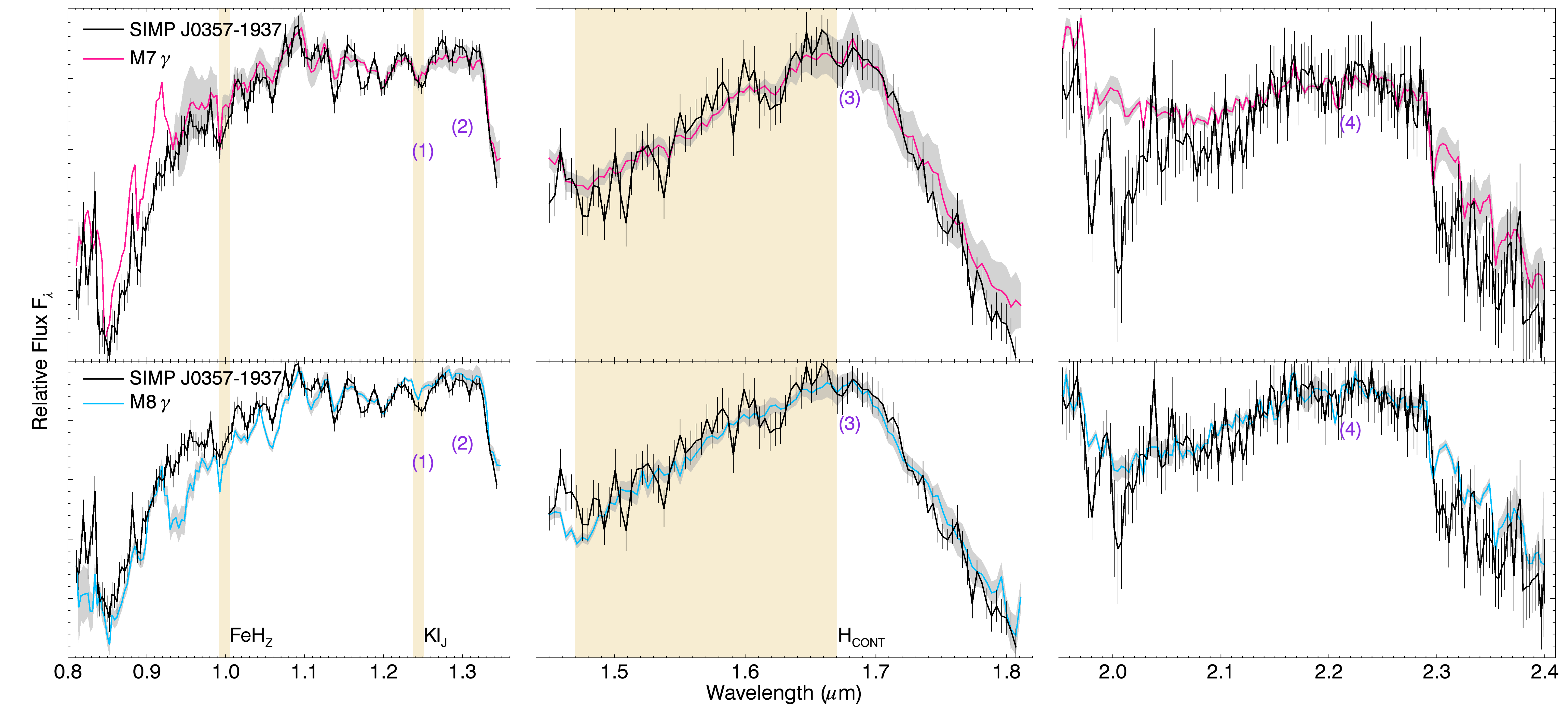}
     \caption{J0357$\relbar$1937 (M7.5:$\,\gamma$)}\label{f6b}
    \end{subfigure}
\caption{NIR spectrum (black) of the SIMP candidates, compared with low-
  and very-low-gravity template spectra. Each band was normalized
 individually. The gray shaded region represents the scatter of the
 individual objects used to create the templates and the black vertical
 lines represent the measurement uncertainty on each bin of the
 observed spectrum. Beige shaded regions correspond to the locations of
 gravity-sensitive spectral indices defined by \citet{all13}. We denote
 regions useful to differentiate between the low- and very-low-gravity
 templates with purple numbers.}\label{f6}
\end{figure}

\begin{figure}
\ContinuedFloat
\centering
    \begin{subfigure}{\textwidth}
     \includegraphics[width=0.995\textwidth]{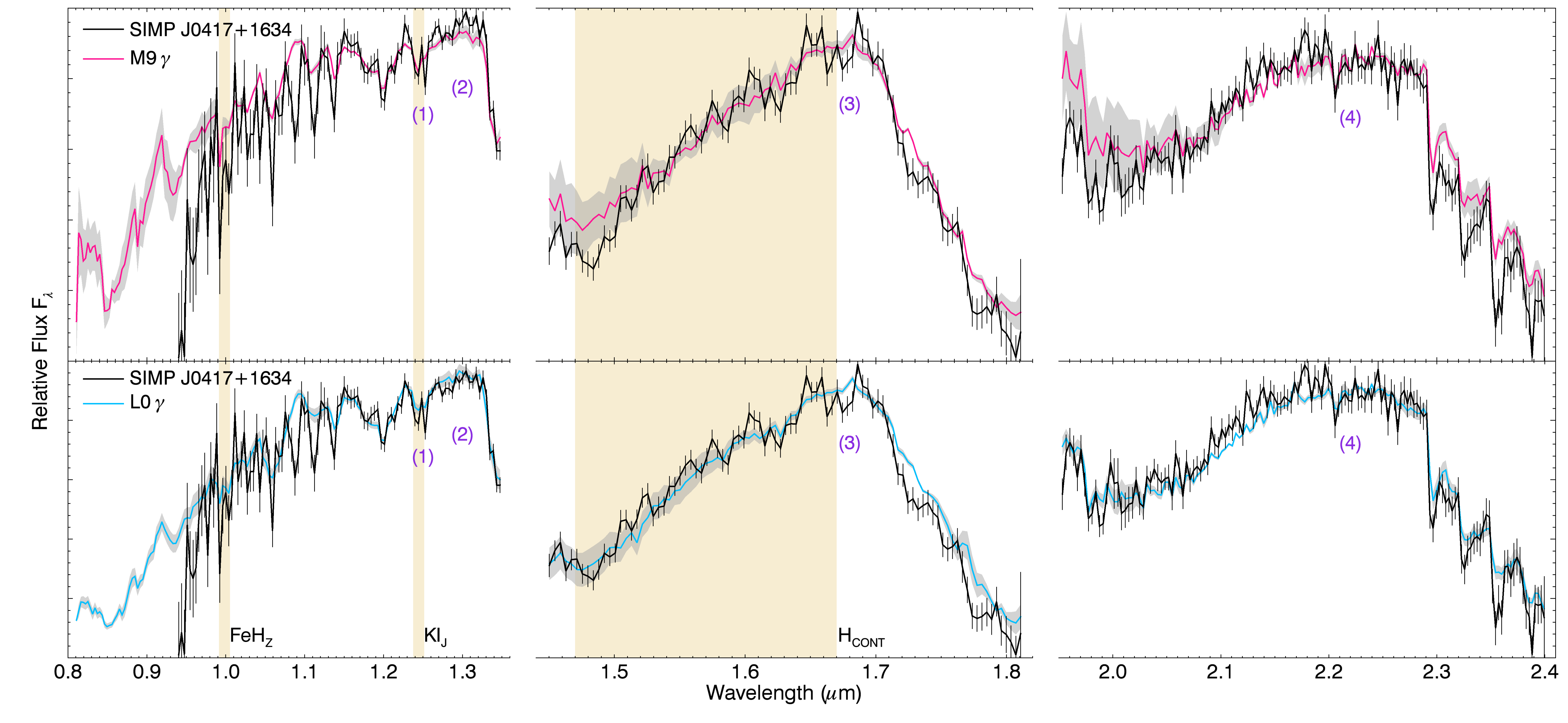}
      \caption{J0417$+$1634 (L0:$\,\gamma$)}\label{f6c}
    \end{subfigure}
\caption{(Continued)}
\end{figure}

\clearpage

\begin{figure}
  \centering
  \begin{subfigure}{\textwidth}
    \includegraphics[width=0.99\textwidth]{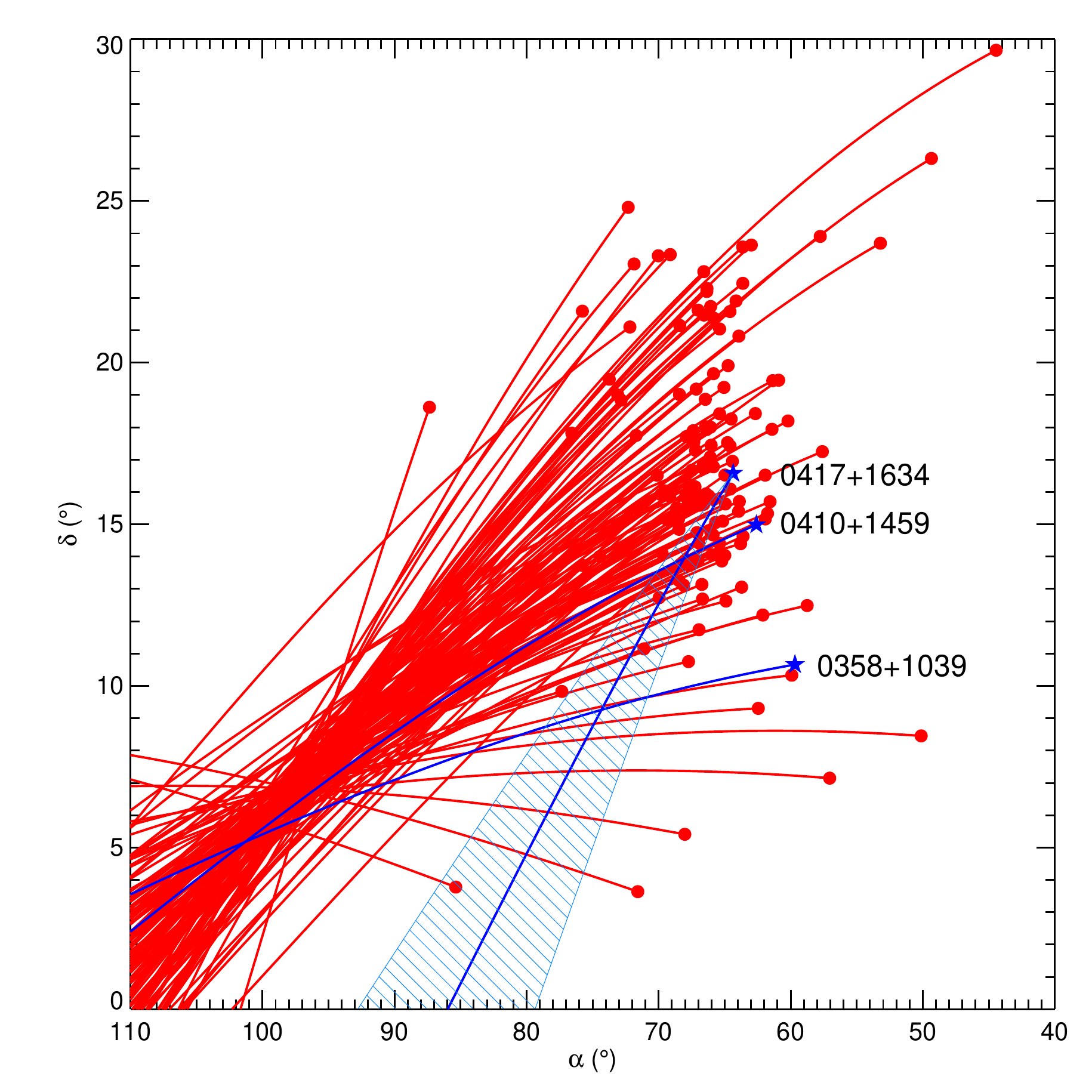}
  \end{subfigure}
  \caption{Proper motion paths for the three SIMP Hyades candidates
    (blue), including the $1\sigma$ uncertainty for SIMPJ0417$+$1634, and for
    the Hyades members taken from \citet[red]{per98} with proper motion from
    \citet{van07}.}\label{f7}
\end{figure}

\clearpage

\begin{figure}
  \centering
  \begin{subfigure}{\textwidth}
    \includegraphics[width=0.99\textwidth]{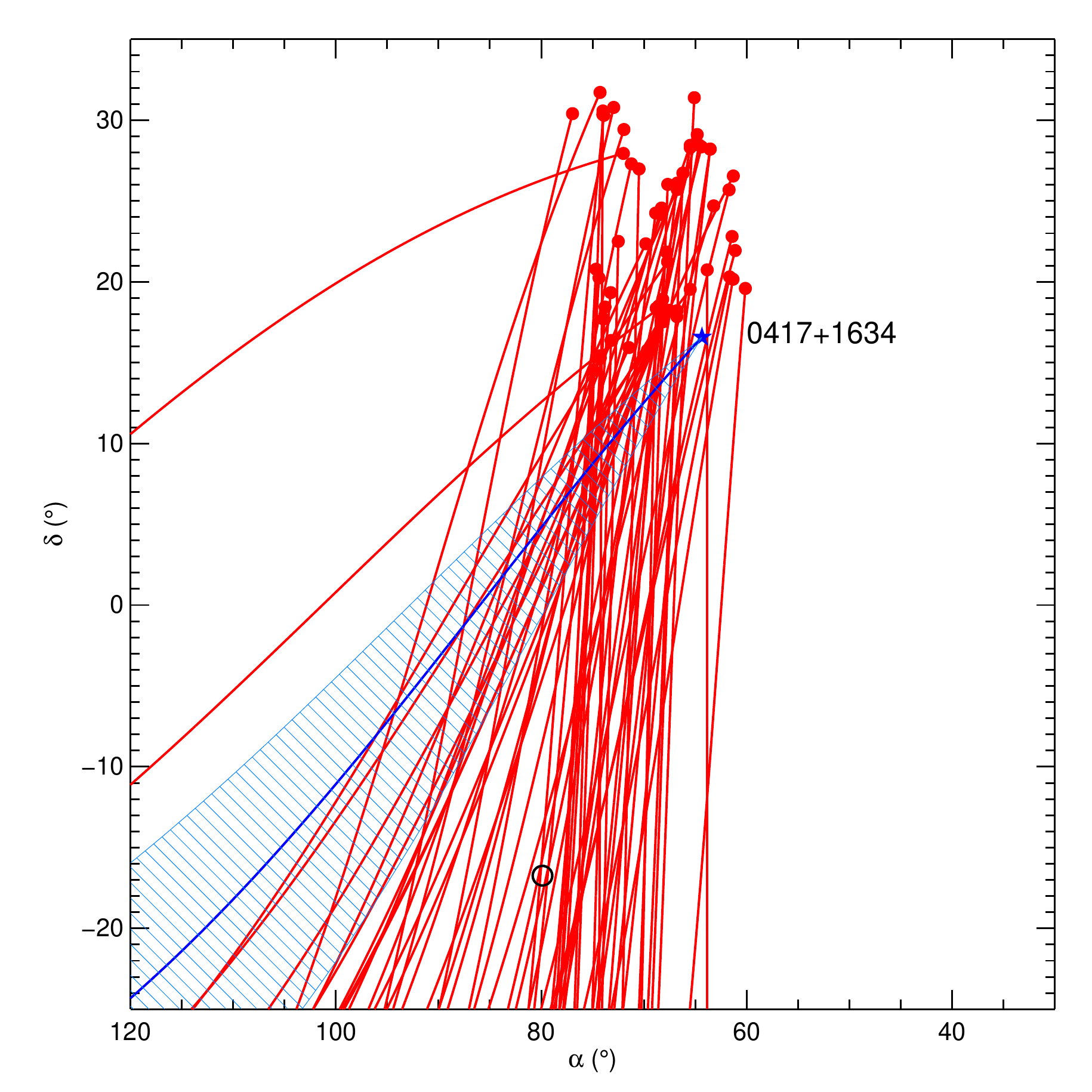}
  \end{subfigure}
  \caption{Proper motion path for SIMPJ0417$+$1634
    (blue, with $1\sigma$ uncertainty) and for the Taurus-Auriga members taken from \citet[red]{fri97}
    with proper motion from \citet{fri97}, \citet{hog00},
    \citet{zac03}, \citet{van07}, \citet{ito08}, and \citet{zac12}. The
    CP (black circle) of a core Taurus moving group was
    defined in \citet{ber06}.}\label{f8}
\end{figure}

\clearpage

\begin{figure}
\centering
     \begin{subfigure}{\textwidth}
     \includegraphics[width=0.495\textwidth]{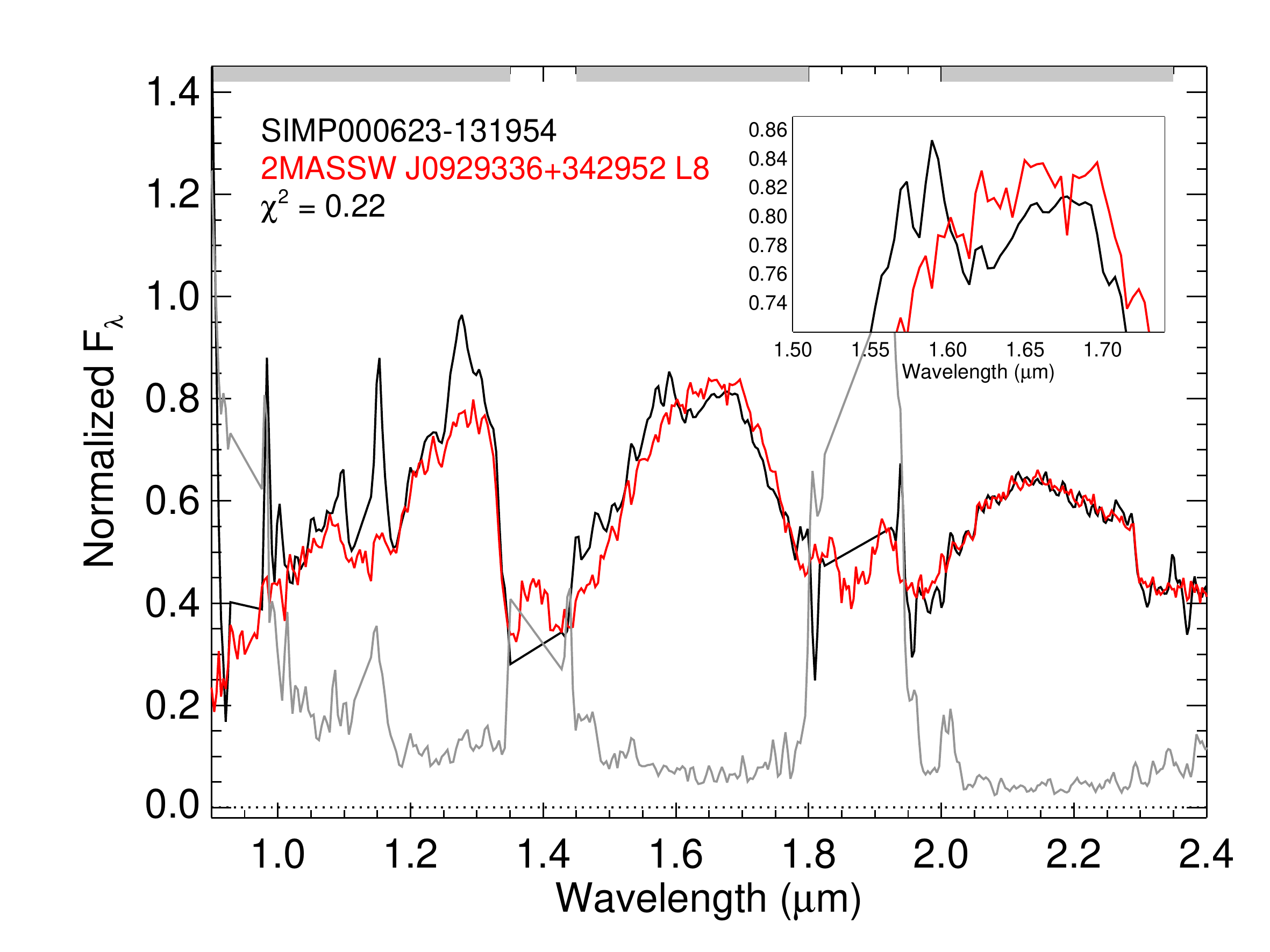}
     \includegraphics[width=0.495\textwidth]{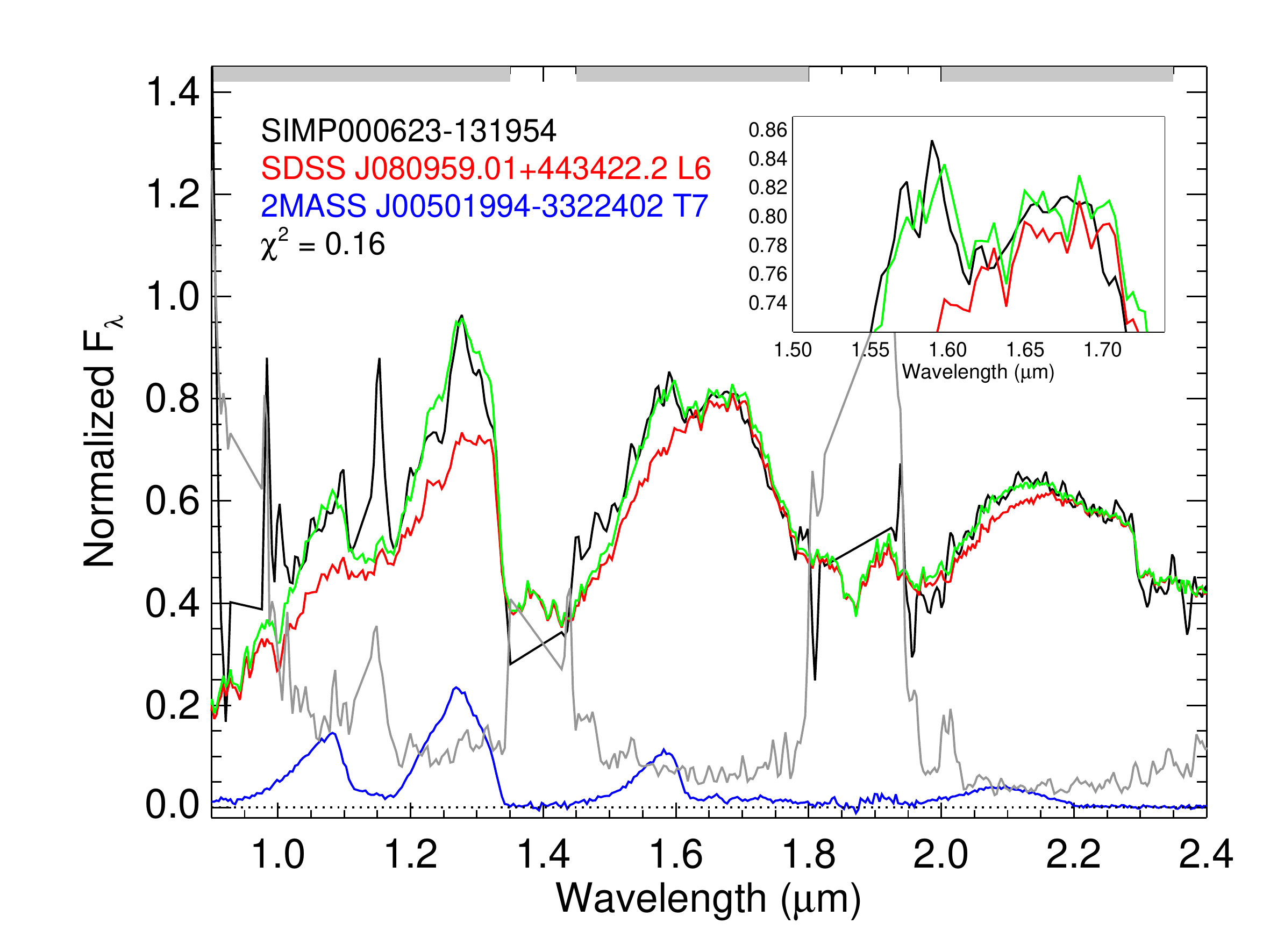}
     \caption{J0006$\relbar$1319 (L6$+$T7). Data for 2MASSW
      J0929336+342952, SDSS J080959.01+443422.2, and 2MASS
      J00501994-3322402 are from \citet{bur10}, A. Burgasser, and
      \citet{bur06}, respectively.}\label{f9a}
    \end{subfigure}
     \begin{subfigure}{\textwidth}
     \includegraphics[width=0.495\textwidth]{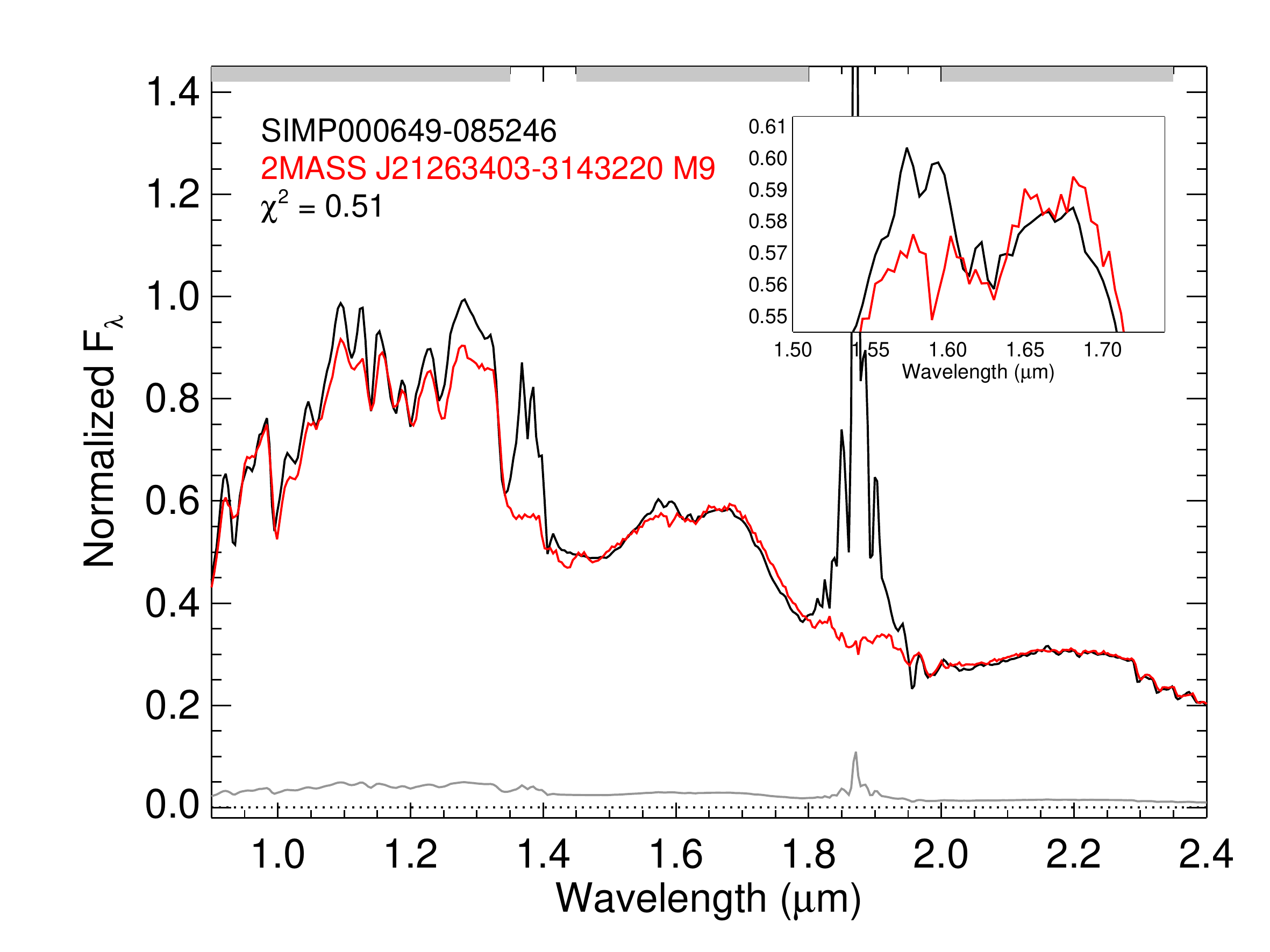}
     \includegraphics[width=0.495\textwidth]{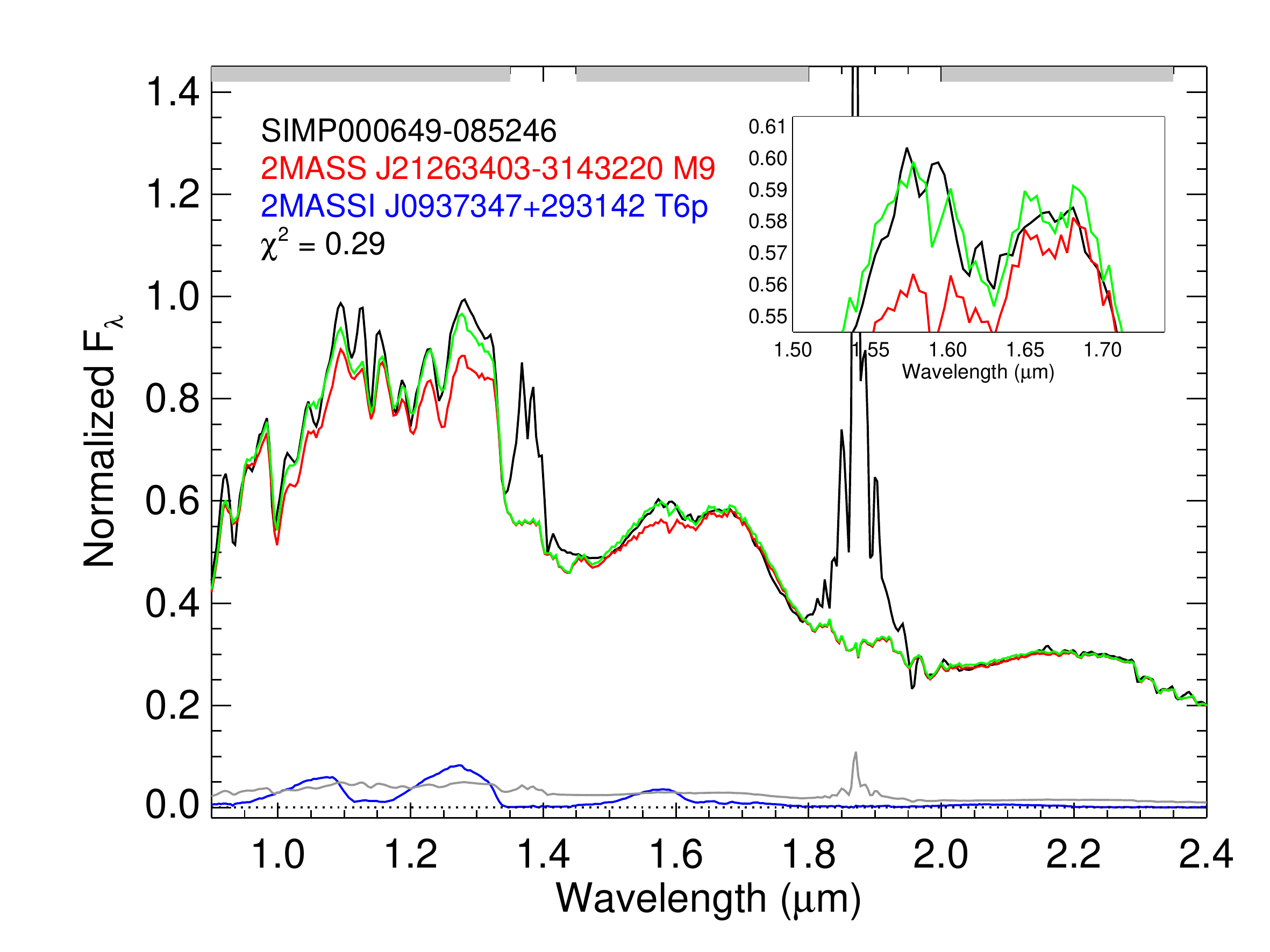}
     \caption{J0006$\relbar$0852 (M9$+$T6). Data for 2MASSW
      J21263403-3143220 and 2MASS J0937347+293142 are from A. Burgasser and
      \citet{bur06}, respectively.}\label{f9b}
    \end{subfigure}
\caption{Best fits to single (left) and binary (right) templates for our
 binary candidates. The black line shows the candidate spectrum. For
 the single fits, the red line is the best single template. For the
 binary fits, the green line is the best binary template, which is the
 addition of the red (primary) and blue (secondary) lines. The gray
 line represents the uncertainty in the candidate spectrum. The gray
 horizontal bars at the top of the figures mark the parts of the
 spectrum being fit, while water absorption dominates the gaps. Even
 with the high uncertainties, a significant fitting improvement is
 observed on the binary fits as compared to the single fits,
 particularly around the methane absorption feature centered at
 1.63$\,\mu$m (see inset).}\label{f9} 
\end{figure}
     
\begin{figure}
\ContinuedFloat
\centering
\begin{subfigure}{\textwidth}
     \includegraphics[width=0.495\textwidth]{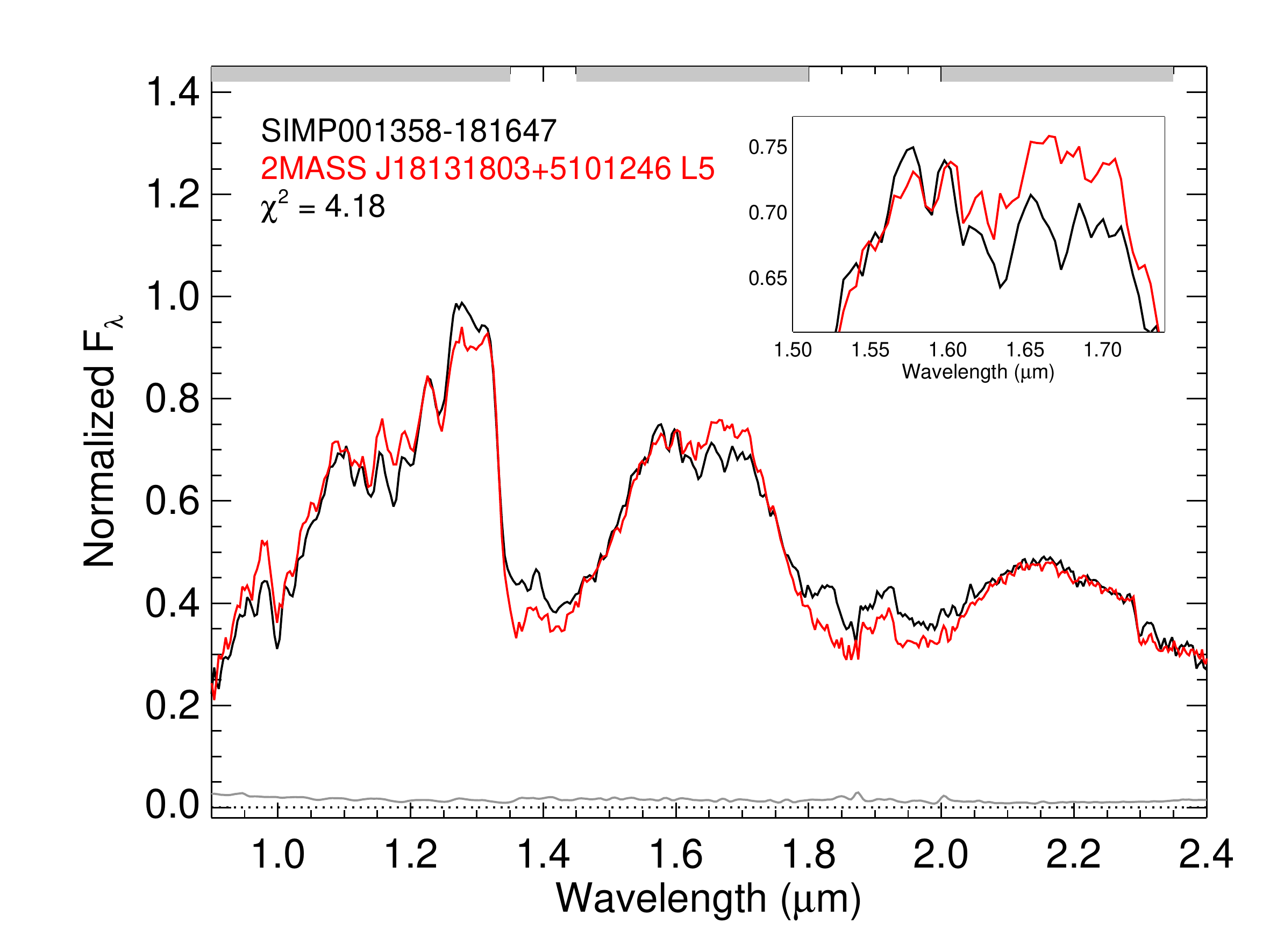}
     \includegraphics[width=0.495\textwidth]{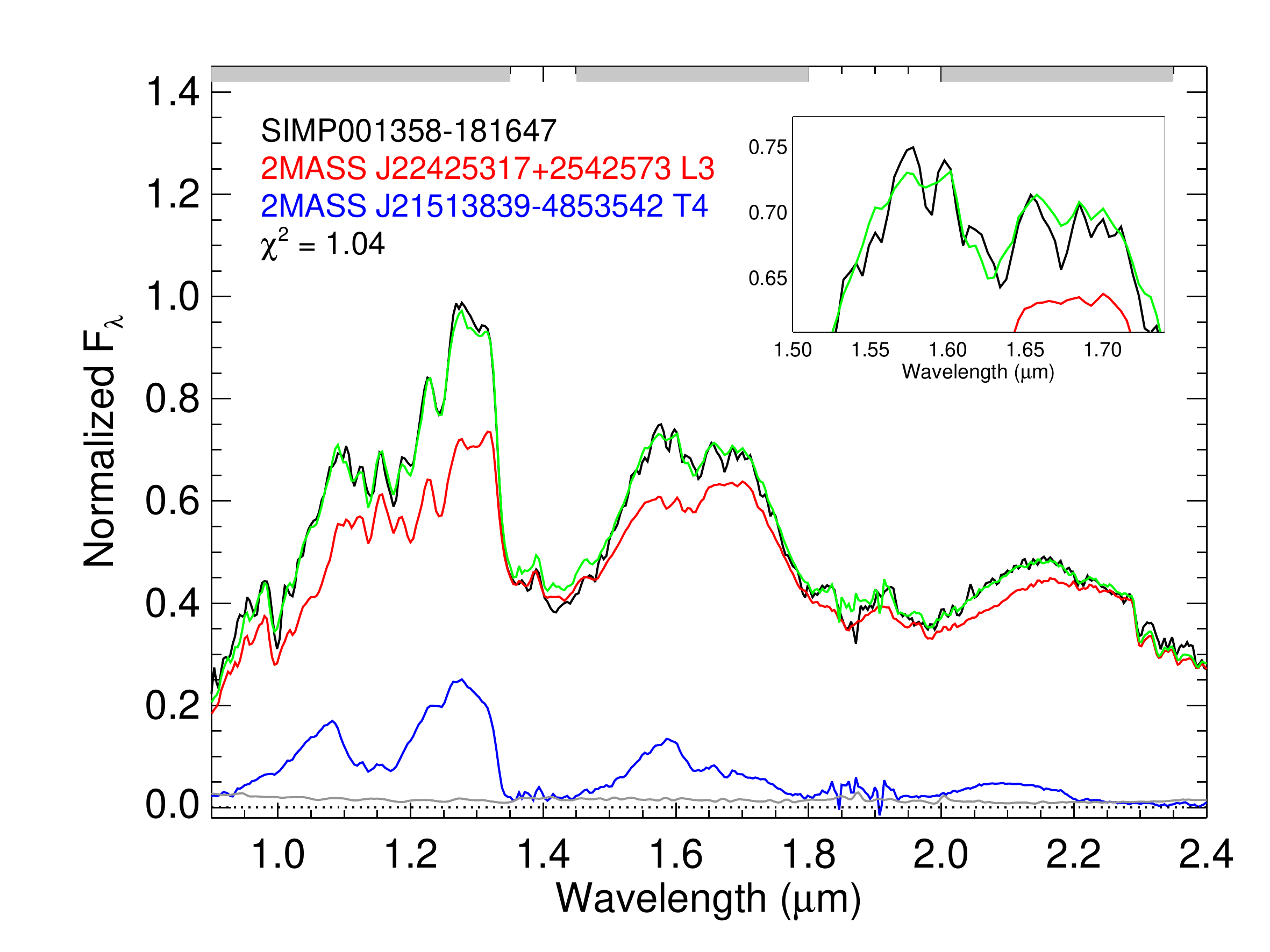}
     \caption{J0013$\relbar$1816 (L3$+$T4). Data for 2MASSW
      J18131803+5101246, SDSS J22425317+2542573, and 2MASS
      J21513839-4853542 are from \citet{kir10}, \citet{bur10}, and
      \citet{bur06b}, respectively.}\label{f9c}
    \end{subfigure}
     \begin{subfigure}{\textwidth}
     \includegraphics[width=0.495\textwidth]{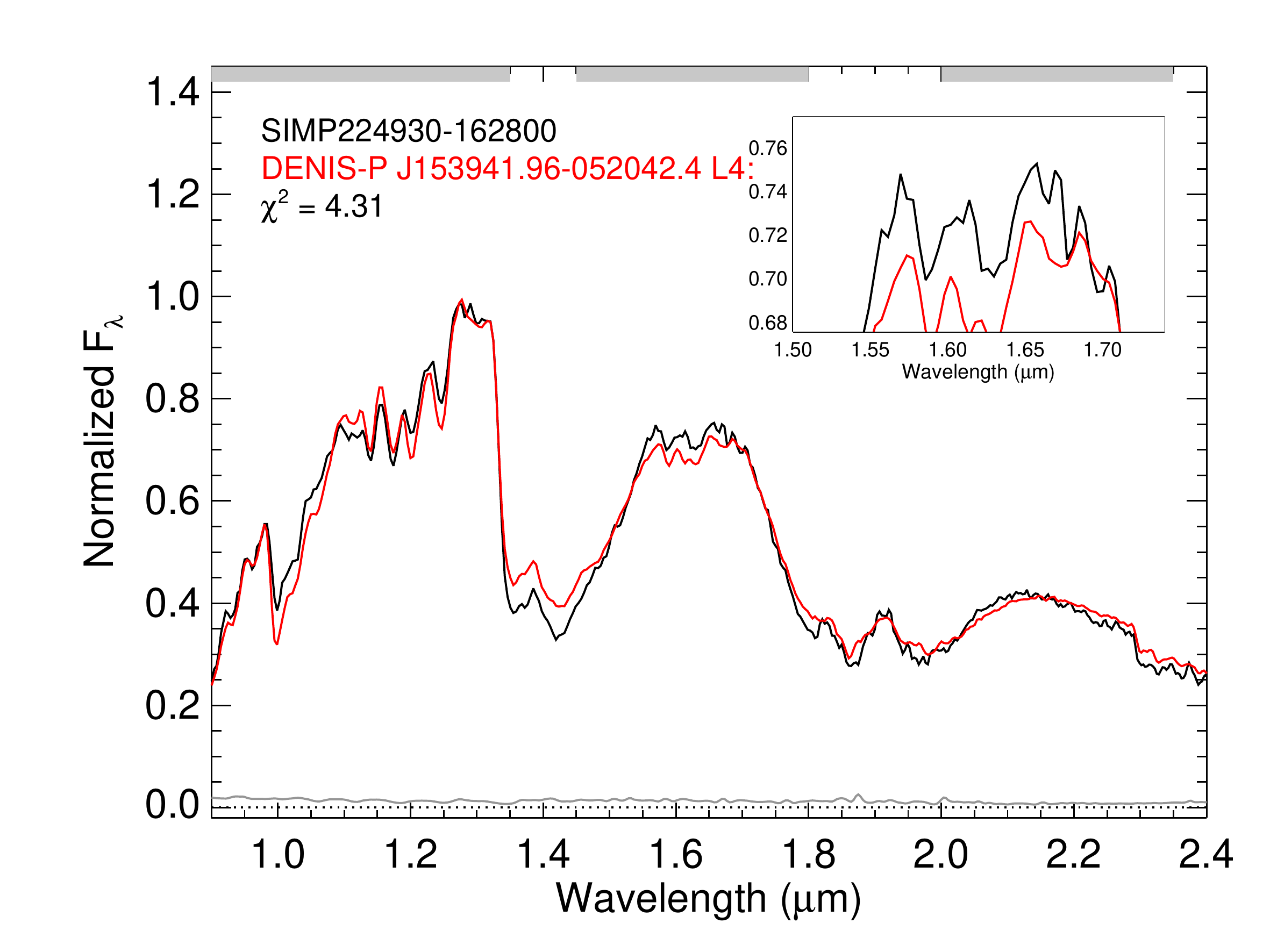}
     \includegraphics[width=0.495\textwidth]{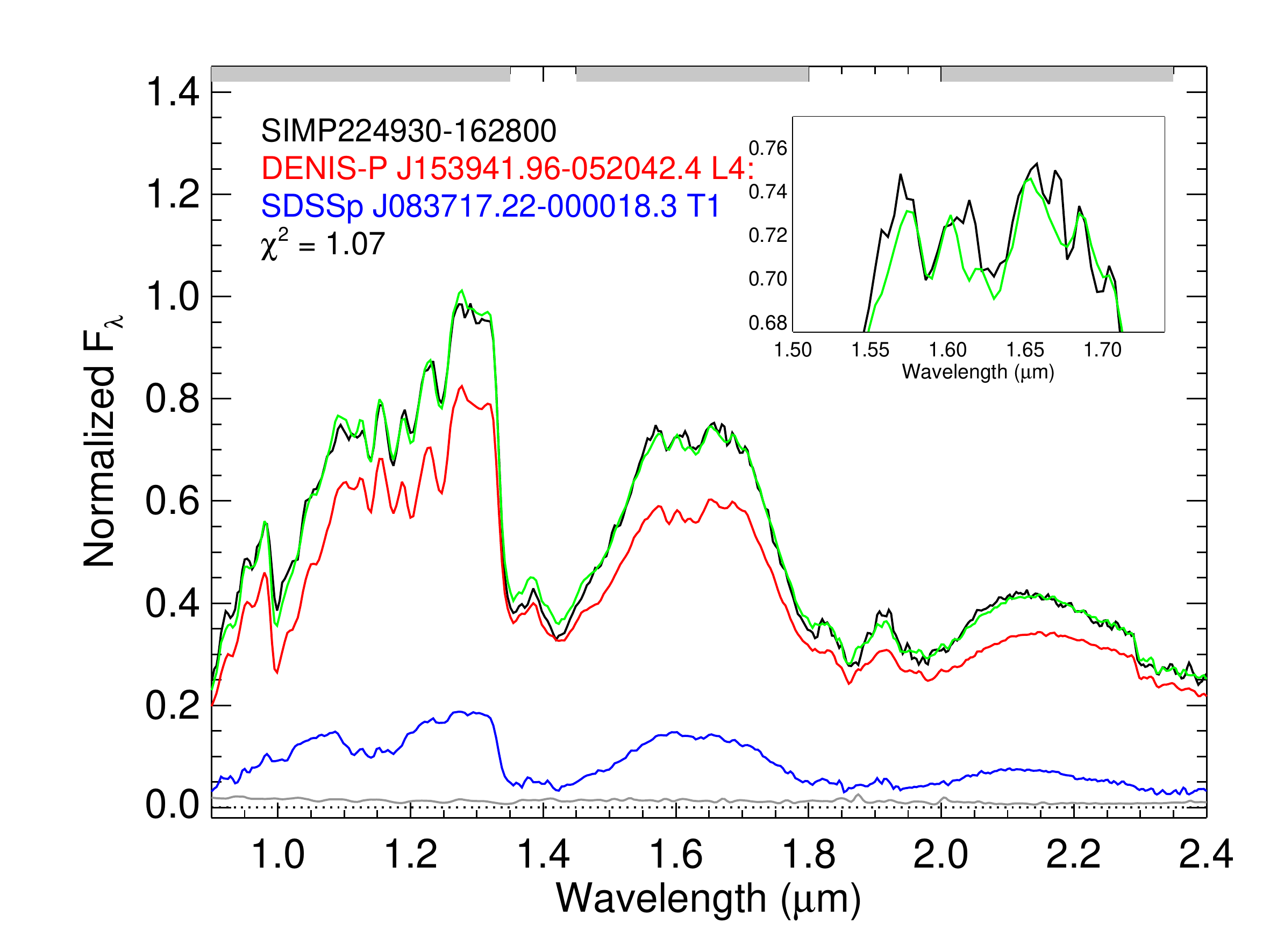}
     \caption{J2249$\relbar$1628 (L4$+$T1). Data for DENIS-P
      J153941.96-052042.4 and SDSSp J083717.22-000018.3 are from
      A. Burgasser and \citet{bur06b}, respectively.}\label{f9d}
    \end{subfigure}
\caption{(Continued)}
\end{figure}

\clearpage

\begin{center}
\renewcommand{\thefootnote}{\alph{footnote}}
\begin{scriptsize}
\setlength{\tabcolsep}{0.0in}
\setlength\LTleft{0in}
\begin{longtable}{@{\extracolsep{0.08in}}cccccccccc}
\multicolumn{10}{c}
{\tablename\ \thetable\ -- Log of SIMP spectroscopic follow-up} \label{t1}\\
\hline\hline \\[-4ex]
SIMP name & Facility/ &
Obs. Date & Int. Time & Mode/ & Standard  & Standard & Weather & Airmass & Comment\tablenotemark{d} \\[-1.5ex]
& Instr. & (UT) & (m) & Slit & Star & Type\tablenotemark{a} & Conditions\tablenotemark{c} & \\
\hline
\endfirsthead
\multicolumn{10}{c}
{\tablename\ \thetable\ -- \textit{Continued}}\\
\hline\hline \\[-4ex]
SIMP name & Facility/ &
Obs. Date & Int. Time & Mode/ & Standard & Standard & Weather & Airmass & Comment\tablenotemark{d} \\[-1.5ex]
& Inst. & (UT) & (m) & Slit & Star & Type\tablenotemark{a} & Conditions\tablenotemark{c} & \\
\hline
\endhead
\hline \\
\endfoot
\hline
\endlastfoot
SIMP J00044599$\relbar$2007410 & G-N/NIRI & 2008 Sep 12 & 25/0/0\tablenotemark{b} & Grism/0\arcsecpoint7 & HD 9399 & A1V & \nodata & 1.32 & \nodata \\
\nodata            & G-N/NIRI & 2008 Sep 30 & 0/7/7\tablenotemark{b} & Grism/0\arcsecpoint7 & HD 9399 & A1V & \nodata & 1.31 & \nodata \\
SIMP J00062287$\relbar$1319503 & OMM/SIMON & 2006 Sep 11 & 40 & Prism/1\arcsecpoint1 & HD 9930 & A1V & Good & 1.98 & 2 \\
\nodata            & G-S/GNIRS & 2007 Jan 06 & 31 & XD/0\arcsecpoint3 & HD 822 & A1V & \nodata & 1.53 & \nodata \\
SIMP J00063158$\relbar$2158225 & G-N/NIRI & 2008 Sep 14 & 20/7/14\tablenotemark{b} & Grism/0\arcsecpoint7 & HD 214172 & A1V & \nodata & 1.48 & \nodata \\
SIMP J00064916$\relbar$0852457 & G-S/GNIRS & 2006 Nov 30 & 18 & XD/0\arcsecpoint3 & HD 1160 & A0V & \nodata & 1.12 & \nodata \\
SIMP J00081285$+$0806441 & OMM/SIMON & 2006 Sep 12 & 12 & Prism/1\arcsecpoint1 & HD 18571 & A0V & Clouds & 1.36 & \nodata \\
SIMP J00132128$+$0841006 & OMM/SIMON & 2006 Sep 10 & 28 & Prism/1\arcsecpoint1 & HD 1160 & A0V & Good & 1.29 & 1,2 \\
\end{longtable}
\begin{flushleft}
\footnotemark[0]{}\\[-12ex]
\footnotemark[0]{\textbf{Note.}}\\
\footnotemark[1]{AT:~Algol type, HB:~Horizontal Branch Star, BL:~beta Lyr type.}\\
\footnotemark[2]{Individual integration times in the $J/H/K$ bands.}\\
\footnotemark[3]{BS:~Bright Sky, HH:~High Humidity, PS:~Poor Seeing.}\\
\footnotemark[4]{1: The target moved out of the slit at some positions on
  the nod pattern due to inaccurate pointing and instrument/telescope
  flexure. 2: The $K$ band is saturated.}
\end{flushleft}
\end{scriptsize}
\renewcommand{\thefootnote}{\arabic{footnote}}
\end{center}

\clearpage

\begin{center}
\renewcommand{\thefootnote}{\alph{footnote}}
\begin{scriptsize}
\setlength{\tabcolsep}{0.0in}
\setlength\LTleft{0in}
\begin{longtable}{@{\extracolsep{0.08in}}cccccc}
\multicolumn{6}{c}
{\tablename\ \thetable\ -- Log of SIMP photometric follow-up at OMM} \label{t2}\\
\hline\hline \\[-4ex]
Object & Instrument & Obs. Date & Int. Time & Weather & Airmass \\[-1.5ex]
& & (UT) & (m)\tablenotemark{a} & Conditions\tablenotemark{b} \\
\hline
\endfirsthead
\multicolumn{6}{c}
{\tablename\ \thetable\ -- \textit{Continued}}\\
\hline\hline \\[-4ex]
Object & Instrument & Obs. Date & Int. Time & Weather & Airmass \\[-1.5ex]
& & (UT) & (m)\tablenotemark{a} & Conditions\tablenotemark{b} \\
\hline
\endhead
\hline \\
\endfoot
\hline
\endlastfoot
J0006$\relbar$1319 & SIMON & 2008 Jan 26 & 7.5/8/8 & Good & 2.42 \\
J0307$+$0852 & SIMON & 2008 Jan 27 & 18/14/16 & Good & 1.25 \\
J0400$\relbar$1322 & SIMON & 2008 Jan 28 & 18/12/14 & Good & 1.93 \\
J0423$+$1212 & SIMON & 2008 Jan 31 & 18/8/18 & Clouds & 1.23 \\
J0429$+$0607 & SIMON & 2008 Jan 28 & 14/11.7/14 & Good & 1.32 \\
J0430$+$1035 & SIMON & 2008 Jan 28 & 18/15/16 & Good & 1.37 \\
J0552$+$0210 & SIMON & 2008 Jan 24 & 18/18/18 & Fog & 1.39 \\
J0851$+$1043 & SIMON & 2008 Jan 24 & 14/14/18 & Clouds & 1.40 \\
\end{longtable}
\begin{flushleft}
\footnotemark[0]{}\\[-12ex]
\footnotemark[0]{\textbf{Note.}}\\
\footnotemark[1]{Individual integration times in the $J/H/K$ band.}\\
\footnotemark[2]{PS:~Poor Seeing}\\
\end{flushleft}
\end{scriptsize}
\renewcommand{\thefootnote}{\arabic{footnote}}
\end{center}

\begin{center}
\begin{landscape}
\renewcommand{\thefootnote}{\alph{footnote}}
\begin{scriptsize}
\setlength{\tabcolsep}{0.0in}
\setlength\LTleft{0in}
\begin{longtable}{@{\extracolsep{0.08in}}cccccccccccc}
\multicolumn{12}{c}
{\tablename\ \thetable\ -- Spectral Indices} \label{t3}\\
\hline\hline \\[-4ex]
\multirow{2}{*}{Object} & \multirow{2}{*}{Inst.\tablenotemark{a}} & Median & \multicolumn{4}{c}{\citet{all13}} &
\multicolumn{5}{c}{\citet{bur06b}} \\
\cline{4-7}\cline{8-12}
& & Type & H$_{2}$O & H$_{2}$OD & H$_{2}$O-1 & H$_{2}$O-2 &
H$_{2}$O-$J$ & H$_{2}$O-$H$ &
CH$_{4}$-$J$ & CH$_{4}$-$H$ &
CH$_{4}$-$K$ \\
\hline
\endfirsthead
\multicolumn{12}{c}
{\tablename\ \thetable\ -- \textit{Continued}}\\
\hline\hline \\[-4ex]
\multirow{2}{*}{Object} & \multirow{2}{*}{Inst.\tablenotemark{a}} & Median & \multicolumn{4}{c}{\citet{all13}} &
\multicolumn{5}{c}{\citet{bur06b}} \\
\cline{4-7}\cline{8-12}
& & Type & H$_{2}$O & H$_{2}$OD & H$_{2}$O-1 & H$_{2}$O-2 &
H$_{2}$O-$J$ & H$_{2}$O-$H$ &
CH$_{4}$-$J$ & CH$_{4}$-$H$ &
CH$_{4}$-$K$ \\
\hline
\endhead
\hline \\
\endfoot
\hline
\endlastfoot
J0004$\relbar$2007 & 3 & L1.5$\pm$1.5 & 1.345 (L3.9) & 1.008 (\nodata) & 0.667 (L1.0) & 0.848 (L0.4) & 0.916 (L1.0) & 0.790 (L2.9) & 0.838 (\nodata) & 1.084 (\nodata) & 1.051 (L2.3) \\
J0006$\relbar$1319 & 1 & L7.0$\pm$1.0 & 1.208 ([L1.0]) & 0.680 (L7.6) & 0.577 (L3.5) & 0.943 ([M7.6]) & 0.718 (L6.7) & 0.683 (L7.1) & 0.911 (\nodata) & 1.117 ([T0.7]) & 1.098 ([L0.1]) \\
\nodata      & 2 & L6.5$\pm$2.0 & 1.382 (\nodata) & 0.626 (\nodata) & 0.497 (\nodata) & 0.846 ([L0.5]) & 0.855 (L2.5) & 0.705 (L6.3) & 0.703 (\nodata) & 0.986 ([T1.0]) & 0.929 (L6.6) \\
J0006$\relbar$2158 & 3 & L0.5$\pm$1.5 & 1.187 (L0.6) & 1.105 (\nodata) & 0.962 (\nodata) & 0.910 (M8.6) & 0.967 (L0.0) & 0.813 (L2.0) & 0.923 (\nodata) & 0.936 ([T1.3]) & 0.981 ([L5.0]) \\
J0006$\relbar$0852 & 2 & L1.0$\pm$1.0 & 1.177 (L0.4) & 0.958 (L1.2) & 0.639 (L1.8) & 0.765 (\nodata) & 0.916 (L1.0) & 0.836 (L1.1) & 0.838 (\nodata) & 0.974 ([T1.1]) & 1.044 (L2.6) \\
J0008$+$0806 & 1 & L2.5$\pm$2.5 & 1.334 (L3.6) & 1.001 (M9.8) & 0.598 (L2.9) & 0.900 (M8.8) & 0.282 ([T4.6]) & 0.717 (L5.9) & 0.962 (\nodata) & 0.944 ([T1.2]) & 1.057 (L2.1) \\
J0013$+$0841 & 1 & L0.5$\pm$2.0 & 1.153 (M9.8) & 0.895 (L2.8) & 0.756 (M8.1) & 0.820 (L1.2) & \nodata & 0.851 (L0.5) & \nodata & 1.114 ([T0.7]) & 1.171 (\nodata) \\
J0013$\relbar$1816 & 4 & L5.5$\pm$2.0 & 1.374 (\nodata) & 0.825 (L4.5) & 0.576 (L3.5) & 0.855 ([L0.2]) & 0.673 (L8.1) & 0.722 (L5.7) & 0.816 (\nodata) & 0.943 ([T1.2]) & 0.956 (L5.8) \\
\end{longtable}
\begin{flushleft}
\footnotemark[0]{}\\[-12ex]
\footnotemark[0]{\textbf{Note.}}\\
\footnotemark[0]{Spectral types in brackets are not used to calculate
  the median type.}\\
\footnotemark[1]{(1) OMM/SIMON; (2) G-S/GNIRS; (3) G-N/NIRI; (4) IRTF/SpeX.}
\end{flushleft}
\end{scriptsize}
\renewcommand{\thefootnote}{\arabic{footnote}}
\end{landscape}
\end{center}

\clearpage

\begin{center}
\newdimen\lchiffre
\newdimen\mchiffre
\setbox0=\hbox{\rm0}
\lchiffre=\wd0
\catcode`?=\active
\def?{\kern\lchiffre}
\setbox0=\hbox{\rm(}
\mchiffre=\wd0
\catcode`!=\active
\def!{\kern\mchiffre}
\renewcommand{\thefootnote}{\alph{footnote}}
\begin{footnotesize}
\setlength{\tabcolsep}{0.0in}
\setlength\LTleft{0in}
\begin{longtable}{@{\extracolsep{0.08in}}ccccccccc}
\multicolumn{9}{c}
{\tablename\ \thetable\ -- Targets Spectral Classification} \label{t4}\\
\hline\hline \\[-4ex]
\multirow{2}{*}{Object} & \multirow{2}{*}{Inst.\tablenotemark{a}} &
Obs. Index & Obs. Visual & Adopted NIR & Previous Opt. & Previous NIR & Ref. \\
& & Type & Type & Type & Type & Type & Opt/NIR \\
\hline
\endfirsthead
\multicolumn{8}{c}
{\tablename\ \thetable\ -- \textit{Continued}}\\
\hline\hline \\[-4ex]
\multirow{2}{*}{Object} & \multirow{2}{*}{Inst.\tablenotemark{a}} &
Obs. Index & Obs. Visual & Adopted NIR & Previous Opt. & Previous NIR & Ref. \\
& & Type & Type & Type & Type & Type & Opt/NIR \\
\hline
\endhead
\hline \\
\endfoot
\hline
\endlastfoot
J0004$\relbar$2007 & 3 & L1.5$\pm$1.5 & !M9$\pm$2 pec & L0:: pec$^2$ & \nodata & \nodata & \nodata \\
J0006$\relbar$1319 & 1 & L7.0$\pm$1.0 & ?L6$\pm$2 pec & L6:+T7: & \nodata & L5 pec\tablenotemark{b} & \nodata/26 \\
\nodata     & 2 & L6.5$\pm$2.0 & ?L6$\pm$2 pec & L6:+T7: & \nodata & L5 pec\tablenotemark{b} & \nodata/26 \\
J0006$\relbar$2158 & 3 & L0.5$\pm$1.5 & M8.5$\pm$2 pec & M9:: pec & \nodata & \nodata & \nodata \\
J0006$\relbar$0852 & 2 & L1.0$\pm$1.0 & !M9$\pm$1 pec & M9:+T6: pec & M9 & M8.5+T5\tablenotemark{c} & 3/2 \\
J0008$+$0806 & 1 & L2.5$\pm$2.5 & L0.5$\pm$2\hphantom{ pec}  & L0.5:: & \nodata & \nodata & \nodata \\
J0013$+$0841 & 1 & L0.5$\pm$2.0 & ?L1$\pm$2\hphantom{ pec} & L1: & \nodata & \nodata & \nodata \\
J0013$\relbar$1816 & 4 & L3.5$\pm$1.5 & L6.5$\pm$2 pec & L3:+T4:? & L1 & L5 pec\tablenotemark{h} & 28/29\\
\end{longtable}
\begin{flushleft}
\footnotemark[0]{}\\[-9ex]
\footnotemark[0]{\textbf{Note.} The adopted NIR types have uncertainties
  of $\pm$0.5 subtype, except for those listed with ``:'' ($\pm$1 subtype)
  or ``::'' ($\geq\pm$1.5 subtypes). pec$^1$ are possibly red objects; pec$^2$ are
  possibly blue objects; pec$^3$ are possibly binary objects.}\\
\footnotemark[1]{(1) OMM/SIMON; (2) G-S/GNIRS; (3) G-N/NIRI; (4) IRTF/SpeX.}\\
\footnotemark[2]{Possible unresolved $\sim$ L5+T5.5 binary noted in \citet{luh14b}.}\\
\footnotemark[3]{Probable very tight binary (M8+T5) part of a triple system
  discovered by \citet{bur12}.}\\
\footnotemark[4]{Very low quality spectrum.}\\
\footnotemark[5]{Resolved binary by \citet{duc13}, combined spectra
  obtained.}\\
\footnotemark[6]{Optical classification based on SDSS colors only.}\\
\footnotemark[7]{Possible unresolved T0+T2 binary from \citet{kel15}. No
peculiarities detected in our spectrum.}\\
\footnotemark[8]{Possible unresolved L+T binary from \citet{bes15}.}\\
\footnotemark[9]{Possible unresolved T2+T7.5 binary from \citet{bes15}.}\\
\footnotemark[10]{Listed as a candidate UCD by \citet{schn16} with NIR classification
  based on 2MASS and AllWISE photometry only.}\\
\footnotemark[0]{\textbf{References.} (1) \citet{art06}, (2) \citet{bur12}, (3) \citet{wes08},
(4) \citet{kir11}, (5) \citet{zha09}: (6) \citet{bur10},
(7) \citet{sch10}, (8) \citet{zha10}, (9) \citet{she09},
(10) \citet{mart10}, (11) \citet{mar13}, (12) \citet{kir10},
(13) \citet{mac13}, (14) \citet{dea11}, (15) \citet{gei11},
(16)  \citet{loo07}, (17) \citet{dea09}, (18) \citet{rei08},
(19) \citet{mau14}, (20) \citet{lod14}, (21) \citet{luh06},
(22) \citet{tho13}, (23) \citet{bar14}, (24) \citet{dea14},
(25) \citet{day13}, (26) \citet{luh14b}, (27) \citet{kel15},
(28) \citet{bar15}, (29) \citet{bes15}, (30) \citet{schn16}.}\\
\end{flushleft}
\end{footnotesize}
\renewcommand{\thefootnote}{\arabic{footnote}}
\end{center}

\begin{center}
\begin{landscape}
\renewcommand{\thefootnote}{\alph{footnote}}
\begin{tiny}
\setlength{\tabcolsep}{0.0in}
\setlength\LTleft{-0.7in}
\begin{longtable}{@{\extracolsep{0.07in}}ccccccccccccccccccc}
\multicolumn{19}{c}
{\tablename\ \thetable\ -- Properties of observed UCDs} \label{t5}\\
\hline\hline \\[-4ex]
\multirow{2}{*}{Object} & SpT & $\mu_{\rm \alpha}\cos\delta$ & $\mu_{\rm \delta}$ &
\multirow{2}{*}{Epochs\tablenotemark{a}} & $V$\textsubscript{tan} &
$d$\textsubscript{phot}\tablenotemark{b} & \multicolumn{2}{c}{SDSS} & \multicolumn{2}{c}{MegaCam} & \multicolumn{3}{c}{2MASS} &
\multicolumn{3}{c}{MKO} & \multicolumn{2}{c}{{\it WISE}}   \\
\cline{8-9}\cline{10-11}\cline{12-14}\cline{15-17}\cline{18-19}
& (NIR) & (mas$\,$yr$^{-1}$) & (mas$\,$yr$^{-1}$) &&
(km$\,$s\textsuperscript{-1}) & (pc) & $i$ & $z$ &
$i$ & $z$ & $J$ & $H$ & $K_s$ & $J$ & $H$ &
$K$ & $W1$ & $W2$\\
\hline
\endfirsthead
\multicolumn{19}{c}
{\tablename\ \thetable\ -- \textit{Continued}}\\
\hline\hline \\[-4ex]
\multirow{2}{*}{Object} & SpT & $\mu_{\rm \alpha}\cos\delta$ & $\mu_{\rm \delta}$ &
\multirow{2}{*}{Epochs\tablenotemark{a}} & $V$\textsubscript{tan} &
$d$\textsubscript{phot}\tablenotemark{b} & \multicolumn{2}{c}{SDSS} & \multicolumn{2}{c}{MegaCam} & \multicolumn{3}{c}{2MASS} &
\multicolumn{3}{c}{MKO} & \multicolumn{2}{c}{{\it WISE}}   \\
\cline{8-9}\cline{10-11}\cline{12-14}\cline{15-17}\cline{18-19}
& (NIR) & (mas$\,$yr$^{-1}$) & (mas$\,$yr$^{-1}$) &&
(km$\,$s\textsuperscript{-1}) & (pc) & $i$ & $z$ &
$i$ & $z$ & $J$ & $H$ & $K_s$ & $J$ & $H$ &
$K$ & $W1$ & $W2$\\
\hline
\endhead
\hline \\
\endfoot
\hline
\endlastfoot
J23152198$+$2235417 & M6:: & 21.7 $\pm$ 29.6 & $\relbar$192.4 $\pm$ 31.3 & 1,2,3,4 & 219 $\pm$ 86 & 238.3 $\pm$ 41.0 & 19.91 $\pm$ 0.03 & 18.71 $\pm$ 0.02 & \nodata & \nodata & 16.74 $\pm$ 0.14 & 16.30 $\pm$ 0.18 & 15.91 $\pm$ 0.29 & \nodata & \nodata & \nodata & 15.71 $\pm$ 0.04 & 15.75 $\pm$ 0.13\\
J04253602$\relbar$1900022 & M6.5: & $\relbar$73.0 $\pm$ 10.5 & $\relbar$91.6 $\pm$ 11.2 & 1,2,4,6 & 72 $\pm$ 21 & 129.5 $\pm$ 21.1 & \nodata & \nodata & 20.48 $\pm$ 0.04 & 18.80 $\pm$ 0.04 & 16.26 $\pm$ 0.07 & 15.41 $\pm$ 0.07 & 15.21 $\pm$ 0.15 & \nodata & \nodata & \nodata & 14.88 $\pm$ 0.03 & 14.55 $\pm$ 0.05\\
J17553598$+$3618117 & M6.5:: & $\relbar$25.9 $\pm$ 30.2 & 37.6 $\pm$ 35.1 & 1,2,4 & 55 $\pm$ 65 & 254.8 $\pm$ 42.8 & \nodata & \nodata & \nodata & \nodata & 17.10 $\pm$ 0.21 & 16.10 $\pm$ 0.17 & 16.10 $\pm$ 0.29 & \nodata & \nodata & \nodata & 16.32 $\pm$ 0.05 & 16.02 $\pm$ 0.10\\
J04182001$+$1121129 & M8: & 182.2 $\pm$ 12.3 & $\relbar$82.4 $\pm$ 8.6 & 1,2,4,6 & 60 $\pm$ 14 & 63.7 $\pm$ 10.3 & \nodata & \nodata & 18.57 $\pm$ 0.03 & 17.17 $\pm$ 0.03 & 14.90 $\pm$ 0.04 & 14.13 $\pm$ 0.05 & 13.78 $\pm$ 0.04 & \nodata & \nodata & \nodata & 13.61 $\pm$ 0.03 & 13.38 $\pm$ 0.03\\
J21430506$\relbar$1544394 & M8: & 183.7 $\pm$ 21.4 & $\relbar$91.7 $\pm$ 21.3 & 1,2,4,6 & 123 $\pm$ 38 & 126.0 $\pm$ 20.9 & \nodata & \nodata & 20.56 $\pm$ 0.03 & 18.97 $\pm$ 0.04 & 16.40 $\pm$ 0.12 & 15.61 $\pm$ 0.13 & 15.39 $\pm$ 0.18 & \nodata & \nodata & \nodata & 15.01 $\pm$ 0.04 & 14.86 $\pm$ 0.08\\
J04144834$+$1529007 & M8.5: & 101.2 $\pm$ 14.5 & 51.1 $\pm$ 14.1 & 1,2,3,4,6 & 52 $\pm$ 18 & 96.5 $\pm$ 15.8 & 20.20 $\pm$ 0.03 & 18.36 $\pm$ 0.02 & 20.14 $\pm$ 0.03 & 18.64 $\pm$ 0.03 & 16.02 $\pm$ 0.08 & 15.56 $\pm$ 0.15 & 14.72 $\pm$ 0.12 & \nodata & \nodata & \nodata & 14.66 $\pm$ 0.03 & 14.40 $\pm$ 0.06\\
\end{longtable}
\begin{flushleft}
\footnotemark[0]{}\\[-12ex]
\footnotemark[0]{\textbf{Note.} pec$^1$ are possibly red objects; pec$^2$ are
 possibly blue objects; pec$^3$ are possibly binary objects.}\\
\footnotemark[1]{Epochs included in the measurements of proper motion:
(1) {\it WISE}; (2) 2MASS; (3) SDSS; (4) SIMP; (5) OMM; (6) CFHT.}\\
\footnotemark[2]{Unless otherwise noted, photometric distances are based on {\it WISE}
 $W2$ band and polynomial fits for $M_{W2}$ from
 \citet{dup12}. Binarie distances are calculated by adding the flux of both
 unresolved components.}\\
\footnotemark[3]{Photometric distance based on 2MASS $H$ band and
 polynomial fits for $M_{H}$ from \citet{dup12}.}\\
\footnotemark[4]{Resolved binary by \citet{duc13}. Rejected as a Hyades
 member by \citet{lod14}.}\\
\end{flushleft}
\end{tiny}
\end{landscape}
\end{center}

\end{document}